\documentclass[
11pt, 
english,
singlespacing, 
headsepline, 
]{MastersDoctoralThesis} 

\usepackage[utf8]{inputenc} 
\usepackage[T1]{fontenc} 
\usepackage{palatino}

\usepackage[backend=bibtex,style=authoryear,natbib=true,url=false,block=space,maxcitenames=3,isbn=false]{biblatex}
\renewbibmacro{in:}{}
\DeclareFieldFormat[article, inbook, incollection, inproceedings, misc, thesis, unpublished]{title}{#1}
\AtEveryBibitem{%
  \clearlist{language}%
}

\usepackage[autostyle=true]{csquotes} 

\usepackage{pdfpages}
\usepackage{amssymb}
\usepackage{amsmath}
\usepackage{xcolor}
\usepackage{subfig}
\usepackage{adjustbox}

\newcommand{\mrm}[1]{\mathrm{#1}}

\newcommand{\sub}[2]{#1_{\mrm{#2}}}

\newcommand\diff{\mathop{}\!\mathrm{d}}
\newcommand{\rp}{\textit{rp}}

\DeclareSIUnit{\nucleon}{nucleon}
\DeclareSIUnit{\erg}{erg}
\DeclareSIUnit{\mevnuc}{\MeV\per\nucleon}
\DeclareSIUnit{\Funits}{\erg\per\second\per\square\cm}
\DeclareSIUnit{\Lunits}{\erg\per\second}
\DeclareSIUnit{\flunits}{\erg\per\square\cm}
\DeclareSIUnit{\yunits}{\gram\per\square\cm}
\DeclareSIUnit{\mdotunits}{\gram\per\square\cm\per\second}
\DeclareSIUnit{\gunits}{\cm\per\square\second}

\newcommand{\msun}{\mrm{M_{\odot}}}

\newcommand{\kepler}{\textsc{Kepler}}
\newcommand{\mesa}{\textsc{MESA}}
\newcommand{\emcee}{\textsc{emcee}}

\newcommand{\shiva}{\textsc{SHIVA}}
\newcommand{\agile}{\textsc{AGILE}}
\newcommand{\pyburst}{\textsc{pyburst}}
\newcommand{\python}{\textsc{python}}
\newcommand{\chainconsumer}{\textsc{ChainConsumer}}

\newcommand{\minbar}{\textsc{MINBAR}}

\newcommand{\saxj}{SAX J1808.4$-$3658}
\newcommand{\gs}{GS 1826$-$238}
\newcommand{\fouru}{4U 1820$-$30}
\newcommand{\ngc}{NGC 6624}
\newcommand{\sixteen}{4U 1636$-$536}

\newcommand{\rxte}{\textit{RXTE}}
\newcommand{\beppo}{\textit{BeppoSAX}}
\newcommand{\integral}{\textit{INTEGRAL}}

\newcommand{\qnuc}{\sub{Q}{nuc}}
\newcommand{\qb}{\sub{Q}{b}}
\newcommand{\qbn}[1]{\sub{Q}{b,#1}}
\newcommand{\mdot}{\dot{m}}
\newcommand{\Mdot}{\dot{M}}
\newcommand{\mdotn}[1]{\sub{\mdot}{#1}}
\newcommand{\Mdotn}[1]{\sub{\Mdot}{#1}}
\newcommand{\mdotedd}{\sub{\mdot}{\eddsymb}}
\newcommand{\Mdotedd}{\sub{\Mdot}{\eddsymb}}
\newcommand{\cno}{Z_\mrm{CNO}}
\newcommand{\hyd}{X_0}
\newcommand{\hel}{Y_0}
\newcommand{\db}{d \sqrt{\xib}}
\newcommand{\xiratio}{\xip / \xib}

\newcommand{\nw}[1]{\sub{#1}{k}}
\newcommand{\gr}[1]{\sub{#1}{g}}
\newcommand{\nwsub}[2]{\sub{#1}{#2, k}}
\newcommand{\grsub}[2]{\sub{#1}{#2, g}}
\newcommand{\nsmass}{M = \SI{1.4}{\msun}}
\newcommand{\nsradius}{R = \SI{10}{km}}
\newcommand{\obslhood}{0}
\newcommand{\obs}[1]{\sub{#1}{\obssymb}}
\newcommand{\obssub}[2]{\sub{#1}{#2, \obssymb}}

\newcommand{\obssymb}{\infty}
\newcommand{\burstsymb}{b}
\newcommand{\perssymb}{p}
\newcommand{\eddsymb}{Edd}
\newcommand{\accsymb}{acc}

\newcommand{\xib}{\sub{\xi}{\burstsymb}}
\newcommand{\xip}{\sub{\xi}{\perssymb}}

\newcommand{\dt}{\Delta t}
\newcommand{\brate}{\nu}
\newcommand{\Lpre}{\sub{L}{pre}}
\newcommand{\Lpeak}{\sub{L}{peak}}
\newcommand{\Fpeak}{\sub{F}{peak}}
\newcommand{\yign}{\sub{y}{ig}}

\newcommand{\Lburst}{\sub{L}{\burstsymb}}
\newcommand{\Fburst}{\sub{F}{\burstsymb}}
\newcommand{\fluence}{\sub{f}{\burstsymb}}
\newcommand{\Eb}{\sub{E}{\burstsymb}}

\newcommand{\Lper}{\sub{L}{\perssymb}}
\newcommand{\Fper}{\sub{F}{\perssymb}}

\newcommand{\fluencep}{\sub{f}{\perssymb}}

\newcommand{\Ledd}{\sub{L}{\eddsymb}}
\newcommand{\Lacc}{\sub{L}{\accsymb}}
\newcommand{\Fedd}{\sub{F}{\eddsymb}}

\newcommand{\tpre}{t_\mrm{pre}}
\newcommand{\tstart}{t_\mrm{start}}
\newcommand{\tpeak}{t_\mrm{peak}}
\newcommand{\tend}{t_\mrm{end}}

\newcommand{\likelihood}{p(D | \theta)}
\newcommand{\prior}{p(\theta)}

\newcommand{\thomson}{\sub{\sigma}{T}}
\newcommand{\massp}{\sub{m}{p}}

\defcitealias{galloway_thermonuclear_2004}{G04}
\defcitealias{heger_models_2007}{H07}
\defcitealias{meisel_consistent_2018}{M18}
\defcitealias{galloway_thermonuclear_2017}{G17}

\thesistitle{Modelling Thermonuclear X-ray Bursts on Accreting Neutron Stars} 
\supervisor{Prof. Alexander \textsc{Heger}} 
\supervisortwo{Assoc. Prof. Duncan \textsc {Galloway}} 
\examiner{} 
\degree{Doctor of Philosophy} 
\author{Zac \textsc{Johnston}} 
\addresses{} 

\subject{Astrophysics} 
\university{\href{}{Monash University}} 
\department{\href{}{School of Physics and Astronomy}} 
\group{\href{}{ }} 
\faculty{\href{}{Faculty of Science}} 

\hypersetup{pdftitle=\ttitle} 
\hypersetup{pdfauthor=\authorname} 
\hypersetup{pdfkeywords=\keywordnames} 

\begin{document}

\frontmatter 
\pagestyle{plain} 

\begin{titlepage}
\begin{center}

\textsc{\LARGE \univname}\\[1.5cm] 
\textsc{\Large Doctoral Thesis}\\[0.5cm] 

\HRule \\[0.4cm] 
{\huge \bfseries \ttitle}\\[0.4cm] 
\HRule \\[1.5cm] 

\begin{minipage}{0.4\textwidth}
    \begin{flushleft} \large
        \emph{Author:}\\
        {\authorname} 
    \end{flushleft}
\end{minipage}
\begin{minipage}{0.5\textwidth}
    \begin{flushright} \large
        \emph{Supervisors:} \\
        \href{https://2sn.org/}{\supname} \\ 
        \href{http://users.monash.edu.au/~dgallow/cgi-bin/blosxom.cgi}{\supnametwo} 
    \end{flushright}
\end{minipage}\\[3cm]
 
\large \textit{A thesis submitted in fulfilment of the requirements\\ for the degree of \degreename}\\
\vfill
\facname\\[0.1cm]
\deptname\\ 
\vfill
{\large 16 February 2020}\\ 
\vfill
\includegraphics[width=4cm]{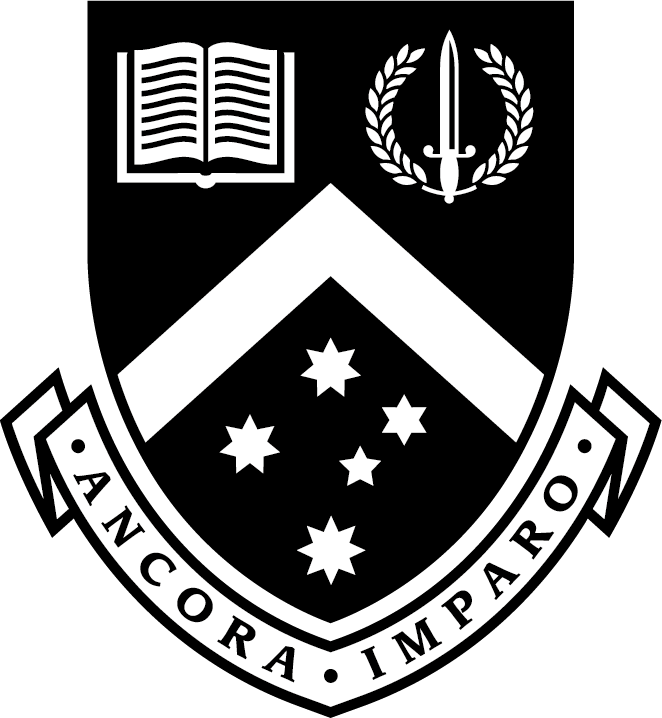} 
\vfill
\end{center}
\end{titlepage}
\cleardoublepage
\begin{declaration}
\addchaptertocentry{\authorshipname}
\section*{Declaration of Authorship}
\noindent I, Zac Johnston, hereby declare that this thesis contains no material which has been accepted for the award of any other degree or diploma at any university or equivalent institution and that, to the best of my knowledge and belief, this thesis contains no material previously published or written by another person, except where due reference is made in the text of the thesis.

This thesis includes one (1) original paper published in a peer reviewed journal and one (1) unpublished publication.
My contribution to these works is listed on the following page.
The core theme of the thesis is the computational modelling of type-I X-ray bursts.
The ideas, development and writing up of the papers in the thesis were the principal responsibility of myself, the student, working within the School of Physics and Astronomy, Monash University, under the supervision of Prof.\ Alexander Heger and Assoc.\ Prof.\ Duncan K.\ Galloway.
The inclusion of co-authors reflects the fact that the work came from active collaboration between researchers and acknowledges input into team-based research.

\vspace{0.5cm}
\noindent Signed,\\

\vspace{2cm}

\noindent Zac Johnston \quad\quad 10 October 2019

\vspace{1cm}
The undersigned hereby certify that the above declaration correctly reflects the nature and extent of the student's and co-authors' contributions to this work.

\vspace{0.5cm}
\noindent Signed,\\

\vspace{2cm}

\noindent Prof.\ Alexander Heger \quad\quad 10 October 2019

\cleardoublepage

\begin{adjustbox}{angle=90}
    \centering
    \begin{tabular}{|l|l|l|l|l|l|}
        \hline
        Chapter & Publication Title & Status & Student's    & Co-authors'  & Monash Student  \\
                &                   &        & Contribution & Contribution & Co-authors \\
        \hline
        5 & Simulating X-ray bursts            & Published & 80\%                  & Alexander Heger: 10\%   & No \\
          & during a transient accretion event &           & Computational models, & Feedback and discussion &    \\
          &                                    &           & data analysis,        & Duncan Galloway: 10\%   &    \\
          &                                    &           & writing manuscript.    & Feedback and discussion &    \\
        \hline
        6 & Multi-epoch X-ray burst modelling:      & Submitted & 80\%                  & Alexander Heger: 10\%   & No \\
          & MCMC with large grids of 1D simulations &           & Computational models, & Feedback and discussion. &    \\
          &                                         &           & data analysis,        & Duncan Galloway: 10\%   &    \\
          &                                         &           & writing manuscript.    & Feedback and discussion. &    \\
        \hline
    \end{tabular}
\end{adjustbox}

\end{declaration}

\cleardoublepage

  \begin{abstract}
\addchaptertocentry{\abstractname}
In low-mass X-ray binaries, the accretion of stellar material onto a neutron star can fuel unstable thermonuclear flashes known as Type I X-ray bursts.
In a matter of seconds, the thin shell of hydrogen and/or helium is converted into heavier elements through nuclear fusion, heating the envelope to $\sim \SI{e9}{K}$.
The burst of thermal emission, dominated by X-rays, lasts $\approx 10$ -- $\SI{100}{s}$ and is observable with satellite-based X-ray telescopes.
The properties of the burst reflect the local conditions of the neutron star surface.
Simulating these events using computational models can provide valuable information about the nature of the accreting system.
Measuring neutron star properties, especially the mass and radius, has been a longstanding objective in astrophysics because it can constrain the equation of state of dense nuclear matter.
One-dimensional (1D) astrophysics codes with large nuclear reaction networks are the current state-of-the-art for simulating X-ray bursts.
These codes can track the evolution of isotopes through thousands of nuclear reaction pathways, to predict the released nuclear energy and final composition of the ashes.
In this thesis, I make extensive use of \kepler{}, a 1D code at the forefront of these efforts.
I first present improvements to the setup and analysis of \kepler{} burst models.
By accounting for nuclear heating in the initial conditions, I shorten the thermal burn-in time, thereby reducing computational expense and producing more consistent burst trains.
To model bursts fueled by transient accretion events, I perform the first such simulations with fully time-dependent accretion rates.
Building upon previous efforts to model the ``Clocked Burster'', \gs{}, I precompute a grid of 3840 simulations and sample the interpolated results using Markov Chain Monte Carlo (MCMC) methods.
By comparing the predictions to multi-epoch observations, I obtain posterior probability distributions for the system parameters.
I then extend these MCMC methods to the pure-helium burster, \fouru{}, using a grid of 168 simulations.
Finally, I discuss potential improvements for future studies, to further develop the computational modelling of accreting neutron stars.

\end{abstract}

  \begin{acknowledgements}
\addchaptertocentry{\acknowledgementname} 

This research was completed within the School of Physics and Astronomy, Monash University, and the Monash Centre for Astrophysics (MoCA).

This research was supported in part by an Australian Government Research Training Program (RTP) Scholarship.

This research was supported in part by the National Science Foundation under Grant No. PHY-1430152 (JINA Center for the Evolution of the Elements).

This work used the astrophysics code \kepler, which is supported by an Australian Research Council (ARC) Future Fellowship (FT120100363).

This work uses preliminary analysis results from the Multi-INstrument Burst ARchive (MINBAR), which has benefited from support by the Australian Academy of Science's Scientific Visits to Europe program, and the Australian Research Council's Discovery Projects (project DP0880369) and Future Fellowship (project FT0991598) schemes.
The MINBAR project has also received funding from the European Union's Horizon 2020 Programme under the AHEAD project (grant agreement no.\ 654215).

This research was supported in part by the Monash eResearch Centre and eSolutions-Research Support Services through the use of the Mon\-ARCH HPC Cluster.

This work was supported in part by Michigan State University through computational resources provided by the Institute for Cyber-Enabled Research (ICER).

This work was performed in part on the OzSTAR national facility at Swinburne University of Technology.
OzSTAR is funded by Swinburne University of Technology and the National Collaborative Research Infrastructure Strategy (NCRIS).

Parts of this work were completed during a six-month research visit to Michigan State University.

This work benefited from student travel support provided by the Astronomical Society of Australia.

I would like to thank my supervisors, Alexander Heger and Duncan Galloway, for their immeasurable feedback, support, and encouragement.

For useful discussions, I would like to thank Adam Jacobs, Laurens Keek, Hendrik Schatz, Ed Brown, Frank Chambers, Zach Meisel, Andrew Casey, Adelle Goodwin, and countless others.

I thank my office mates and close friends, Conrad Chan, David Liptai, and Hayley Macpherson, for keeping me sane.

I thank Belinda Jude for her crucial support over the years.

I thank my parents, Carolyn and Peter Johnston, for their love and support since the very beginning, and for their unwavering encouragement throughout my studies.

I thank Sara Hugentobler, for her love and support.
I couldn't have done it without you.

\end{acknowledgements}

\tableofcontents

  \begin{abbreviations}{ll} 
\addchaptertocentry{\abbrevname}

\textbf{1D} & 1-Dimensional \\
\textbf{2D} & 2-Dimensional \\
\textbf{3D} & 3-Dimensional \\
\textbf{AMXP} & Accreting Millisecond X-ray Pulsar \\
\textbf{GR} & General Relativity \\
\textbf{LMXB} & Low Mass X-ray Binary \\
\textbf{MCMC} & Markov Chain Monte Carlo \\
\textbf{NICER} & Neutron Star Interior Composition Explorer \\
\textbf{NS} & Neutron Star \\
\textbf{PNS} & Proto-Neutron Star \\
\textbf{PRE} & Photospheric Radius Expansion \\
\textbf{RXTE} & Rossi X-ray Timing Explorer \\
\textbf{XRB} & X-Ray Burst \\

\end{abbreviations}


  \begin{symbols}{lll} 
\addchaptertocentry{\symbolsname}

$d$         & Distance                & \si{cm} \\
$\Eb$       & Burst energy            & \si{erg} \\
$\fluence$  & Burst fluence           & \si{\flunits} \\
$\fluencep$ & Persistent fluence      & \si{\flunits} \\
$\Fburst$   & Burst flux              & \si{\Funits} \\
$\Fedd$     & Eddington flux          & \si{\Funits} \\
$\Fper$     & Persistent flux         & \si{\Funits} \\
$\Fpeak$    & Peak burst flux         & \si{\Funits} \\
$g$         & Gravitational Acceleration & \si{\gunits} \\
$i$         & Inclination             & \si{deg} \\
$\Lacc$     & Accretion luminosity    & \si{\Lunits} \\
$\Lburst$   & Burst luminosity        & \si{\Lunits} \\
$\Ledd$     & Eddington luminosity    & \si{\Lunits} \\
$\Lper$     & Persistent luminosity   & \si{\Lunits} \\
$\Lpeak$    & Peak burst luminosity   & \si{\Lunits} \\
$M$         & Stellar mass            & \si{\msun} \\
$\Mdot$     & Global accretion rate   & $\Mdotedd$ \\
$\Mdotedd$  & Global Eddington accretion rate    & \si{\msun.yr^{-1}} \\
$\mdot$     & Local accretion rate    & $\mdotedd$ \\
$\mdotedd$  & Local Eddington accretion rate     & \si{\mdotunits} \\
$\qb$       & Crustal heating rate    & \si{\mevnuc} \\
$\qnuc$     & Nuclear heating rate    & \si{\mevnuc} \\
$R$         & Stellar radius          & \si{km} \\
$\tpre$     & Pre-burst time          & \si{s} \\
$\tstart$   & Burst start time        & \si{s} \\
$\tpeak$    & Burst peak time         & \si{s} \\
$\tend$     & Burst end time          & \si{s} \\
$\hyd$      & Hydrogen mass fraction  & -- \\
$\hel$      & Helium mass fraction    & -- \\
$y$         & Column depth            & \si{\yunits} \\
$\yign$     & Ignition column depth   & \si{\yunits} \\
$\cno$      & CNO mass fraction       & -- \\
$z$         & Gravitational redshift  & -- \\

\addlinespace 

$\alpha$    & Alpha ratio             & -- \\
$\dt$       & Recurrence time         & \si{s} \\
$\brate$    & Burst rate              & \si{s^{-1}} \\
$\kappa$    & Opacity                 & \si{cm^{2}.g^{-1}} \\
$\xi$       & GR-corrected radius ratio  & -- \\
$\xib$      & Burst anisotropy        & -- \\
$\xip$      & Persistent anisotropy   & -- \\
$\tau$      & Autocorrelation time    & -- \\
$\varphi$   & GR-corrected mass ratio & -- \\

\end{symbols}


\mainmatter 
\pagestyle{thesis} 

\chapter{Introduction and Background}
\label{ch:background}
Type I X-ray bursts are recurring thermonuclear flashes on accreting neutron stars.
They are distinct from Type II X-ray bursts, which are caused by sporadic accretion \citep{hoffman_dual_1978}.

Throughout this work, we will simply use ``bursts'' to refer to Type I X-ray bursts.
Previous reviews have been provided by \citet{lewin_x-ray_1993, bildsten_thermonuclear_1997, strohmayer_new_2006, galloway_thermonuclear_2017-1}.

In this chapter, we provide a brief overview of Type I X-ray bursts.
We describe the first detections and modelling efforts (Section~\ref{sec:background_history}), the catalogues of burst observations (Section~\ref{sec:background_observations}), the mechanisms of burst ignition (Section~\ref{sec:background_mechanism}), and previous works with computational burst modelling (Section~\ref{sec:background_modelling}).

In Chapter~\ref{ch:methods}, we describe the astrophysical code used throughout this thesis, \kepler{}, and the process of extracting burst properties from the models and predicting observable quantities.
In Chapter~\ref{ch:results}, we present improvements to the setup of \kepler{} burst models, and direct comparisons between \kepler{} and \mesa{} burst models.
In Chapter~\ref{ch:paper1}, we present the first burst simulations with time-dependent accretion rates.
In Chapter~\ref{ch:paper2}, we present the application of Markov chain Monte Carlo methods to large grids of \kepler{} models.
In Chapter~\ref{ch:4u1820}, we present the extension of these MCMC methods to a hydrogen-poor system, \fouru{}.
In Chapter~\ref{ch:conclusion}, we summarise the work presented in this thesis, and discuss potential improvements for future work.

\section{Early History}
\label{sec:background_history}
The thin shell instability was discovered by \citet{schwarzschild_thermal_1965}, in which thermonuclear burning restricted to a thin shell ($\Delta R \ll R$) can undergo a thermal runaway due to its inability to expand and cool.
\citet{hansen_steady-state_1975} modelled the thermonuclear stability of accreted hydrogen and helium on neutron stars, and found that most configurations were subject to the thin shell instability.
Concurrent to this work, Type I X-ray bursts were discovered independently by \citet{babushkina_hard_1975, grindlay_discovery_1976, belian_discovery_1976}, shortly followed with further detections by \citet{lewin_discovery_1976} and \citet{clark_recurrent_1976}.

\citet{woosley_-ray_1976, maraschi_x-ray_1977} independently attributed the newly-discovered bursts to a thermonuclear origin -- the unstable burning regime first uncovered by \citet{hansen_steady-state_1975}.
The thermonuclear model was further developed by \citet{joss_x-ray_1977, lamb_nuclear_1978, taam_nuclear_1978}.
Alongside these efforts, \citet{sugimoto_general_1978, fujimoto_asymptotic_1979} examined the analogous phenomenon of helium shell flashes on accreting white dwarfs.
Shortly thereafter, \citet{fujimoto_shell_1981} presented a foundational work placing the various ignition conditions of hydrogen and helium under a consistent framework.

The first detailed numerical models were produced by \citet{joss_helium-burning_1978}, who simulated helium bursts using a modified one-dimensional stellar evolution code, \textsc{ASTRA} \citep{rakavy_carbon_1967}.
The characteristic properties of the observed bursts were successfully reproduced, including the onset and decay timescales, overall luminosities, and recurrence times.
These models marked the beginning of the computational modelling of X-ray bursts (Section~\ref{sec:background_modelling}).

\section{Observational Catalogues}
\label{sec:background_observations}
New generations of X-ray telescopes were launched in the following decades, and the growing ``zoo'' of bursting behaviour offered multiple pathways into the study of thermonuclear burning on neutron stars.
Large catalogues of these observations enable population studies of bursting systems, and the identification of global patterns in bursting behaviour.
In particular, the collections of burst data serve as test beds for computational models.

An early collection was compiled from the literature by \citet{van_paradijs_relation_1988}, containing 45 bursts from ten systems.
They found that the burst duration was anti-correlated with the persistent flux, hinting at a common relationship between systems.

\citet{cornelisse_six_2003} compiled 1823 bursts from nine systems observed with \beppo, covering the years 1996 to 2002.
Their analysis found global transitions between bursting regimes.
When the persistent luminosity increased to $\Lper \approx \SI{2e37}{\Lunits}$, the sources transitioned from long, frequent, and quasi-periodic bursts, to short and infrequent bursts.
Above this limit, the bursts grew more irregular until ceasing altogether above $\Lper \approx \SI{5.5e37}{\Lunits}$.

\citet{galloway_thermonuclear_2008} provided another extensive catalogue, with observations from the X-ray telescope, \rxte{}.
Bringing together 1187 individual bursts from 48 accreting neutron stars, this catalogue further enabled the study of diverse bursting patterns.
The phenomena included burst oscillations, short-waiting time bursts, photospheric radius-expansion bursts, and the unexpected ``turn over'' in burst rate at high accretion rates.
A successor to this catalogue, the \textit{Multi-INstrument Burst ARchive} (\minbar)\footnote{\url{http://burst.sci.monash.edu/minbar}}, extends the library to over 7000 bursts from 85 source, with data from multiple instruments, including \rxte{}, \beppo{}, and \integral{}.
Its unprecedented size makes \minbar{} the most comprehensive burst catalogue to date, and we make use of its data throughout this thesis.

\section{The Burst Mechanism}
\label{sec:background_mechanism}
Thermonuclear X-ray bursts are the result of stellar material accreting onto neutron stars.
In the idealised picture, the accreted material, rich in hydrogen and/or helium, spreads evenly over the surface and forms a shell only $\sim \SI{1}{m}$ thick.
The base of this envelope is steadily buried deeper as new material is accreted on top, and is compressed to higher pressures and temperatures.
Under hydrostatic equilibrium, the pressure at the base is given by the weight of the overlying fuel,
\begin{equation}
    P = y g,
\end{equation}
where $y \sim 10^7$--$\SI{e8}{\yunits}$ is the column depth and $g \sim \SI{e14}{\gunits}$ is the neutron star surface gravity.

When the base of the envelope reaches temperatures of $T \sim \SI{e8}{K}$, nuclear burning can become thermally unstable.
An increase in the rate of nuclear burning increases the temperature, in turn accelerating fusion in an unstable feedback loop.
This thermal runaway occurs when the temperature sensitivity of nuclear heating exceeds that of radiative cooling.
For a fixed pressure, this threshold is given by
\begin{equation}
    \label{eq:background_ignition}
    \frac{d \sub{\epsilon}{nuc}}{d T} = \frac{d \sub{\epsilon}{cool}}{d T},
\end{equation}
where $\sub{\epsilon}{nuc}$ is the specific nuclear heating rate, and $\sub{\epsilon}{cool}$ is the specific radiative cooling rate.

\begin{figure}
    \centering
    \includegraphics[width=0.9\textwidth]{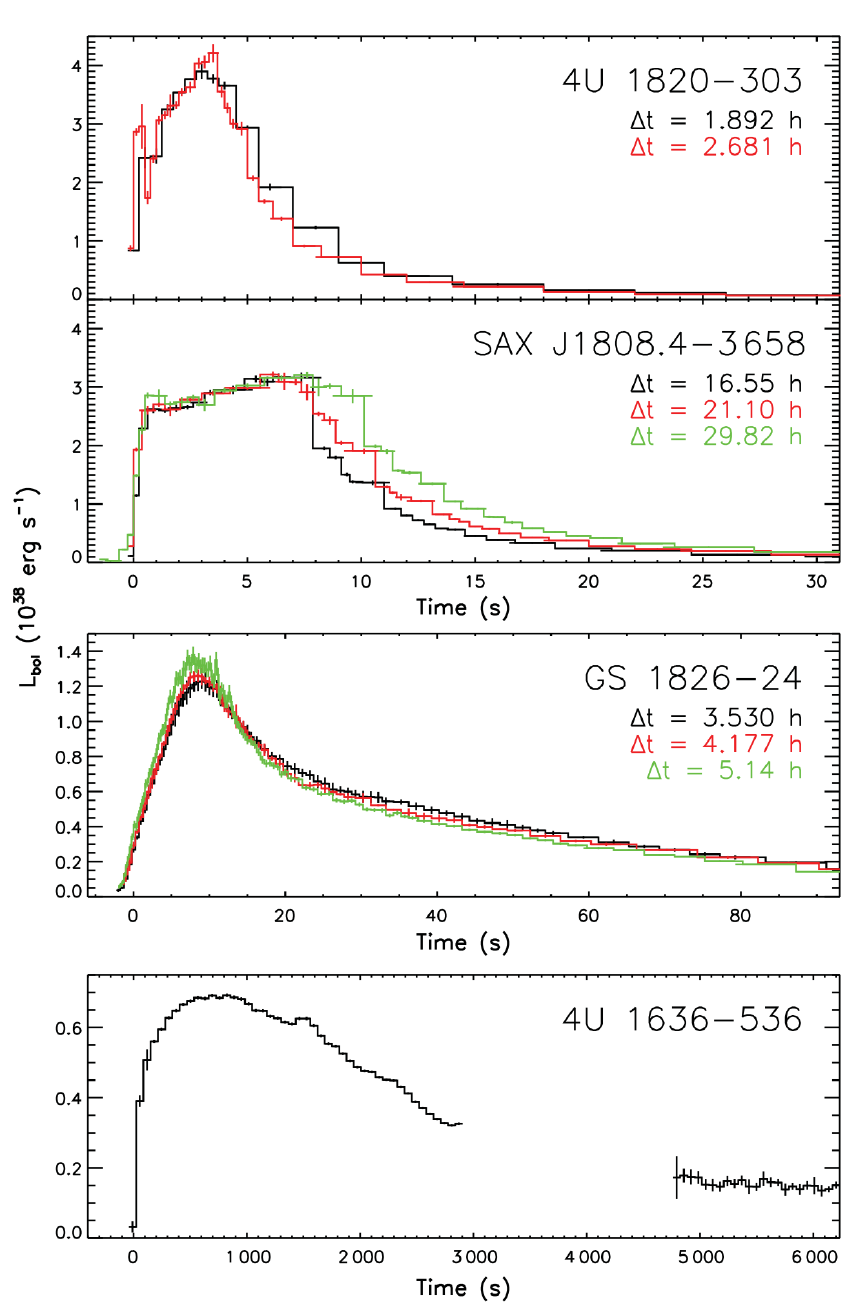}
    \caption{The variety of lightcurves observed from four bursting sources \citep[figure from][]{galloway_thermonuclear_2017}.
    \fouru{} is an ultra-compact binary accreting almost pure helium, and exhibits short PRE bursts at relatively high accretion rates ($\mdot \approx 0.2\, \mdotedd$).
    In \saxj{}, the low accretion rates ($\mdot \lesssim 0.05\, \mdotedd$) allow hydrogen to be depleted through hot CNO burning, leading to powerful PRE bursts with a characteristic ``plateau'' at the Eddington luminosity.
    \gs{} accretes roughly solar material, $\hyd \approx 0.7$, where the hydrogen fuels \rp-process burning and leads to long burst tails.
    The persistent accretor, \sixteen, in addition to hydrogen-rich bursts similar to \gs{}, has exhibited four superbursts -- extremely long bursts with recurrence times of $\dt \sim \SI{1}{yr}$.
    }
    \label{fig:background_refset}
\end{figure}

Thermonuclear burning spreads throughout the envelope as a convective ``flame'', consuming most of the available fuel in $\lesssim \SI{1}{s}$.
The nuclear flash releases $\sim \SI{e39}{erg}$, heating the envelope to $\sim \SI{e9}{K}$.
Due to the extreme gravitational potential of $\approx \SI{200}{\mevnuc}$, in contrast to $\approx \SI{5}{\mevnuc}$ from nuclear burning, the vast bulk of the accreted material remains bound to the surface.
As the outflowing thermal energy reaches the neutron star photosphere, it radiates strongly in X-rays.
The surface luminosity peaks within $\approx 1$--$\SI{10}{s}$, and then decays to pre-burst levels as envelope cools over the following $\approx 10$--$\SI{100}{s}$.
Examples of observed lightcurves are shown in Figure~\ref{fig:background_refset}.

After a burst, the accreted envelope has been processed into ``ashes'' -- the leftover products of nuclear fusion.
Fresh material is accreted on top, burying the ashes and eventually incorporating them into the neutron star crust.
The composition of these ashes impacts the thermal properties of the crust that is formed \citep[e.g.,][]{brown_ocean_1998, gupta_heating_2007}.

\subsection{Nuclear pathways}
\label{subsec:background_cno}
The dominant nuclear pathways on accreting neutron stars include the hot ($\beta$-limited) CNO cycle, the $3\alpha$ (triple-$\alpha$) process, the $\alpha$-process, the $\alpha p$-process, and the \textit{rp}-process.
Detailed descriptions of these pathways and their role in bursts can be found in \citet{lewin_x-ray_1993, woosley_models_2004, fisker_explosive_2008, jose_hydrodynamic_2010}.

For temperatures of $T > \SI{8e7}{K}$, hydrogen is converted to helium through the hot CNO cycle \citep[e.g.,][]{strohmayer_new_2006, wiescher_cold_2010},
\begin{equation}
\label{eq:background_cno_reactions}
    ^{12}\mrm{C} (p,\gamma) ^{13}\mrm{N} (p,\gamma) ^{14}\mrm{O} (\beta^{+}) ^{14}\mrm{N} (p,\gamma) ^{15}\mrm{O} (\beta^{+}) ^{15}\mrm{N} (p,\alpha) ^{12}\mrm{C},
\end{equation}
where the net result is $4p \rightarrow \alpha$.
In this catalytic cycle, the rate is limited by slow $\beta$-decays, and the burning of hydrogen becomes independent of temperature.
When active, steady CNO burning between bursts dominates the heating of the accreted layers.
For a given parcel of accreted material, the time to deplete hydrogen \citep[e.g., as derived in][]{lampe_influence_2016} is given by
\begin{equation}
\label{eq:background_cno_time}
    \sub{t}{CNO} \approx \SI{20}{h} (\frac{\hyd}{0.7}) (\frac{\cno}{0.01})^{-1},
\end{equation}
where $\hyd$ is the initial hydrogen mass fraction and $cno$ is the mass fraction of CNO isotopes.
Equation~\ref{eq:background_cno_time} is only approximate because $\hyd$ and $\cno$ are mass fractions, not number fractions, and depend on the chosen distribution of CNO isotopes.

The $3\alpha$-process converts helium into carbon with a net reaction of $3\alpha \rightarrow ^{12}\mrm{C}$.
Due to its strong temperature sensitivity, the $3\alpha$-process is thermally unstable across typical accretion rates, and serves as the ignition reaction for most burst regimes (described further in \S~\ref{subsec:background_regimes}).

For temperatures above $T \gtrsim \SI{1e9}{K}$, successive $alpha$ captures overtake $3\alpha$ as the main source of energy \citep{fujimoto_shell_1981}, building progressively heavier elements,
\begin{equation}
\label{eq:background_alpha_process}
    ^{12}\mrm{C} (\alpha,\gamma) ^{16}\mrm{O} (\alpha,\gamma) ^{20}\mrm{Ne} (\alpha,\gamma) ^{24}\mrm{Mg}  (\alpha,\gamma) ^{28}\mrm{Si} \ldots
\end{equation}

If hydrogen is present in the accreted fuel, a complex network of reactions can build heavy isotopes to the iron-group and beyond \citep{wallace_explosive_1981, hanawa_thermal_1984}.
For temperatures of $T \gtrsim \SI{4e8}{K}$, the break-out reactions of $^{14}\mrm{O} (\alpha,p) ^{17}\mrm{F}$ and $^{15}\mrm{O} (\alpha,\gamma) ^{19}\mrm{Ne}$ destroy catalysts of the CNO cycle, and interrupt the burning of hydrogen to helium \citep{lewin_x-ray_1993}.
This breakout paves the way for the $\alpha p$-process, whereby sequences of $(\alpha,p)$ and $(p,\gamma)$ reactions proceed to iron-group nuclei.

The presence of hydrogen at these high temperatures also allows the rapid proton (rp) process to occur.
Successive proton-captures and $\beta^+$-decays proceed along the proton drip-line far from the valley of stability, and produce heavy nuclei with atomic mass numbers of $A \approx 60$--$100$ \citep{wallace_explosive_1981, schatz_rp-process_1998, koike_rapid_1999, schatz_end_2001, koike_final_2004}
The rp-process is relatively slow in comparison to the $\alpha$ reactions of the burst ignition.
This delayed energy release results in extended burst tails which are characteristic of mixed hydrogen/helium bursts (e.g.,\ \gs{} in Figure~\ref{fig:background_refset}).

The nuclear pathways described above span thousands of reactions, the vast majority of which have not been measured experimentally.
When nuclear reaction rates are manually varied within their uncertainty, burst models predict altered lightcurves and ashes \citep{koike_rapid_1999, parikh_effects_2008, parikh_impact_2009}.
Existing uncertainties in reaction rates thus contribute to burst model uncertainties, and limit their ability to constrain observed systems.
Sensitivity studies are used to determine which reactions have the strongest influence on burst properties, and thus which nuclear experiments to prioritise in future \citep{cyburt_dependence_2016, meisel_influence_2019}.

\subsection{Bursting regimes}
\label{subsec:background_regimes}
Different regimes of bursting are predicted to occur depending on the local conditions.
The composition of the accreted fuel, the accretion rate, and the thermal structure of the envelope all influence the thermonuclear path to ignition.
The transitions between these regimes were first laid out by \citet{fujimoto_shell_1981}, and summarised again by \citet{bildsten_thermonuclear_1997}.
Most recently, \citet{galloway_thermonuclear_2017} revised the list to a total of seven classifications.

We provide here a brief summary of these regimes as predicted by models, adapted from \citet{galloway_thermonuclear_2017}.
The transitions assume an accreted composition of $\hyd = 0.73$ and $\cno = 0.02$, and a crustal heating strength of $\qb = \SI{0.1}{\mevnuc}$.
The accretion rates are given as a fraction of the canonical Eddington rate for solar composition, $\mdotedd = \SI{8.775e4}{g.cm^{-2}.s^{-1}}$ (equivalent to a global rate of $\Mdotedd = \SI{1.75e-8}{\msun.yr^{-1}}$ for a neutron star mass and radius of $M = \SI{1.4}{\msun}$ and $R = \SI{10}{km}$, and assuming spherically-symmetric accretion).
Note that Cases I and II have not yet been observed, and that the transition to stable hydrogen/helium burning is observed to occur at roughly half the predicted accretion rate listed here.

\renewcommand{\labelenumi}{\Roman{enumi}}
\begin{enumerate}
    \item For $\mdot \lesssim 0.001\, \mdotedd$ ($T \lesssim \SI{7e8}{K}$), the CNO cycle is thermally unstable, and hydrogen ignition occurs. The hydrogen burst also triggers unstable helium burning.
    \item For $0.001 \lesssim \mdot \lesssim 0.004\, \mdotedd$, the hydrogen burst of Case I is too shallow to ignite helium, due to sedimentation of helium and the CNO isotopes \citep{peng_sedimentation_2007}. The accreted helium continues to build a column of fuel below the hydrogen bursts, eventually reaching unstable helium ignition on its own.
    \item For $0.004 \lesssim \mdot \lesssim 0.08\, \mdotedd$ ($T \gtrsim \SI{7e8}{K}$), hydrogen burning is stable, and burns completely to helium via the hot CNO cycle. A pure helium layer accumulates below the hydrogen-burning region, and eventually reaches unstable helium ignition.
    \item For $\mdot \approx 0.1\, \mdotedd$, no bursts occur because helium is stably burned to carbon before it can ignite, although this regime may lead to carbon-fuelled superbursts \citep{keek_carbon_2016}.
    \item For $0.1 \lesssim \mdot \lesssim 1.0\, \mdotedd$, hydrogen is steadily burned to helium, as in Case III, but the helium ignites before the hydrogen is depleted. The mixed hydrogen/helium burst can produce heavy ashes through a combination of $3\alpha$, $\alpha p$ and \textit{rp}-processes \citep{wallace_explosive_1981}.
    \item For $\mdot \approx 1.0\, \mdotedd$, the accreted fuel undergoes oscillatory burning, producing a ``marginally stable'' burning regime \citep{heger_millihertz_2007}.
    \item For $\mdot \gtrsim 1.0\, \mdotedd$, both hydrogen and helium burning are stable, and no bursts occur. This model-predicted transition to stable hydrogen/helium burning occurs much later than inferred from observations, which stabilise around $0.1 \lesssim \mdot \lesssim 0.3\, \mdotedd$ \citep{van_paradijs_relation_1988, cornelisse_six_2003}.
\end{enumerate}

These regimes produce two broad categories of observed bursts: helium (He), and hydrogen/helium (H/He) bursts.

Helium bursts are typically characterised by short rise times ($< \SI{1}{s}$), short overall durations ($\lesssim \SI{10}{s}$), and photospheric radius-expansion \citep[PRE;][]{tawara_very_1984, lewin_precursors_1984}.
During PRE, spectral fits to the bursts yield an expanded blackbody radius, a reduced effective temperature, and approximately constant bolometric luminosity.
These features are thought to indicate that the bursts reach the Eddington luminosity -- the limit at which radiation pressure balances gravitational pressure.
Example lightcurves of observed PRE bursts are shown in the upper two panels of Figure~\ref{fig:background_refset}.

In contrast, mixed hydrogen/helium bursts are characterised by relatively long rise times ($\approx 1$--$\SI{5}{s}$), broad sub-Eddington peaks, and long tails ($\sim \SI{e2}{s}$).
Their burst profiles are understood to result from a more extended nuclear energy release than He bursts, leading to longer tails.
Example lightcurves of H/He bursts are shown in the third panel of Figure~\ref{fig:background_refset}, observed from the famous ``clocked burster'', \gs{}.

\section{Computational Models}
\label{sec:background_modelling}
Much of our current understanding of the bursting mechanism can be attributed to computational models.
In the decades since the first efforts (Section~\ref{sec:background_history}), models of X-ray bursts have progressed from relatively simple sets of analytic equations to state-of-the-art astrophysics codes.
These codes are now capable of simulating dozens of sequential bursts, while tracking the nuclear reactions of thousands of isotopes.
We briefly summarise here previous modelling efforts, with a focus on one-dimensional burst codes, in particular \kepler{}, which we make extensive use of in this thesis.

One-dimensional (1D) burst codes approximate the neutron star envelope as a spherically symmetric shell.
While 1D models lack the inherently multi-dimensional (multi-D) effects of convection and flame spreading \citep[e.g.,][]{zingale_comparisons_2015, cavecchi_fast_2016}, multi-D models are constrained by computational expense to a simulation time of $\lesssim \SI{1}{s}$.
Computationally cheaper methods, such as semi-analytic models \citep[e.g.,][]{cumming_models_2003} and one-zone models \citep[e.g.,][]{schatz_end_2001}, allow for extensive parameter explorations, but lack feedback from nuclear burning on the radial temperature and density profiles, and ultimately their influence on the burst lightcurves.
1D burst codes, therefore, remain the best tools currently available for performing parameter studies which self-consistently follow the mixing and burning of accreted material.

\agile{} is an implicit 1D general relativistic code \citep{liebendorfer_adaptive_2002}.
The code was used by \citet{fisker_importance_2006} to model X-ray bursts and test the lower limit of the $^{15}\mrm{O} (\alpha,\gamma) ^{19}\mrm{Ne}$ reaction rate, using a reaction network of 298 isotopes.
\agile{} was subsequently used to model the extent of the \textit{rp}-process in mixed H/He bursts, using a network with 304 isotopes \citep{fisker_explosive_2008}.

\shiva{} is a 1D astrophysics code that was originally applied to classical novae on accreting white dwarfs \citep{jose_nucleosynthesis_1998}.
The code was used by to model X-ray bursts with a nuclear network of 324 isotopes \citep{jose_hydrodynamic_2010}, computing a simulation with a setup similar to the ``ZM'' \kepler{} model from \citet{woosley_models_2004}.
\shiva{} predicted recurrence times that were twice as long as \kepler{}, which is likely at least partly due to the incorrect opacities previously used in the latter (see Section~\ref{sec:results_opacity}).

\mesa{} is another 1D astrophysics code capable of simulating bursts \citep{paxton_modules_2015}.
Compositions of burst ashes predicted using \mesa{} were used in \citet{meisel_constraints_2017} to examine their impact on crust cooling.
\mesa{} was not used for large-scale burst modelling until the multi-epoch models of \gs{} by \citet[][in Chapter~\ref{ch:paper2} we present multi-epoch \kepler{} models using the same reference dataset]{meisel_consistent_2018}.
A follow-up study also investigated the influence of reaction rate uncertainties on the inferred neutron star properties \citep{meisel_influence_2019}.

\subsection{\kepler}
\label{subsec:background_kepler}
\kepler{} \citep{weaver_presupernova_1978} is a 1D stellar astrophysics code which has been used to model multiple aspects of the bursting process.
In Chapter~\ref{ch:methods}, we discuss the setup and execution of these models.

\kepler{} was first applied to bursts by \citet{wallace_thermonuclear_1982}, who produced four simulations of hydrogen/helium fuel, using only a 19-isotope nuclear network.
The first use of \kepler{} in its modern burst configuration was performed by \citet{woosley_models_2004}.
They introduced a fully adaptive nuclear network, which could simulate the extended nuclear reactions of the rp-process up to the proton drip line, for the first time in 1D.\

Expanding on this work, \citet{heger_models_2007} performed seven models and compared the predictions with the system \gs{} -- the Clocked Burster.
One of these models, labelled \textit{A3}, matched the observed lightcurve morphology with surprising accuracy, and is now a common reference point for burst models.

\citet{keek_multi-zone_2011} applied \kepler{} to the superburst regime, modelling the ignition of deep carbon oceans and the resulting hours-long bursts.
In a follow-up study, \citet{keek_superburst_2012} accreted helium-rich fuel onto a carbon ocean already close to ignition, to simulate a superburst during regular bursts.
\citet{keek_carbon_2016} discovered a stable helium burning regime (Case IV in Section~\ref{subsec:background_regimes}), which only occurred for a narrow range of accretion rates around $\mdot \approx 0.1\, \mdotedd$ and which could explain the production of carbon oceans as superburst fuel.

\citet{lampe_influence_2016} presented the largest grid of 1D models to date.
They explored the dependence of burst properties on accretion rate and metallicity, and compared the results to observed trends.

\citet{cyburt_dependence_2016} used \kepler{} to explore the sensitivity of burst models to uncertainties in nuclear reaction rates.
By varying key reaction rates in the nuclear network, they ranked rates by their influence on the burst lightcurves and properties.
They found that the CNO breakout reaction $^{15}\mrm{O} (\alpha,\gamma) ^{19}\mrm{Ne}$ had the strongest sensitivity.

\chapter{Methods: Modelling X-ray Bursts}
\label{ch:methods}
For the research presented in this thesis, we use the \kepler{} code to simulate X-ray bursts.
In this chapter, we provide an overview of the \kepler{} model for bursting (Section~\ref{sec:methods_kepler}), the methods used to extract burst properties from the model output (Section~\ref{sec:methods_extracting}), and the general relativity (GR) corrections applied to the Newtonian \kepler{} quantities (Section~\ref{sec:methods_gr}).

\section{\kepler: A 1D Hydrodynamic Burst Code}
\label{sec:methods_kepler}
\kepler{} was developed in the 1970s for modelling the pre-supernova evolution of massive stars~\citep{weaver_presupernova_1978}.
In the decades since, the code has been applied to regimes of stellar evolution and explosive nucleosynthesis \citep[e.g.,][]{woosley_evolution_2002, rauscher_nucleosynthesis_2002, heger_how_2003, woosley_models_2004}.
In this thesis, we will focus on the application of \kepler{} to thermonuclear X-ray bursts.
A selection of previous burst studies are described in Section~\ref{subsec:background_kepler}.

Previous descriptions of \kepler{} for burst modelling can be found in \citet{woosley_models_2004, keek_multi-zone_2011}
A \kepler{} burst model consists of a Lagrangian grid of zones in the radial direction.
The zones span a thin shell of material at the neutron star surface, extending down from column depths of $y \sim \SI{e3}{\yunits}$ to the deep ocean at $y \sim \SI{e12}{\yunits}$.
The neutron star crust is located below this lower boundary.
Zones are added and removed according to zoning parameters, which can be tuned to control the grid resolution.
Convection is parametrised in 1D using mixing length theory, where the diffusion coefficient is set by the estimated convective velocity.
\kepler{} uses an adaptive nuclear network which can simulate all the nuclear processes described in Section~\ref{subsec:background_cno}.
Isotopes are dynamically added and removed from the network during the simulation as they are created and destroyed.
A public reaction rate library has been maintained at REACLIB\footnote{\url{https://reaclib.jinaweb.org/}} \citep{cyburt_jina_2010}, and we use this library for the \kepler{} models presented in Chapter~\ref{ch:paper2} and Chapter~\ref{ch:4u1820}.

\subsection{Input Parameters}
\label{subsec:methods_kepler_input}
Parameters can be used to adjust the properties and behaviour of the modelled envelope.
Resolution and zoning parameters control the numerical structure, and are tuned to ensure convergence of the results.
The particular system being modelled is represented by physical parameters, such as the accretion rate and fuel composition.
These variables can be explored and modified to determine the likely values which best reproduce the observed data.
We can thus use these parameters to understand and constrain the physical characteristics of observed bursting systems.

\subsubsection{Mass, Radius, and Gravity}
The parameter for the neutron star mass, $M$, sets the gravitational mass interior to the base of the model.
Along with the radius, $R$, this determines the gravitational acceleration, $g$, experience by each zone.
The thin shell of the envelope, $\Delta R \ll R$, results in a gravitational acceleration that differs by ${\lesssim} \SI{1}{\percent}$ across the model domain.

\subsubsection{Chemical Composition}
The accreted composition is set with the parameters of the hydrogen mass fraction, $\hyd$, and the CNO mass fraction (``metallicity''), $\cno$.
For simplicity, the CNO metallicity is accreted into \kepler{} as $^{14}\mrm{N}$, which is the dominant isotope from stellar CNO burning.
The remainder of the fuel is put into helium ($^{4}\mrm{He}$): $\hel = 1 - \hyd - \cno$.

\subsubsection{Accretion Rate}
The mass accretion rate, $\Mdot$, is the dominant parameter for modelling X-ray bursts.
It sets the rate at which nuclear fuel is added to the envelope, and thus how quickly an explosive layer can be built up.
It also determines the total rate of crustal heating, $\qb$.
The accretion rate of a \kepler{} model can be constant to represent persistent accretors, or time-varying to represent transient accretors (see Chapter~\ref{ch:paper1}).
The accretion rate is used during both model initialisation (Section~\ref{subsec:methods_kepler_setup}) and execution (Section~\ref{subsec:methods_kepler_execution}).
It is typically expressed as a fraction of the Eddington-limited accretion rate for solar composition, $\Mdotedd = \SI{1.75e-8}{\msun.yr^{-1}}$.
Although the actual $\Mdotedd$ depends on the composition and neutron star gravity, the canonical value serves as a common reference point between models and codes.

\subsubsection{Crustal Heating}
The crustal heating parameter, $\qb$, controls the heat flowing into the envelope from the crust below.
It is effectively a lower boundary condition, setting the heat flux at the innermost zone.
During accretion, the weight of new material compresses the neutron star crust, inducing electron captures and pycnonuclear (density-driven) reactions \citep{haensel_models_2008}.
The total energy yield from this process is 1--2 $\si{\mevnuc}$ \citep{haensel_models_2008}, but the net flux reaching the envelope is typically assumed to be $\approx \SI{0.1}{\mevnuc}$ \citep[e.g.,][]{heger_models_2007}.

\subsubsection{Assumptions}
No model is without assumptions and limitations.
\kepler{}, being a 1D code, inherently assumes spherical symmetry.
The only spatial degree of freedom is in the radial direction, meaning that energy, heat, and chemical composition can only move ``up'' or ``down'', but not across the surface.
To a certain degree this is an adequate approximation on the global scale, but it cannot truly capture the multidimensional nature of accretion, convection, turbulence, and the lateral spreading of the nuclear flame across the neutron star surface \citep[e.g.,][]{shara_localized_1982, zingale_comparisons_2015, cavecchi_flame_2013}.
These effects likely play important roles on the ignition conditions, the rise of the burst lightcurve, and quasi-periodic oscillations \citep{watts_thermonuclear_2012}.

\kepler{} burst models do not include rotation or magnetic fields, which can also influence the dynamics of flame spreading \citep{cavecchi_rotational_2015, cavecchi_fast_2016}.

\subsection{Setup and Initialisation}
\label{subsec:methods_kepler_setup}
The setup phase of a model involves multiple steps.
First, an approximate thermal and chemical profile is constructed.
The composition profile consists of a heavy substrate lying below a lighter envelope.
The substrate, representing the heavy ashes of previous burning, is composed of $^{54}\mrm{Fe}$ and acts as an inert thermal buffer for the heating generated during the model.
New material is then accreted, without nuclear burning, to build the outer layers to the thin surface.
With accretion and nuclear reactions disabled, the thermal profile is relaxed, with the base heat flux at the inner boundary determined by the $\qb$ parameter.
The envelope then relaxes into thermal equilibrium, where the flux leaving the surface is equal to the flux entering the base.
In Section~\ref{sec:results_preheating}, we present an improvement to this method, where we include a heat source representing nuclear burning, which improves the model equilibrium and reduces burn-in.
The model is now initialised and ready for the full simulation to begin.

\subsection{Execution and Output}
\label{subsec:methods_kepler_execution}
\begin{figure}
    \centering
    \includegraphics[width=\textwidth]{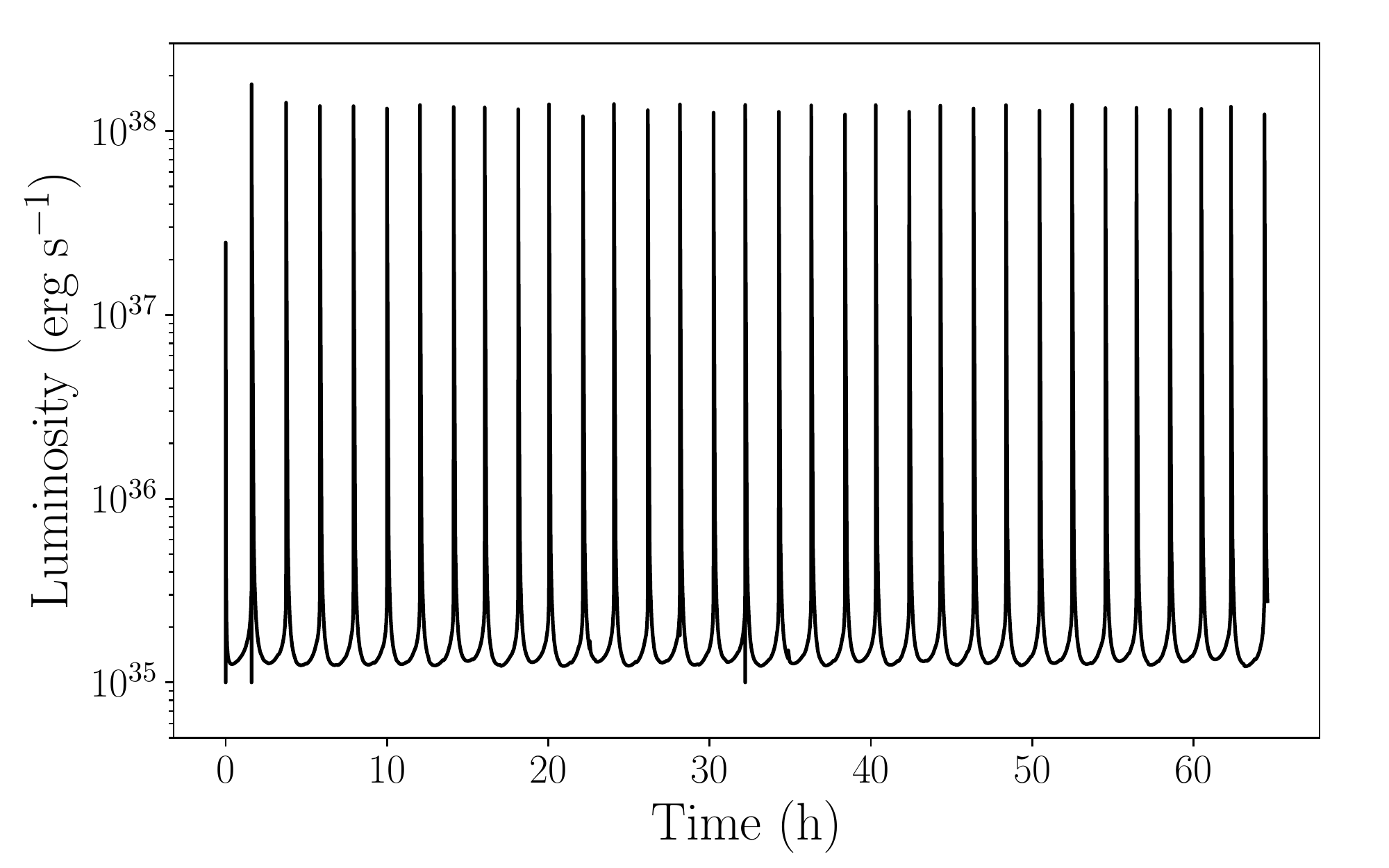}
    \includegraphics[width=\textwidth]{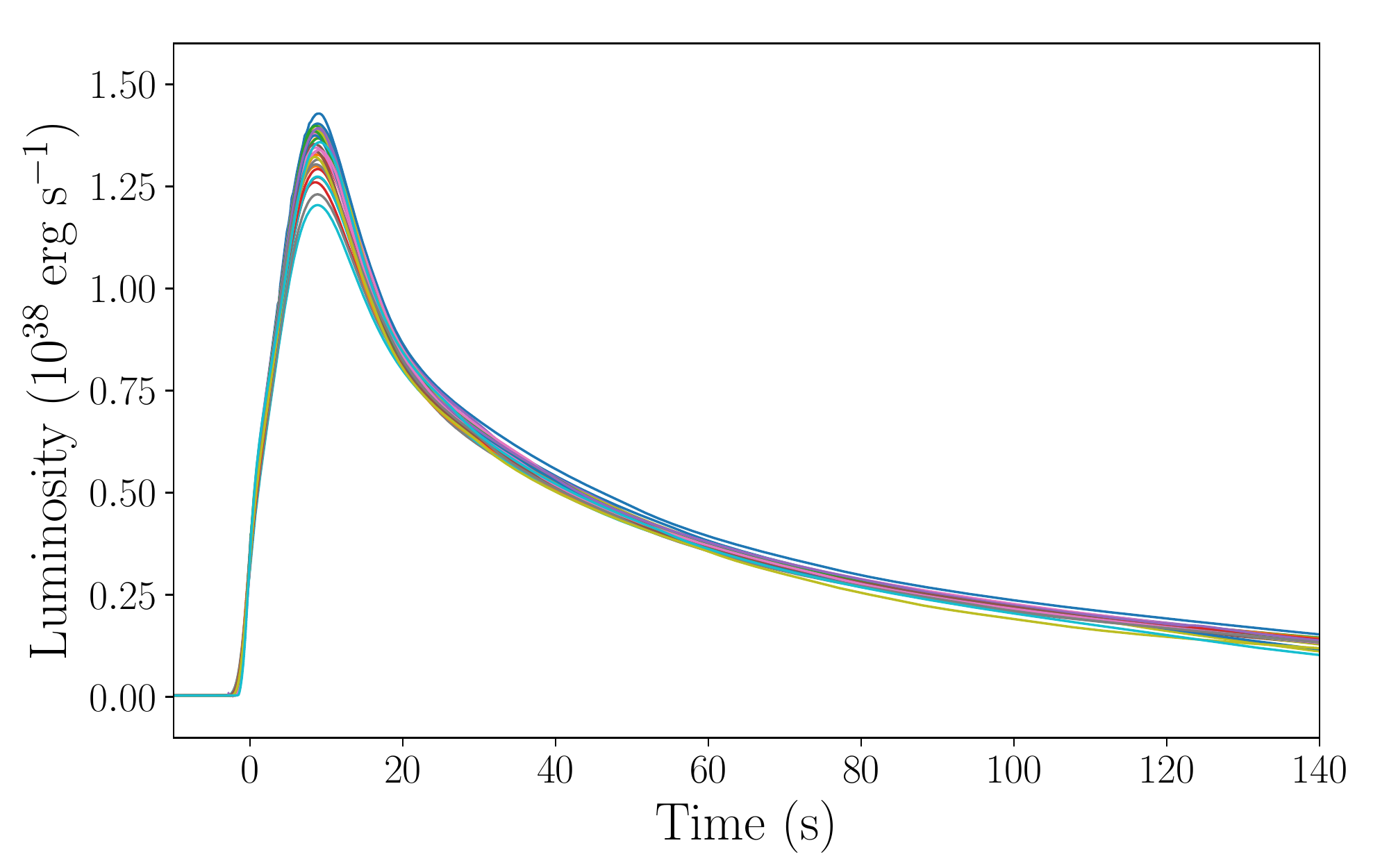}
    \caption{Example lightcurves from a \kepler{} simulation.
An entire model is plotted in the upper panel, and the individual extracted bursts stacked in the lower panel.
This figure also appears in \citet{johnston_multi-epoch_2019} (Chapter~\ref{ch:paper2}).
    }
    \label{fig:methods_model_lightcurve}
\end{figure}

The full simulation is started by switching on accretion and the nuclear reaction network.
The model accretes material at the specified rate and composition, and nuclear reactions begin processing the fuel during steady burning until the unstable ignition of a burst occurs (Section~\ref{sec:background_mechanism}).
The bursting process repeats, producing a ``train'' of sequential bursts.
We typically choose the total simulation time to generate the desired number of bursts, using a prediction of the recurrence time.
The primary model output we use in this thesis is the lightcurve -- the bolometric surface luminosity as a function of time.
A model lightcurve with a train of bursts is shown in Figure~\ref{fig:methods_model_lightcurve}.
The individual bursts can then be sliced out (as described below), and the burst properties calculated.

\section{Extracting Bursts from Models}
\label{sec:methods_extracting}
For the extraction of burst properties from \kepler{} models, we have developed software tools over the course of this project, which we have collected under \href{https://github.com/zacjohnston/pyburst}{\pyburst}, a \python{} library available on the open-source platform github\footnote{\url{https://github.com/zacjohnston/pyburst}}.
We describe here the general procedure of the model analysis.

Given a model lightcurve (Figure~\ref{fig:methods_model_lightcurve}), we identify each burst and extract its properties.
We then average the properties over the burst sequence (excluding some number of initial bursts, Section~\ref{subsec:methods_average}) to obtain summary properties for the model.
The standard deviation of these quantities indicates the inherent burst-to-burst variation, which we take as the model uncertainty (note that this does not include systematic uncertainties, for example due to reaction rate uncertainties).

To analyse a model, \pyburst{} follows a pipeline that proceeds roughly as follows:

\begin{enumerate}
    \item Identify local maxima in the model lightcurve.
    \item Discard non-burst maxima, such as numerical spikes or bumps.
    \item Identify the start and end of each burst lightcurve.
    \item Calculate the properties of each burst.
    \item Average the properties across all bursts.
\end{enumerate}

From this procedure, we obtain a collection of bursts with their individual and averaged properties, along with associated uncertainties.

\subsection{Calculating burst properties}
\label{subsec:methods_extracting_bprops}
The most common burst properties of interest are the recurrence time, $\dt$, the peak luminosity, $\Lpeak$, the burst energy (sometimes called fluence, see below), $\Eb$, and the alpha ratio, $\alpha$.

\subsubsection{Identifying bursts}
Firstly, the maxima in the model lightcurve are found.
We save computational time, and avoid false-positives from small bumps, by imposing a minimum luminosity threshold.
If we are confident that the modelled bursts will always peak above $\SI{e37}{\Lunits}$, for example, we need only consider these sections of the lightcurve.
Once we have identified the maxima in a lightcurve, we filter out any that are not deemed to be bursts.
Once the candidate burst peaks, $\Lpeak$, have been verified, \pyburst{} proceeds to calculate the remaining burst properties.

\subsubsection{Recurrence time and burst rate}
The recurrence time, $\dt$, is the elapsed time since the previous burst.
Specifically, \pyburst{} defines it as the time between burst peaks.
Note that the recurrence time is (by definition) undefined for the first burst in the sequence.
The burst rate is simply $\brate = 1 / \dt$.

\subsubsection{Ignition column}
A burst ignites when the base of the accreted layer reaches unstable thermonuclear conditions.
Given a burst with a recurrence time of $\dt$ and a constant accretion rate of $\mdot$, the mass of the accreted layer is then simply
\begin{equation}
    \sub{M}{\accsymb} = 4 \pi R^2 \mdot \dt,
\end{equation}
and the ignition depth is
\begin{equation}
    \yign = \mdot \dt.
\end{equation}

\subsubsection{Lightcurve points}
\begin{figure}
    \centering
    \includegraphics[width=\textwidth]{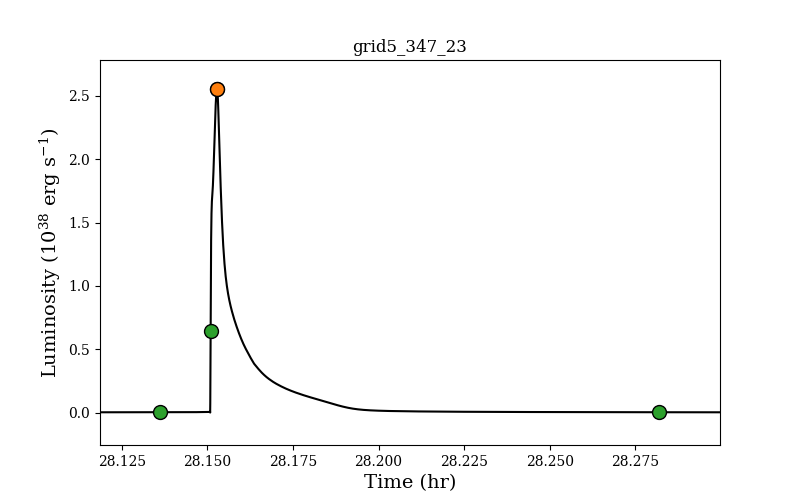}
    \caption{An example \kepler{} burst lightcurve with key points as identified by \pyburst.
The green circles indicate, from left to right, $\tpre$, $\tstart$, and $\tend$.
The orange circle indicates the burst peak, $(\tpeak, \Lpeak)$.}
    \label{fig:methods_lightcurve}
\end{figure}

To extract the burst energetics, we determine the start and end points of the burst lightcurve (Figure~\ref{fig:methods_lightcurve}).
We define a reference point $t_\mathrm{pre}$, that is a set time interval prior to the peak, chosen to ensure the entire burst rise is captured, $\SI{30}{s}$, for example.
The start of the burst rise, $\tstart$, is defined as the point where the lightcurve has reached some fraction of the peak luminosity, $\SI{25}{\percent}$, for example.
The burst end, $\tend$, is defined as the point where the luminosity has decayed to a given fraction of the peak luminosity, $\SI{1}{\percent}$, for example.

\subsubsection{Burst energetics}
The burst energetics are calculated using the above lightcurve reference points.
The burst energy, $\Eb$, is obtained by integrating the luminosity between $\tpre$ and $\tend$, after subtracting a baseline luminosity, $\Lpre$, which is taken at $\tpre$.
Some emission prior to the burst may be included in this integration, but the contribution is negligible ($< \SI{0.1}{\percent}$) if $\Lpre$ is subtracted.

Note that the burst energy here is often called the burst fluence, which is used interchangeably with the observed quantity of time-integrated burst flux ($\int \Fburst \diff t$).
Although these quantities are indeed related (Section~\ref{subsec:methods_observables_fluence}), the term ``fluence'' specifically refers to time-integrated flux, and to avoid confusion we will here maintain the distinction between the burst energy, $\Eb$, and the fluence, $\fluence$.

\subsubsection{$\alpha$ (alpha) ratio}
We then obtain $\alpha$, the ratio between the persistent energy and burst energy:
\begin{equation}
    \alpha = \frac{-\phi \dt \Mdot}{\Eb},
\end{equation}
where $\phi = -z c^2 / (1+z)$ is the gravitational potential at the neutron star surface.
This ratio represents the relative efficiency of accretion compared to thermonuclear burning, and can be used to infer the dominant composition of the accreted fuel.
A small $\alpha$ ratio of ${\approx} 40$ corresponds to strong nuclear energy efficiency, indicating hydrogen-rich fuel.
A large $\alpha$ ratio of $\gtrsim 100$ corresponds to relatively inefficient nuclear energy generation, indicating fuel that is hydrogen-deficient (i.e.,\ helium-rich).

\subsection{Average Model Properties}
\label{subsec:methods_average}
Having calculated the above properties for each burst in the sequence, we then calculate summary quantities for the model as a whole.
Because the initial bursts are often much more energetic than the subsequent train \citep{woosley_models_2004}, they're usually treated as a ``burn-in'' phase of the simulation, and excluded from analysis.
In previous studies, only the first one or two bursts were excluded  \citep[e.g.,][]{heger_models_2007, cyburt_jina_2010, lampe_influence_2016}.
We discovered the presence of extended burn-in (Section~\ref{sec:results_preheating}).

We can also calculate average burst lightcurves for comparison to observations.
The most common approach is to stack the burst lightcurves, aligned by peak, for example, and calculate the average and standard deviation of the luminosity along the curve.
With this method, a mean burst lightcurve can be obtained and compared with observations.

\section{Correcting for General Relativity}
\label{sec:methods_gr}
Due to its origin as a stellar evolution code, \kepler{} uses Newtonian gravity, which is sufficiently accurate for regular stars.
X-ray bursts, however, occur in a highly-relativistic environment\footnote{A canonical neutron star, with mass $\nsmass$ and radius $\nsradius$, has a gravitational acceleration at the surface of $g \approx \SI{2.4e14}{cm.s^{-2}}$ (i.e.~${\approx} 8000$ times the speed of light per second).}.
To accurately model bursts, it is crucial to account for the effects of General Relativity (GR) when comparing models with observations.

Fortunately, the results of Newtonian \kepler{} models require only straightforward corrections, due to the thin-shell nature of the envelope.
These corrections have been described in detail previously \citep{woosley_models_2004, keek_multi-zone_2011, lampe_influence_2016}, and we will provide here a brief summary.
This section largely follows the notation and conventions used in Appendix~B of \citet{keek_multi-zone_2011}.

In this section we will signify Newtonian \kepler quantities with the subscript '$k$', GR quantities with the subscript '$g$', and quantities in the frame of a distant observer with the subscript '$\infty$'.
These corrections rely on the assumption of a thin shell at the surface of a neutron star.

\subsection{Definitions}
\label{subsec:methods_gr_definitions}
To resolve the discrepancy between the Newtonian gravity of \kepler{} and the GR gravity of actual neutron stars, we first note that gravitational acceleration is approximately constant throughout the thin shell of an accreted envelope.
For a 10 m thick envelope on the surface of a canonical neutron star, the acceleration differs by only ${\approx} \SI{2}{\percent}$ from top to bottom.

Under the approximation that gravity is constant, a Newtonian \kepler{} model with a mass and radius ($\nw{M}$, $\nw{R}$) is equivalent to an ``actual'' neutron star with a different mass and radius ($\gr{M}$, $\gr{R}$) if it has the same acceleration under GR.\@

The acceleration under Newtonian gravity is given by
\begin{equation}
    \label{eq:methods_gnw}
    \nw{g} = \frac{G M}{R^2},
\end{equation}
where $G$ is the gravitational constant.
The acceleration under GR is instead given by
\begin{equation}
    \label{eq:methods_ggr}
    \gr{g} = \frac{G M (1+z)}{R^2},
\end{equation}
where the gravitational redshift is given by
\begin{equation}
    \label{eq:methods_redshift}
    1+z = \frac{1}{\sqrt{1 - \frac{2 G M}{c^2 R} }}.
\end{equation}
In other words, $\gr{g} = (1+z) \nw{g}$ for a given mass and radius\footnote{For a canonical neutron star, $1+z \approx 1.26$.}.
If we impose the requirement that the acceleration is equal under both the Newtonian and GR regimes, we obtain
\begin{equation}
    \label{eq:methods_gr_acceleration}
    \begin{split}
        \nw{g} &= \gr{g}, \\
        \frac{G \nw{M}}{\nw{R}^2} &= \frac{G \gr{M} (1+z)}{\gr{R}^2}, \\
        \left( \frac{\gr{R}}{\nw{R}} \right)^2 &= \frac{\gr{M}}{\nw{M}} (1+z).
    \end{split}
\end{equation}
Note that from here on, $(1+z)$ from Equation~\eqref{eq:methods_redshift} is always calculated with the GR mass and radius.
If we define the ratios\footnote{Note that the radius ratio $\xi$ is unrelated to the anisotropy factors $\xib$ and $\xip$, which we will introduce later.}
\begin{equation}
    \label{eq:gr_xi_phi}
    \varphi = \frac{\gr{M}}{\nw{M}}, \quad \xi = \frac{\gr{R}}{\nw{R}},
\end{equation}
we can write Equation~\eqref{eq:methods_gr_acceleration} as
\begin{equation}
    \label{eq:xi_phi_redshift}
    \xi^2 = \varphi (1+z).
\end{equation}
If we define the gravitational radius for the Newtonian case
\begin{equation}
    \label{eq:gr_zeta}
    \zeta = \frac{G \nw{M}}{c^2 \nw{R}},
\end{equation}
we can also write Equation~\eqref{eq:methods_redshift} as
\begin{equation}
    \label{eq:redshift_zeta}
    1+z = \frac{1}{\sqrt{1 - \frac{2 \varphi \zeta}{\xi}}}.
\end{equation}
Substituting Equation~\eqref{eq:redshift_zeta} into Equation~\eqref{eq:xi_phi_redshift} we obtain
\begin{equation}
    \label{eq:gr_fourth_order}
    \varphi^2 + 2 \varphi \xi^3 \zeta - \xi^4 = 0.
\end{equation}
Thus, a Newtonian \kepler{} model with a given mass and radius ($\nw{M}$, $\nw{R}$) corresponds to any pair ($\gr{M}$, $\gr{R}$) which satisfies Equation~\eqref{eq:gr_fourth_order}.
Note that a given gravitational acceleration corresponds to a contour of mass-radius pairs.
Depending on which quantities are already known, we can solve the above problem for the remaining variables.

\subsubsection*{Solving for Mass}
If the actual neutron star radius ($\gr{R}$) is known, we can determine the mass ($\gr{M}$) by solving Equation~\eqref{eq:gr_fourth_order} for the mass ratio:
\begin{equation}
    \label{eq:gr_phi_solved}
    \varphi = \zeta \xi^3 \left( \sqrt{1 + \zeta^{-2} \xi^{-2}} - 1 \right),
\end{equation}
and then using $\gr{M} = \varphi \nw{M}$.

\subsubsection*{Solving for Radius}
If instead the actual neutron star mass ($\gr{M}$) is known, we can find the radius ($\gr{R}$) by solving Equation~\eqref{eq:gr_fourth_order} for the radius ratio:
\begin{equation}
    \label{eq:gr_xi_solved}
    \xi = \frac{\zeta \varphi}{2} \left( 1 + \sqrt{1 - A} + \sqrt{2 + A + \frac{2}{\sqrt{1 - A}}} \right),
\end{equation}
where we have defined
\begin{equation}
    \begin{split}
        A &= \sqrt[3]{\frac{2}{9}} \frac{\left( \frac{B^2}{\varphi^2} - 2 \sqrt[3]{6} \right)}{B \zeta^2}, \\ \\
        B &= \sqrt[3]{9 \zeta^2 \varphi^4 + \sqrt{3} \varphi ^3 \sqrt{16 + 27 \zeta^4 \varphi^2}}.
    \end{split}
\end{equation}
We can then simply use $\gr{R} = \xi \nw{R}$

\subsection{Correcting local Newtonian quantities}
\label{subsec:methods_gr_bprops}
In the previous section, we solved for the GR neutron star masses and radii ($\gr{M}$, $\gr{R}$) which are equivalent to a Newtonian \kepler{} model.
The physical quantities predicted by the model are then corrected to be consistent with the actual mass and radius.\@

\subsubsection*{Accretion rate}
The Newtonian \kepler{} model is equivalent to a GR-corrected neutron star with the same local accretion rate, $\mdot$.
From Equation~\eqref{eq:gr_xi_phi}, the ratio of the GR-corrected neutron star surface area to the Newtonian surface area is equal to $\xi^2$.
The global accretion rate, $\Mdot = 4 \pi R^2 \mdot$, is thus scaled according to
\begin{equation}
    \label{eq:gr_mdot}
    \begin{split}
    \gr{\Mdot} &= \xi^2 \nw{\Mdot} \\
               &= \varphi (1+z) \nw{\Mdot},
    \end{split}
\end{equation}
where we have used the relation $\xi^2 = \varphi(1+z)$ from Equation~\eqref{eq:xi_phi_redshift}.

\subsubsection*{Luminosity}
Similarly, the Newtonian luminosities are scaled by the surface area,
\begin{equation}
    \label{eq:gr_lum_correction}
    \begin{split}
    \gr{L} &= \xi^2 \nw{L} \\
           &= \varphi (1+z) \nw{L}.
    \end{split}
\end{equation}

\subsubsection*{Accretion luminosity}
The accretion luminosity ($\Lacc$) originates from material accreting at a certain rate ($\Mdot$) through a gravitational potential ($\phi$),
\begin{equation}
    \label{eq:methods_gr_accretion}
    \Lacc = - \Mdot \phi.
\end{equation}
The potentials for the Newtonian and GR regimes are given by
\begin{equation}
    \label{eq:gr_potentials}
    \nw{\phi} = - \frac{ G \nw{M} }{\nw{R}}, \quad \gr{\phi} = - \frac{c^2 z}{1+z}.
\end{equation}
Using Equations~\eqref{eq:gr_mdot}~and~\eqref{eq:gr_potentials}, we obtain
\begin{equation}
    \label{eq:methods_gr_Lacc}
    \begin{split}
    \sub{L}{\accsymb,g} &= - \gr{\Mdot} \gr{\phi} \\
                   &= - \varphi (1+z) \nw{\Mdot} \gr{\phi} \\
                   &= \varphi c^2 z \nw{\Mdot}.
    \end{split}
\end{equation}

\subsubsection*{Burst energy (fluence)}
The burst energy ($\Eb$) is the total energy radiated during the burst,
corresponding to the time-integrated burst luminosity ($\int \Lburst \diff t$).
Using the area ratio ($\xi^2$), we obtain
\begin{equation}
    \label{eq:gr_energy}
    \begin{split}
        \grsub{E}{\burstsymb} &= \xi^2 \nwsub{E}{\burstsymb} \\
                              &= \varphi (1+z) \nwsub{E}{\burstsymb}.
    \end{split}
\end{equation}

\subsubsection*{Eddington luminosity}
The Eddington luminosity (for a spherically-symmetric object) is found by balancing the gravitational force with the radiation pressure, and is given by
\begin{equation}
    \Ledd = \frac{4 \pi R^2 g c}{\kappa},
\end{equation}
where $\kappa$ is the opacity.
Substituting $g$ from Equations~\eqref{eq:methods_gnw}~and~\eqref{eq:methods_ggr}, we have

\begin{equation}
    \label{eq:methods_gr_Ledd}
    \sub{L}{\eddsymb,g} = \frac{ 4 \pi G c (1+z) \gr{M} }{\kappa}, \quad \sub{L}{\eddsymb,k} = \frac{ 4 \pi G c \nw{M} }{\kappa}.
\end{equation}

Noting that $\gr{M} = \varphi \nw{M}$ and $1+z = \xi^2 / \varphi$, we obtain

\begin{equation}
    \begin{split}
        \sub{L}{\eddsymb,g} &= \frac{ 4 \pi G c \xi^2 \nw{M} }{\kappa} \\
                            &= \xi^2 \sub{L}{\eddsymb,k}
    \end{split}
\end{equation}

We could also have noted that $g$ is equal in both cases (by definition), leaving a factor of $\gr{R}^2 / \nw{R}^2 = \xi^2$.

\subsection{Transforming to an observer frame}
\label{subsec:methods_gr_observer}
The GR-corrected quantities are then converted from the local reference frame of the neutron star surface to the frame of a distant observer, which we signify with the subscript ``$\obssymb$''.

\subsubsection*{Timescales}
Timescale quantities are time-dilated by
\begin{equation}
    \label{eq:methods_gr_observer_time}
    \obs{t} = (1+z) \gr{t} = (1+z) \nw{t}.
\end{equation}

\subsubsection*{Luminosity}
Photons are redshifted to lower energies upon leaving the gravitational potential of the neutron star, in addition to their rate of arrival becoming time-dilated.
Combined with Equation~\eqref{eq:gr_lum_correction}, the luminosity for an observer is given by
\begin{equation}
    \label{eq:methods_gr_observer_lum}
    \obs{L} = \frac{\gr{L}}{(1+z)^2} = \frac{\varphi \nw{L}}{1+z}.
\end{equation}

\subsubsection*{Burst energy}
Because the burst energy ($\Eb$) is integrated over time, the time-dilation from Equation~\eqref{eq:methods_gr_observer_time} is removed, giving
\begin{equation}
    \label{eq:methods_gr_observer_energy}
    \obssub{E}{\burstsymb} = \frac{ \grsub{E}{\burstsymb} }{1+z} = \varphi \nwsub{E}{\burstsymb},
\end{equation}
where we have used Equation~\eqref{eq:gr_energy}, and $\xi^2 = \varphi (1+z)$ from Equation~\eqref{eq:xi_phi_redshift}.

\section{Predicting Observable Burst Properties}
\label{sec:methods_observables}
The extracted burst properties (Section~\ref{sec:methods_extracting}), once corrected to an equivalent neutron star surface under GR (Section~\ref{sec:methods_gr}), can then be used to calculate ``observables'' as measured with Earth-based instruments.

\begin{figure}
    \centering
    \includegraphics[width=\textwidth]{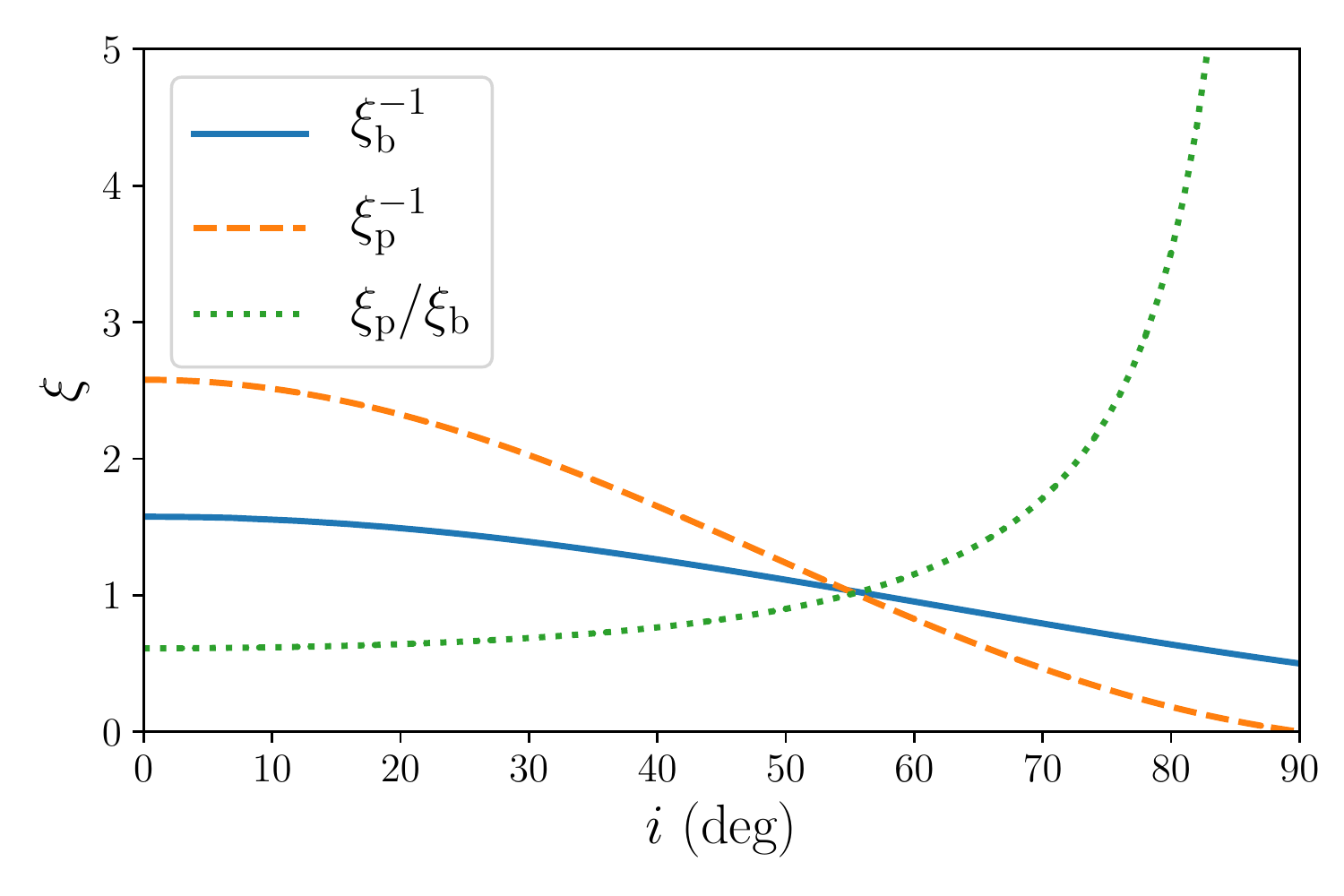}
    \caption{Anisotropy factors versus inclination angle, as predicted by \citet{he_anisotropy_2016} with their \textit{disc a} model for a thin, flat accretion disc.
    The factors alter the apparent isotropic flux given by Equation~\eqref{eq:methods_observables_anisotropy}.
    }
    \label{fig:methods_anisotropy}
\end{figure}

\subsection{Burst rate and recurrence time}
\label{subsec:methods_observables_rate}
The burst rate ($\brate$) is time-dilated according to Equation~\eqref{eq:methods_gr_observer_time},
\begin{equation}
    \label{eq:methods_observables_rate}
    \obs{\brate} = \frac{\gr{\brate}}{1+z} = \frac{\nw{\brate}}{1+z}.
\end{equation}
The recurrence time is simply the inverse,
\begin{equation}
    \label{eq:methods_observables_dt}
    \obs{\dt} = (1+z) \gr{\dt} = (1+z) \nw{\dt}.
\end{equation}

\subsection{Flux -- accounting for anisotropy}
\label{subsec:methods_observables_flux}
X-ray telescopes count photons within a given energy band, from which the incident bolometric flux ($F$) can be inferred.
Assuming a uniform (i.e.,\ isotropic) angular distribution for the source of radiation, the flux is given by
\begin{equation}
    \label{eq:methods_observables_flux_iso}
    F = \frac{ \obs{L} }{4 \pi d^2},
\end{equation}
where $d$ is the distance to the source.
Note that for visual clarity, from here on we assume that fluxes are always in the reference frame of an observer, and omit the subscript `$\obssymb$'.

\kepler{} models predict the local bolometric luminosity ($\nw{L}$), which is GR-corrected ($\gr{L}$) and shifted to the frame of a distant observer ($\obs{L}$) using Equation~\eqref{eq:methods_gr_observer_lum}.
If we assume the radiation is isotropic, this luminosity corresponds to the luminosity in Equation~\eqref{eq:methods_observables_flux_iso}, and the observed flux can be directly predicted.

In reality, however, the neutron star is surrounded by an accretion disc, which can intercept, scatter, and obscure photons, resulting in a non-uniform (i.e.,\ anisotropic) angular distribution of radiation \citep[e.g.][]{lapidus_angular_1985, sztajno_constraints_1987, fujimoto_angular_1988, he_anisotropy_2016}.
The apparent luminosity for an observer depends on the inclination, $i$, of the line of sight as measured from the rotation axis of the disc.

The angular distribution of radiation is dependent on the morphology and radiative properties of the accretion disc.
These properties, including the inclination itself, are often highly uncertain.
Anisotropic effects are thus typically represented with generalised factors, given by $\xi$ \citep[introduced by][]{sztajno_constraints_1987}:
\begin{equation}
    \label{eq:methods_observables_anisotropy}
    \Fburst = \frac{\obssub{L}{\burstsymb} }{4 \pi d^2 \xib}, \quad \Fper = \frac{ \obssub{L}{\perssymb} }{4 \pi d^2 \xip},
\end{equation}
where the the burst (`b') and persistent (`p') emission are treated separately.
Because $\xib$ and $\xip$ are degenerate with $d$, inferred distances are typically reported in the form $d \sqrt{\xi}$.
Models which predict the dependence of $\xib$ and $\xip$ on the inclination \citep[e.g.][Figure~\ref{fig:methods_anisotropy}]{he_anisotropy_2016} can then be used to constrain the absolute distance, $d$.

\subsection{Persistent accretion flux}
\label{subsec:methods_persistent}
The persistent flux ($\Fper$) is the steady emission observed between bursts.
This quantity is thought to originate primarily from the luminosity generated by accretion ($\Lacc$) given in Equation~\eqref{eq:methods_gr_accretion}.
Assuming that the contribution of steady nuclear burning is negligible (i.e.,\ $\Lper = \Lacc$), we have, from Equation~\eqref{eq:methods_observables_anisotropy},
\begin{equation}
    \label{eq:methods_observables_fper}
    \Fper = \frac{ \obssub{L}{\accsymb} }{ 4 \pi d^2 \xip }.
\end{equation}
From Equations~\eqref{eq:methods_gr_Lacc}~and~\eqref{eq:methods_gr_observer_lum}, we obtain
\begin{equation}
    \label{eq:methods_observables_Lacc_obs}
    \obssub{L}{\accsymb} = \frac{ c^2 z \gr{\Mdot} }{ (1 + z)^3 } = \frac{ c^2 z \varphi \nw{\Mdot} }{ (1 + z)^2 }.
\end{equation}
Equation~\eqref{eq:methods_observables_fper} then becomes
\begin{equation}
    \label{eq:methods_observables_Fper-obs}
    \Fper = \frac{ c^2 z \gr{\Mdot} }{ 4 \pi d^2 \xip (1 + z)^3 } = \frac{ c^2 z \varphi \nw{\Mdot} }{ 4 \pi d^2 \xip (1 + z)^2 }.
\end{equation}
Thus, the persistent accretion flux can be predicted directly with the model parameters.

\subsection{Burst fluence}
\label{subsec:methods_observables_fluence}
The burst fluence ($\fluence$) is the time-integrated burst flux ($\int \Fburst \diff t$), and is the observable equivalent of the burst energy ($\Eb$).
From Equation~\eqref{eq:methods_observables_anisotropy} we obtain
\begin{equation}
    \label{eq:methods_observables_fluence-general}
    \fluence = \frac{ \obssub{E}{\burstsymb} }{4 \pi d^2 \xib}.
\end{equation}
Substituting Equation~\eqref{eq:methods_gr_observer_energy} gives
\begin{equation}
    \label{eq:methods_observables_fluence-obs}
    \fluence = \frac{ \grsub{E}{\burstsymb} }{4 \pi d^2 \xib (1+z)} = \frac{ \varphi \nwsub{E}{\burstsymb} }{4 \pi d^2 \xib}.
\end{equation}

\subsection{Persistent fluence}
\label{subsec:methods_observables_per_fluence}
The persistent fluence ($\fluencep$) is the time-integrated persistent flux ($\int \Fper \diff t$) since the previous burst.
Assuming that $\Lacc$ is constant between bursts, the persistent fluence is given by
\begin{equation}
    \label{eq:methods_observables_fluence_p_general}
    \fluencep = \frac{ \obssub{L}{\accsymb} \obs{\dt} }{  4 \pi d^2 \xip }.
\end{equation}
From Equations~\eqref{eq:methods_observables_rate}~and~\eqref{eq:methods_observables_Fper-obs}, we obtain
\begin{equation}
    \label{eq:methods_observables_fluence_p}
    \fluencep = \frac{ c^2 z \gr{\Mdot} \gr{\dt} }{ 4 \pi d^2 \xip (1 + z)^2 } = \frac{ c^2 z \varphi \nw{\Mdot} \nw{\dt} }{ 4 \pi d^2 \xip (1 + z) }
\end{equation}

\subsection{Alpha ratio}
\label{subsec:methods_observables_alpha}
A common measure of burst energetics is the ratio of the persistent to burst fluence (Section~\ref{subsec:methods_extracting_bprops}),
\begin{equation}
    \label{eq:methods_observables_alpha_define}
    \alpha = \frac{\fluencep}{\fluence}.
\end{equation}
From Equations~\eqref{eq:methods_observables_fluence-general}~and~\eqref{eq:methods_observables_fluence_p_general}, we obtain
\begin{equation}
    \label{eq:methods_observables_alpha_general}
    \alpha = \frac{\xib}{\xip} \cdot \frac{ \obssub{L}{\accsymb} \obs{\dt} }{ \obssub{E}{\burstsymb} }.
\end{equation}
Thus, $\alpha$ is independent of distance, and can also be used to infer the anisotropy ratio $\xib / \xip$.
Using Equations~\eqref{eq:methods_observables_dt}~and~\eqref{eq:methods_observables_Lacc_obs}, we obtain

\begin{equation}
    \label{eq:methods_observables_alpha_full}
    \alpha = \frac{ \xib c^2 z \gr{\Mdot} \gr{\dt} }{ \xip (1+z) \grsub{E}{\burstsymb} }   =   \frac{ \xib c^2 z \nw{\Mdot} \nw{\dt} }{ \xip \varphi(1+z) \nwsub{E}{\burstsymb} }.
\end{equation}

\subsection{Eddington flux}
\label{subsec:methods_observables_eddington}
The Eddington-limited flux ($\Fedd$) is typically inferred from the peak of PRE bursts, which are thought to reach the local Eddington luminosity ($\Ledd$).
Once again, from Equation~\eqref{eq:methods_observables_anisotropy} we have
\begin{equation}
    \label{eq:methods_observables_Fedd}
    \Fedd = \frac{ \obssub{L}{\eddsymb} }{ 4 \pi d^2 \xib }.
\end{equation}
Using Equations~\eqref{eq:methods_gr_Ledd}~and~\eqref{eq:methods_gr_observer_lum}, we obtain
\begin{equation}
    \label{eq:methods_observables_Ledd}
        \obssub{L}{\eddsymb} = \frac{ 4 \pi G c \gr{M} }{ \kappa (1+z) } = \frac{ 4 \pi G c \varphi \nw{M} }{ \kappa (1+z) }.
\end{equation}
Equation~\eqref{eq:methods_observables_Fedd} then becomes
\begin{equation}
    \label{eq:methods_observables_Fedd-general}
    \Fedd = \frac{ G c \gr{M} }{ \kappa d^2 \xib (1+z) } = \frac{ G c \varphi \nw{M} }{ \kappa d^2 \xib (1+z) }.
\end{equation}

For ionised material, the radiation pressure is exerted on electrons via Thomson scattering, whereas the mass is dominated by nucleons.
For hydrogen, we can make the approximation that the opacity is given by $\kappa = \sub{\sigma}{T} / \sub{m}{p}$, where $\sub{\sigma}{T}$ is the Thomson scattering cross section, and $\sub{m}{p}$ is the proton mass.

Additionally, if we assume that the accreted material is a mixture of hydrogen and helium, we can introduce a factor to account for the composition, $2 / (1 + \hyd)$, where $\hyd$ is the hydrogen mass fraction.
Compared to pure hydrogen ($\hyd = 1.0)$, pure helium ($\hyd = 0.0$) has double the mass per charge, resulting in double the Eddington limit.

Applying both of these values, Equation~\eqref{eq:methods_observables_Fedd-general} becomes
\begin{equation}
    \label{eq:methods_observables_Fedd-obs}
    \Fedd = \frac{ 2 G \massp c \gr{M} }{ \thomson d^2 \xib (1+z) (1 + \hyd) } = \frac{ 2 G \massp c \varphi \nw{M} }{ \thomson d^2 \xib (1+z) (1 + \hyd) }.
\end{equation}

\chapter{Results: Improvements to \kepler{} Burst Models}
\label{ch:results}
In this chapter we outline the main improvements to \kepler{} burst models used for this research.

In Section~\ref{sec:results_opacity}, we discuss a previous error in the opacities, which caused an artificially hot envelope and increased burst rate compared to other codes -- a discrepancy already noticed by others in the literature.
This correction is applied to all burst models presented in this thesis.

In Section~\ref{sec:results_preheating}, we describe the inclusion of a mock nuclear heat source during the model setup, which improved issues with thermal ``burn-in'' of the envelope.
This treatment was applied to the models presented in Chapters~\ref{ch:paper2} and \ref{ch:4u1820}, but was developed after the publication of Chapter~\ref{ch:paper1}.

Finally, in Section~\ref{sec:results_mesa}, we compare \kepler{} models to another one-dimensional (1D) burst code, \mesa{}, after including the above improvements.
This test is the closest direct comparison of 1D burst codes to date, and we demonstrate that the improvements made to \kepler{} reduce the discrepancy between the models.

\section{Corrected Opacities}
\label{sec:results_opacity}
During the preparation of models for \citet{johnston_simulating_2018} (Chapter~\ref{ch:paper1}), it was discovered\footnote{by Adam Jacobs, Michigan State University, pers.\ comm.} that the \kepler{} burst models mistakenly included an opacity multiplication factor of $\approx 1.5$.

The opacity modification originated from tests for an alternative to the GR-corrections described in Section~\ref{sec:methods_gr}.
The boosted opacity was intended to slow heat transport such that the time dilation effects of GR would be replicated.
The approach, however, was abandoned in favour of applying GR-corrections in post-processing.
The multiplication factor, however, was mistakenly left in the model setup files, transmitting the error to all subsequent studies, including \citet{woosley_models_2004, heger_models_2007, cyburt_dependence_2016, lampe_influence_2016, galloway_thermonuclear_2017}.

\begin{figure}
    \centering
    \includegraphics[width=0.8\textwidth]{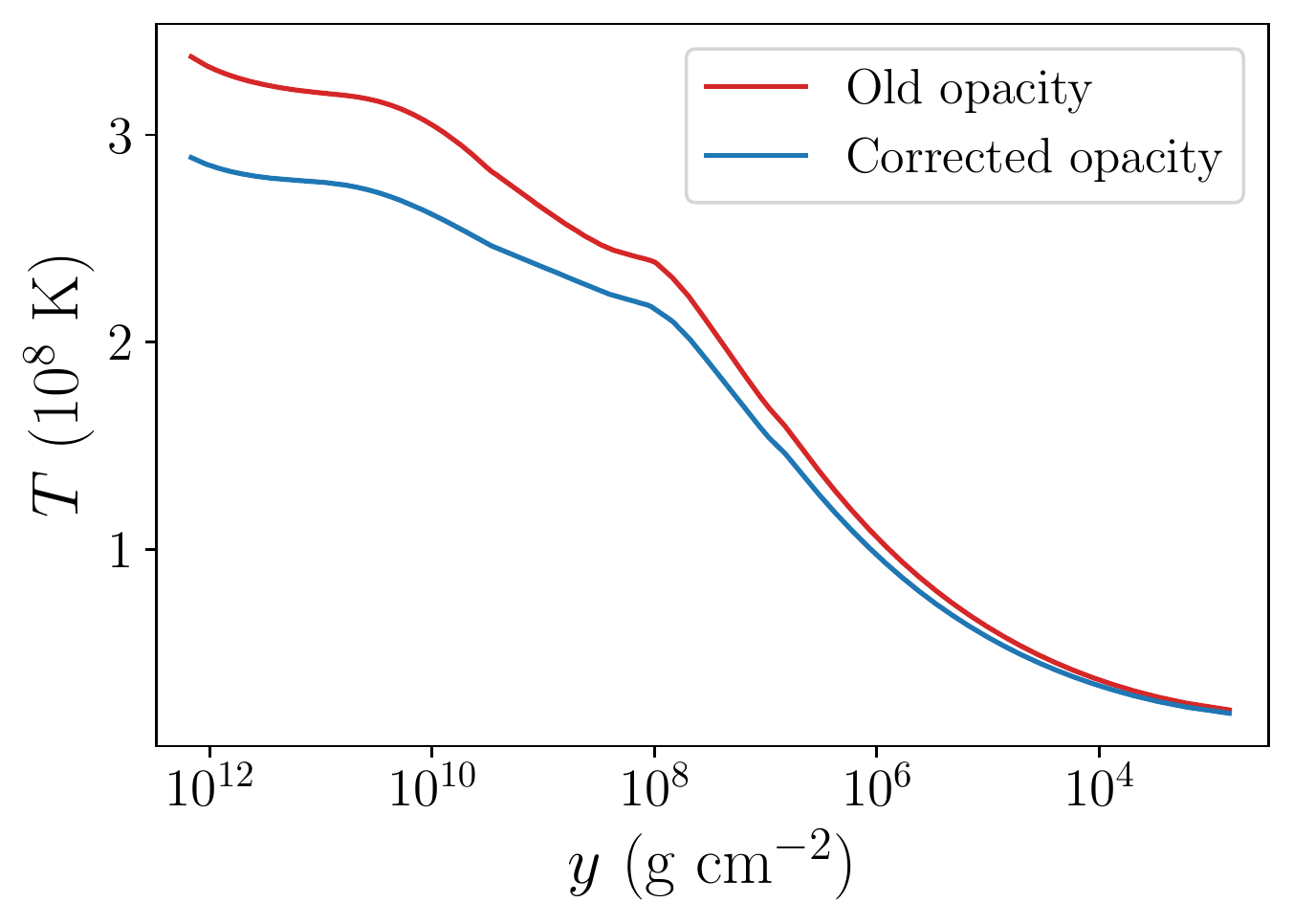}
    \caption{
    \kepler{} temperature profiles, for a model with the erroneous opacity multiplier of $\approx 1.5$ (red curve), and with the multiplier removed (blue curve).
    The larger opacities reduce the efficiency of thermal transport, producing a hotter envelope and shorter burst recurrence times (see also Table~\ref{tab:results_opacity}).
    This error was corrected for the models in \citet[][(Chapter~\ref{ch:paper1})]{johnston_simulating_2018}, and is present in all previously published \kepler{} burst models.
    }
    \label{fig:results_opacity}
\end{figure}

To test the effect of the corrected opacity, we computed two \kepler{} models for comparison.
We used the parameters from the popular reference model \textit{A3} from \citet{heger_models_2007}.
The parameters used were a hydrogen mass fraction of $\hyd = 0.7048$, CNO mass fraction of $\cno = 0.02$, crustal heating rate of $\qb = \SI{0.1}{\mevnuc}$, surface gravity of $g = \SI{1.858e14}{\gunits}$, and accretion rate of $\mdot = 0.0903\, \mdotedd$, where $\mdotedd = \SI{8.775e4}{\mdotunits}$.
The opacity multiplier was left in place for the first model, and removed from the second.
The models were run for $\approx \SI{100}{\hour}$ in the model rest frame, producing $\approx 30$ bursts each.

The temperature profiles are compared in Figure~\ref{fig:results_opacity}.
The snapshots are taken near the end of each model, prior to the ignition of the next burst, to allow the envelopes sufficient time to settle into a steady bursting state.
The larger opacity produces systematically higher temperatures, which alters the conditions for burst ignition.
The average burst properties for each model, after excluding the first 10 bursts, are listed in Table~\ref{tab:results_opacity}.
The predictions of the two models are inconsistent, and the original opacity multiplier results in recurrence times of $\dt = 2.71 \pm \SI{0.06}{\hour}$, compared to $3.38 \pm \SI{0.11}{\hour}$ when it is removed.

Because the increased opacity produces an artificially hotter envelope, the conditions for burst ignition are reached at shallower depths, and thus earlier in time for a given accretion rate.
The result is larger burst rates, $\brate$, and reduced energetics, such as the peak luminosity and burst energy.

Indeed, discrepancies between the burst codes had already been noticed.
Using the 1D code \shiva{}, \citet{jose_hydrodynamic_2010} found that their burst models predicted recurrence times a factor of $\approx 2$ longer than \kepler{} models from \citet{woosley_models_2004}.
Similarly, with the 1D code \mesa{}, \citet{paxton_modules_2015} required accretion rates a factor of $\approx 2$ larger to achieve a similar $\dt$ to \citet{heger_models_2007}.
This inconsistency with \mesa{} was reproduced by \citet{meisel_consistent_2018}, with comparisons to \kepler{} models from \citet{lampe_influence_2016}.

This systematic error in previous \kepler{} models should be taken into account when comparing to the previous studies.
Without directly recomputing the original models, however, it is difficult to apply a straightforward correction to the previous results.
As a rule of thumb, the original models can be considered equivalent to models with larger accretion rates or base heating.

We perform a more direct comparison of the updated \kepler{} models with \mesa{} in Section~\ref{sec:results_mesa}.

\renewcommand{\arraystretch}{1.3}
\begin{table}
    \centering
    \caption{
    Burst properties from models with and without the erroneous multiplier on opacity.
    Each value is an average over the burst sequence, excluding the initial 10 bursts,
    where the uncertainties are $1\sigma$ standard deviations.
    The artificially-large opacity in the original model results in systematic differences in the burst predictions.
    All values are in the local Newtonian frame of the \kepler{} model, i.e.,\ \textit{not} corrected for GR (see Section~\ref{sec:methods_gr}).
    }
    \label{tab:results_opacity}
    \begin{tabular}{llcc}
        \hline
        \hline
        && \multicolumn{2}{c}{Opacity Multiplier} \\
        \cline{3-4}
                 &                       & Old             &  Corrected \\
        \hline
        N bursts & --                    & 38              & 30              \\
        $\dt$    & (h)                   & $2.71 \pm 0.06$ & $3.38 \pm 0.11$ \\
        $\brate$ & (day$^{-1}$)          & $8.9 \pm 0.2$   & $7.1 \pm 0.2$   \\
        $\Lpeak$ & $(\SI{e38}{\Lunits})$ & $1.25 \pm 0.08$ & $2.04 \pm 0.17$ \\
        $\Eb$    & $(\SI{e39}{erg})$     & $4.08 \pm 0.10$ & $4.92 \pm 0.11$ \\
        \hline
    \end{tabular}
\end{table}

\section{Nuclear Preheating and Model Burn-in}
\label{sec:results_preheating}
The thermal history of the neutron star envelope can shape its bursting behaviour \citep{taam_x-ray_1980}.
Bursts produce nuclear heating and leftover ashes, which determine the thermal and compositional state of the envelope for subsequent bursts.
This thermal and compositional ``inertia'' necessitates the simulation of many consecutive bursts, in order to reach a quasi-periodic limit cycle \citep{woosley_models_2004}.
Time-dependent burst models, therefore, are subject to an initial ``burn-in'' phase, which is
then excluded from our analysis.

In previous \kepler{} studies, typically only the first 1--2 bursts were discarded \citep[e.g.,][]{heger_models_2007, cyburt_jina_2010, lampe_influence_2016}.
The models were assumed to have reached a steady state by that point, particularly because the first burst is such an energetic outlier by comparison \citep{woosley_models_2004}.
During our model tests, however, we discovered systematic trends in the burst properties which can persist for tens of bursts.
Because 10--30 bursts are usually simulated per model \citep[e.g.,][]{heger_models_2007, lampe_influence_2016}, this extended burn-in can potentially impact the entire model sequence.

A possible contribution to model burn-in is that nuclear heating, $\qnuc$, is not accounted for when setting up the envelope.
The base flux from crustal heating, $\qb$, is used as a lower boundary condition at $y \approx \SI{e12}{\yunits}$, and the envelope is relaxed to thermal equilibrium.
Mass accretion and nuclear burning are then switched on, and the full simulation begins.
Because the thermal profile is dominated by heat from the crust, only the crustal heating rate, $\qb \approx \SI{0.1}{\mevnuc}$, is included.
Nuclear heating in the shallower layers of $y \sim 10^7$--$\SI{e8}{\yunits}$ was assumed to stabilise within the first few bursts, and have little impact on the overall thermal profile.
If nuclear heating does significantly contribute to the long-term thermal structure of the envelope, then the existing models are out-of-equilibrium.

We tested the influence of nuclear heating on the model setup and subsequent simulation.
In addition to $\qb$, we include a heat source of $\qnuc = \SI{5}{\mevnuc}$ at a depth of $y = \SI{8e7}{\yunits}$, distributed with a Gaussian width of $\sigma = \SI{8e6}{\yunits}$.
The envelope is relaxed to equilibrium, and when the simulation begins, the $\qnuc$ source is switched off and the full nuclear network calculations are enabled.

We tested this implementation with three sets of model parameters.
The first set was for mixed hydrogen/helium (H/He) bursts, such as those observed from \gs{} \citep[as modelled by][]{heger_models_2007}.
The second and third sets were for ``pure'' helium (He) bursts, such as those observed from \fouru{} \citep[as modelled by][]{cumming_models_2003}, which exhibit photospheric-radius expansion (PRE).
For the H/He set, we used an accreted hydrogen mass fraction of $\hyd = 0.73$, a CNO mass fraction of $\cno = 0.005$, a crustal heating of $\qb = \SI{0.05}{\mevnuc}$, a surface gravity of $g = \SI{2.654e14}{\gunits}$, and an accretion rate of $\mdot = 0.2\, \mdotedd$, where $\mdotedd = \SI{8.775e4}{\mdotunits}$.
For the pure He sets, we used $\hyd = 0.0$, $\cno = 0.015$, $\qb = \SI{0.1}{\mevnuc}$, $g = \SI{1.858e14}{\gunits}$, and accretion rates of $\mdot = 0.2$ and $0.4\, \mdotedd$, respectively.
For each set, we computed one model with the original $\qb$-only setup, and another with the nuclear ``preheating'' setup.

\begin{figure}
    \centering
    \subfloat{\includegraphics[width=0.8\textwidth]{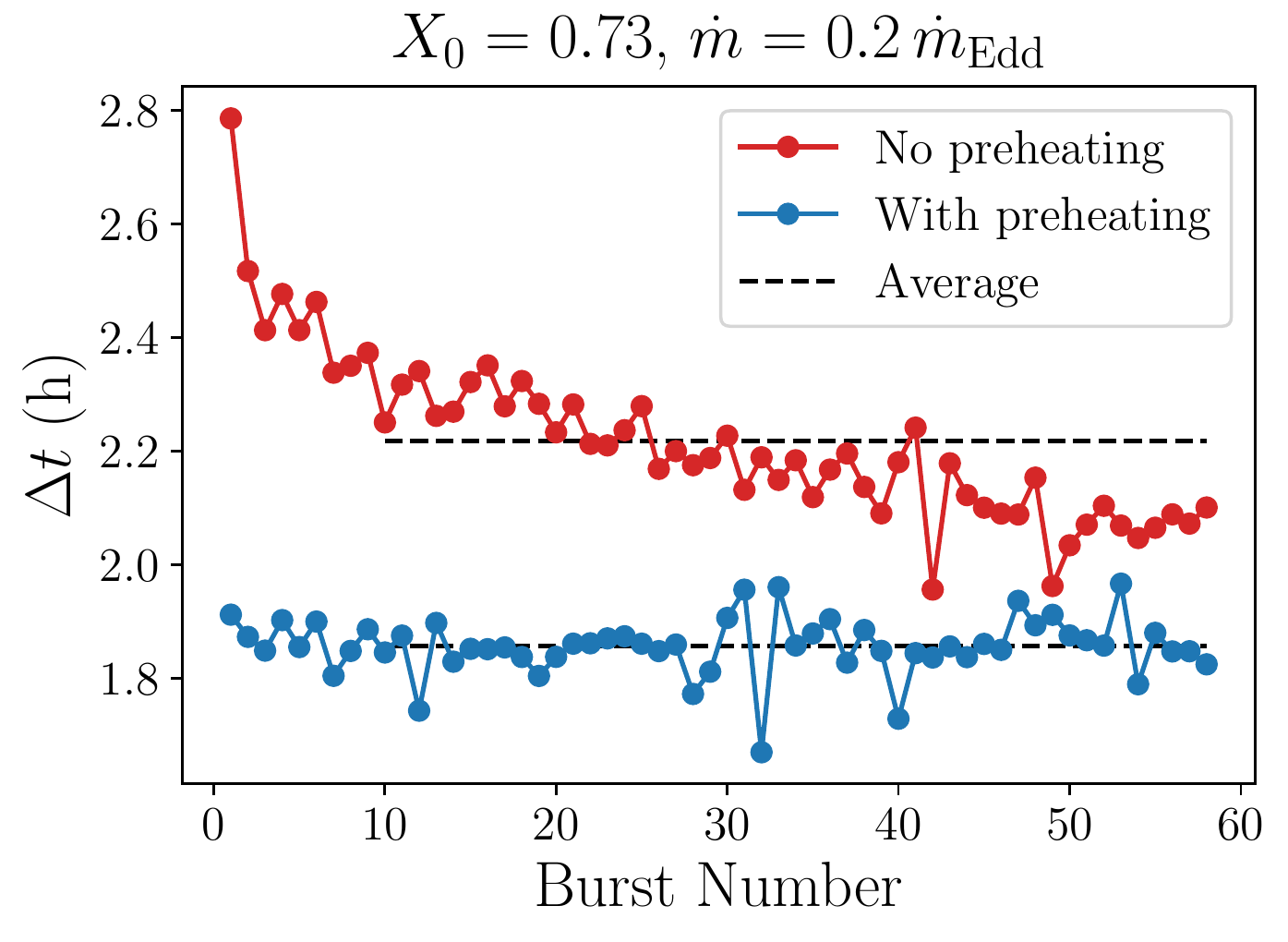}}

    \subfloat{\includegraphics[width=0.8\textwidth]{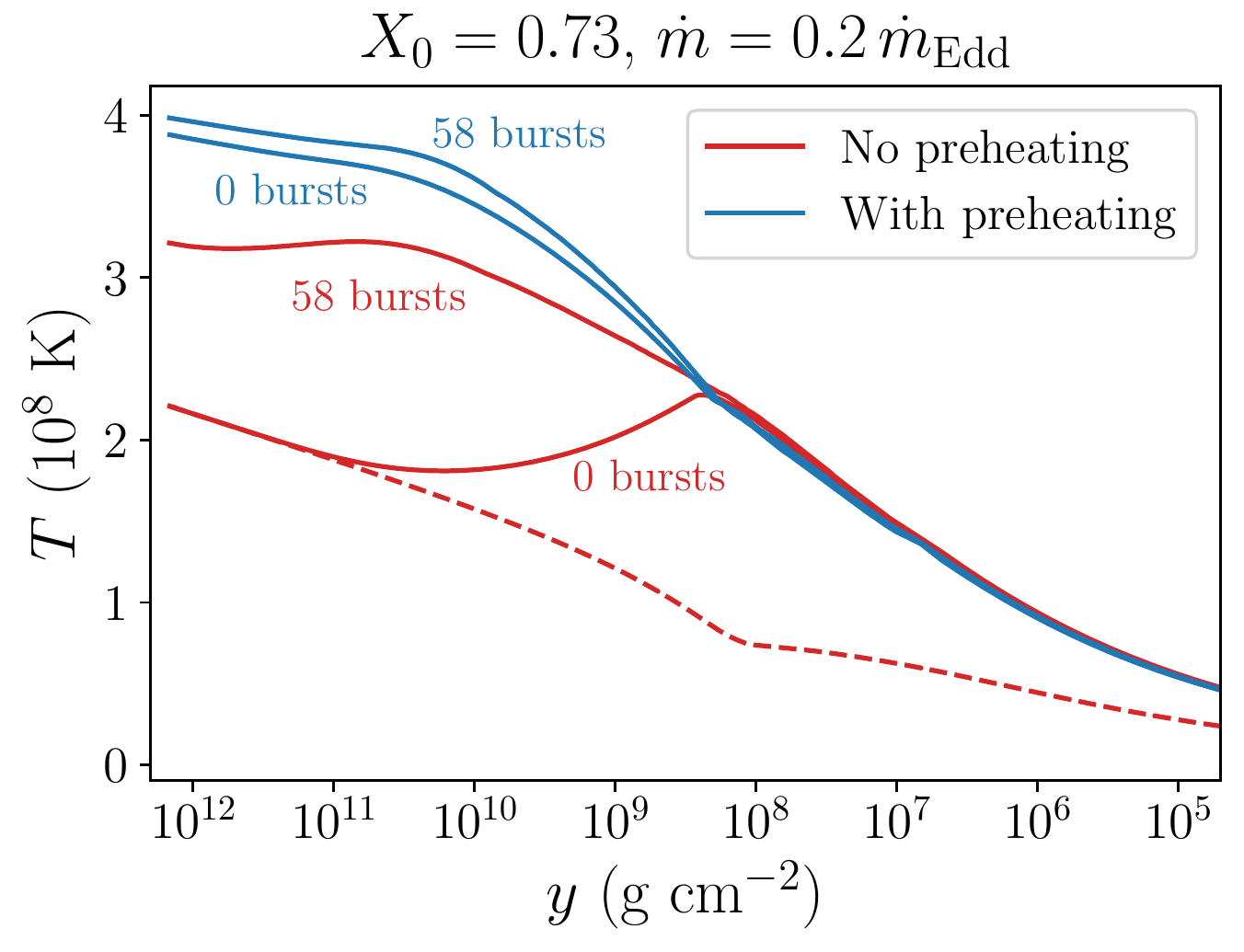}}
    \caption{The effect of including mock nuclear heating during the setup of mixed hydrogen/helium models.
    (\textit{Upper panel}): The recurrence times, $\dt$, of model burst sequences both with and without nuclear preheating.
    The horizontal dashed lines are the average values, after excluding the first 10 bursts.
    (\textit{Lower panel}): The corresponding temperature profiles for the two models, at selected points along the burst sequence.
    The profiles are taken shortly before the following burst ignites.
    The coloured text associated with each curve indicates the number of bursts that have elapsed.
    The red dashed curve is at $t = 0$, and the adjacent red ``0 bursts'' curve is taken before the first burst ignites.
    A version of this figure appears in \citet[][Chapter~\ref{ch:paper2}]{johnston_multi-epoch_2019}.}
    \label{fig:results_preheating_gs}
\end{figure}

\subsection{Mixed Hydrogen/Helium Models}
\label{subsec:results_preheating_gs}
In Figure~\ref{fig:results_preheating_gs}, the resulting burst sequences (upper panel) and thermal profiles (lower panel) for the H/He models are shown.
A clear trend of decreasing recurrence time, $\dt$, can be seen for the original setup without preheating (red points).
This trend continues even up to $\approx 60$ bursts -- roughly double the length of typical \kepler{} simulations \citep[e.g.,][]{heger_models_2007, lampe_influence_2016}.
In contrast, when nuclear preheating is included, the simulation has reached a steady bursting state within the first few bursts.
As indicated by the dashed lines, an average taken from the original setup would overestimate $\dt$ by $\approx \SI{20}{\%}$.

The burn-in is further illustrated by the temperature profiles (Figure~\ref{fig:results_preheating_gs}, lower panel).
Without accounting for nuclear heating, the envelope begins $\approx \SI{50}{\%}$ colder than equilibrium, and steadily heats up once the full nuclear calculations of bursts are included (red curves).
After 58 bursts, the temperature of the deeper layers ($y \gtrsim \SI{e10}{\yunits}$) has increased by $\approx \SI{50}{\%}$.
With nuclear preheating, the simulation begins much closer to equilibrium, and the temperature of the deeper layers has only increased by $\approx \SI{2}{\%}$ after 58 bursts.
This small increase in temperature suggests that equilibrium has not yet been achieved, although a slight offset is expected given the relatively crude treatment for nuclear heating.
Nevertheless, the effect of this leftover burn-in on $\dt$ appears to be smaller than the inherent burst-to-burst variation.

\begin{figure}
    \centering
    \subfloat{\includegraphics[width=0.8\textwidth]{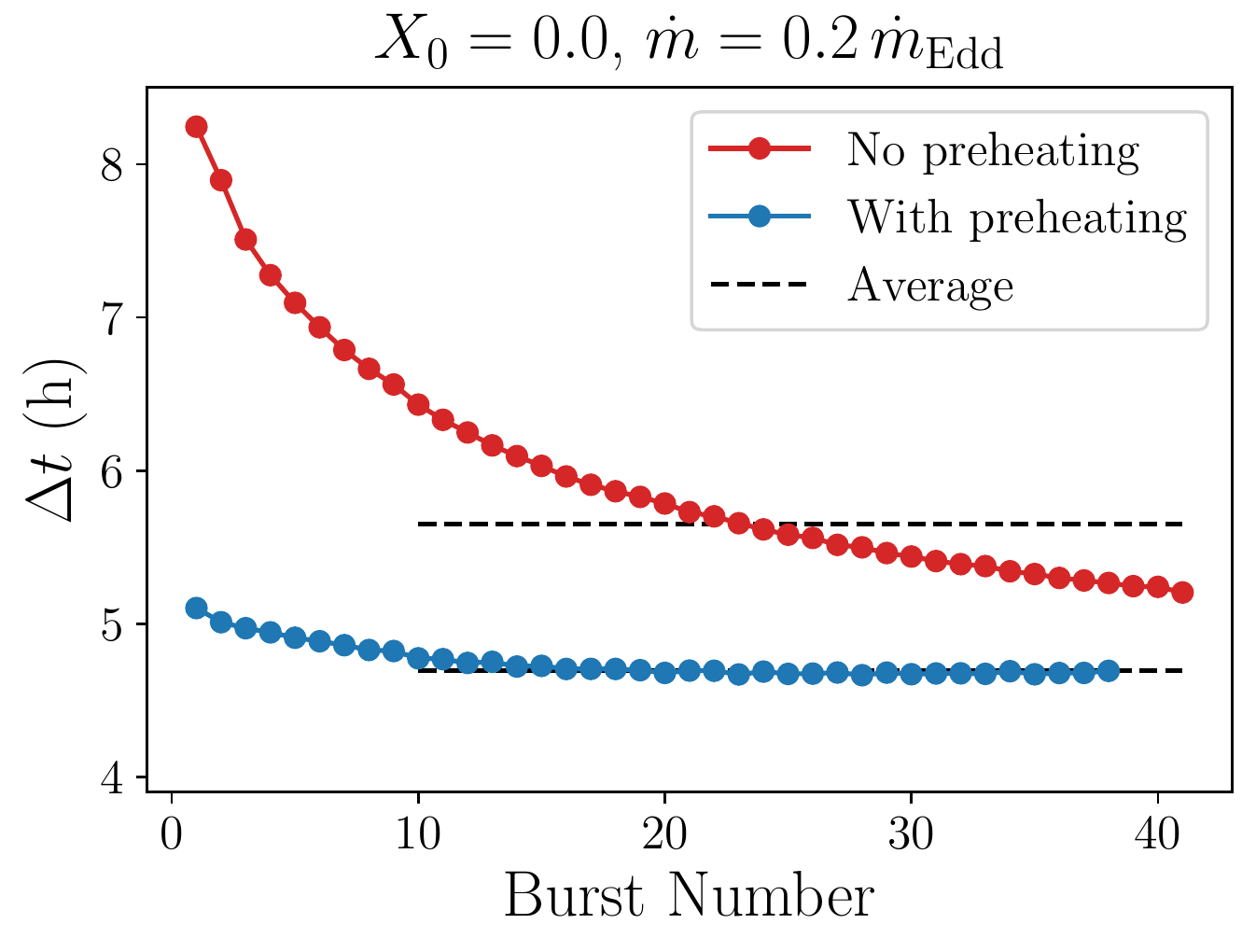}}

    \subfloat{\includegraphics[width=0.8\textwidth]{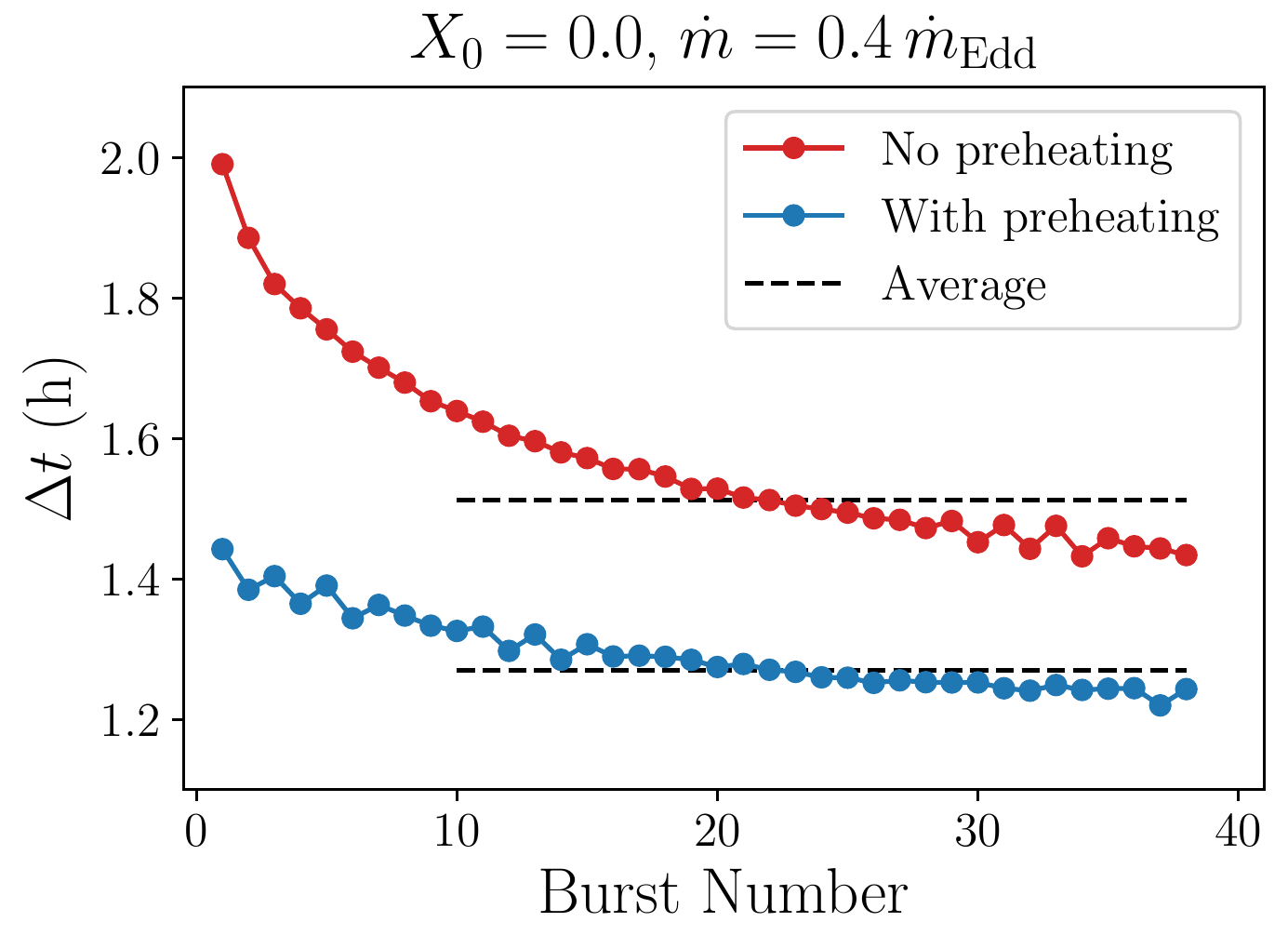}}
    \caption{The same as the upper panel of Figure~\ref{fig:results_preheating_gs}, but for two pure helium models with $\mdot = 0.2$ (\textit{upper panel}) and $0.4\, \mdotedd$ (\textit{lower panel}).
    The preheating implementation is identical to the H/He models, with a heat source of $\qnuc = \SI{5}{\mevnuc}$ centred at a depth of $y = \SI{8e7}{\yunits}$.
    }
    \label{fig:results_preheating_4u_train}
\end{figure}

\subsection{Pure Helium Models}
\label{subsec:results_preheating_4u}
\begin{figure}
    \centering
    \subfloat{\includegraphics[width=0.8\textwidth]{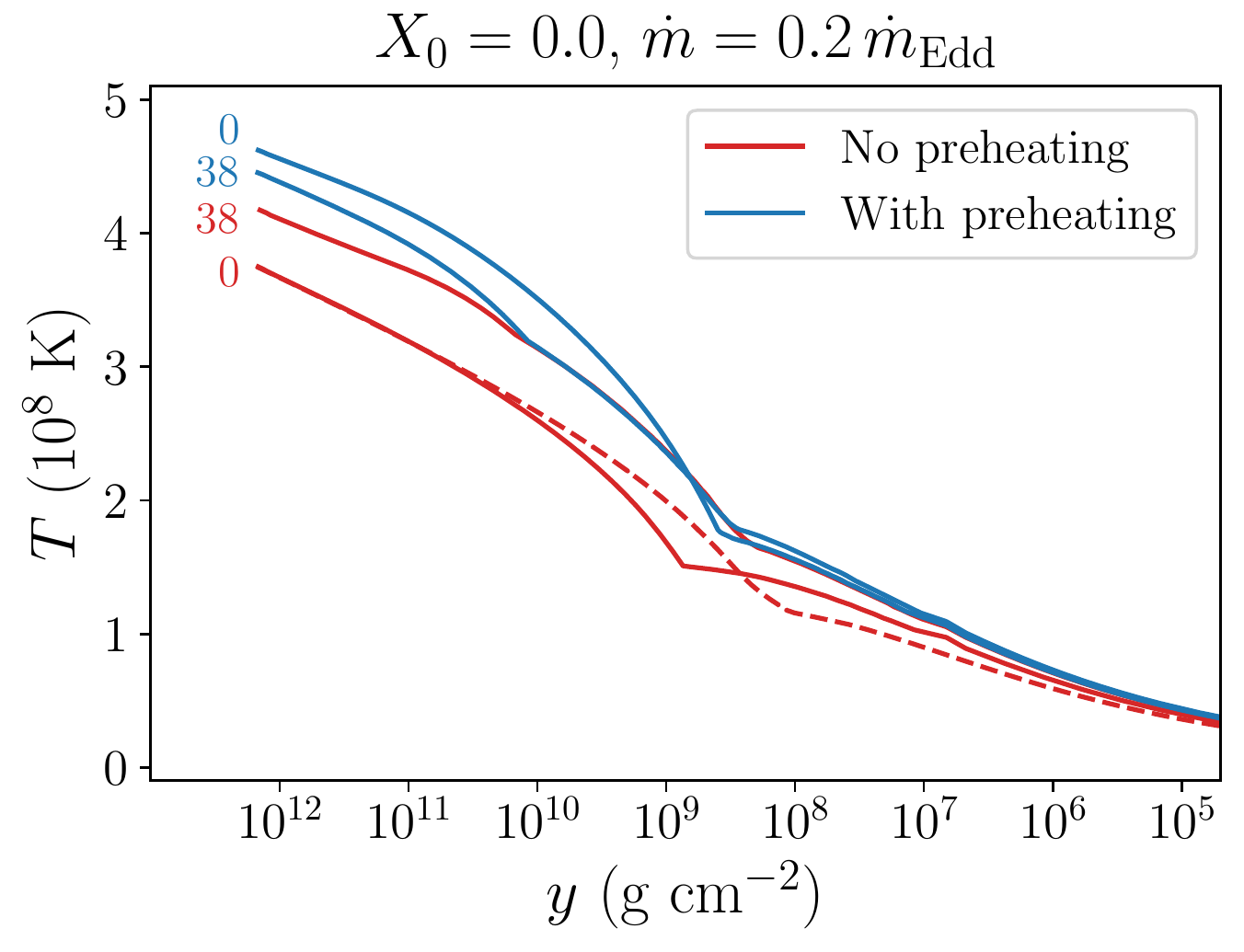}}

    \subfloat{\includegraphics[width=0.8\textwidth]{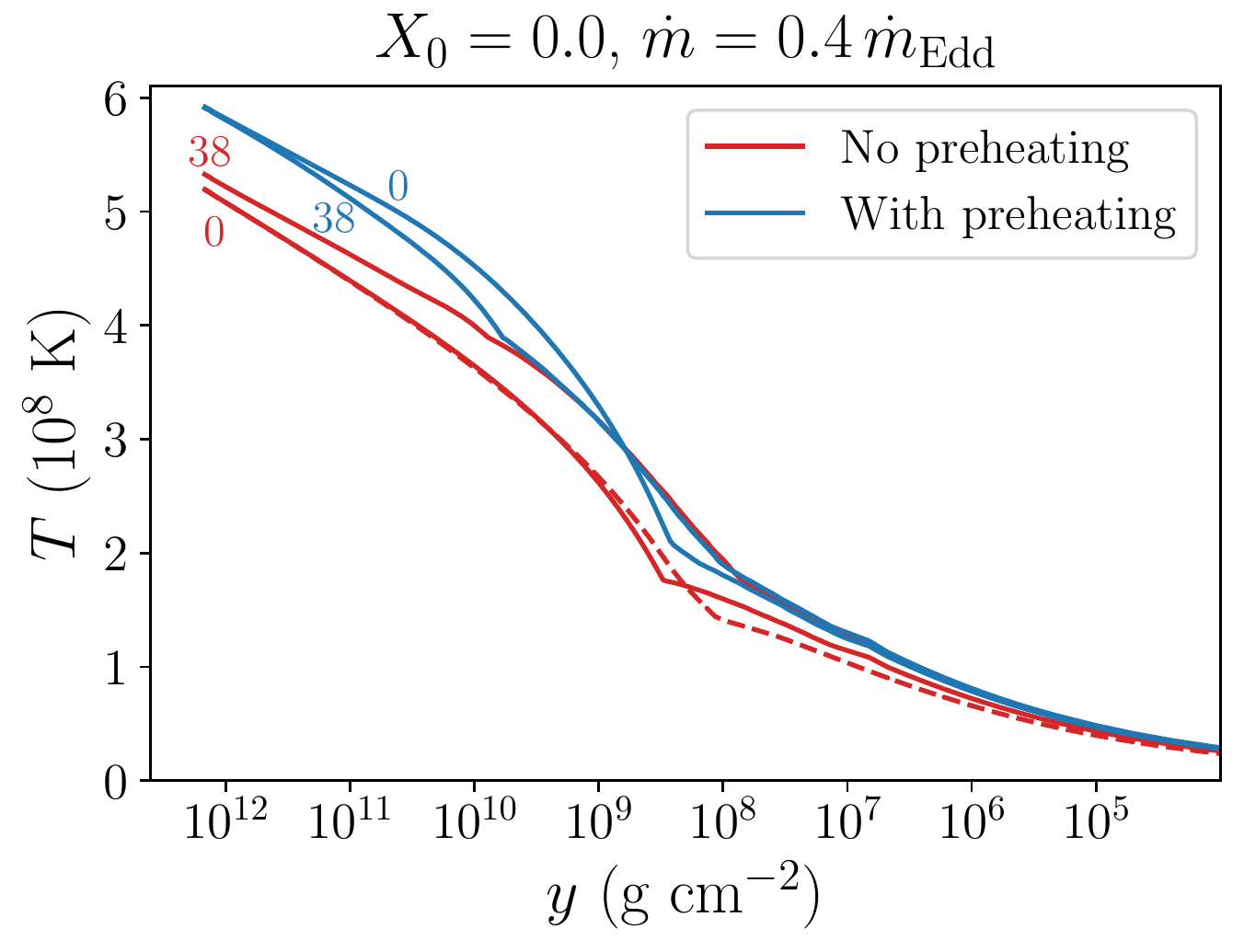}}
    \caption{The same as the lower panel of Figure~\ref{fig:results_preheating_gs}, but for the corresponding pure He models in Figure~\ref{fig:results_preheating_4u_train}.
    The coloured numbers next to each curve signify the number of bursts elapsed.
    The red dashed curve is $t = 0$.
    Note the inversion of heating/cooling in the deeper layers, at $y \gtrsim \SI{e9}{\yunits}$.
    The kink in each curve corresponds to the transition in composition from the accreted material to the inert iron substrate.
    }
    \label{fig:results_preheating_4u_temp}
\end{figure}

The burst sequences for the pure He models are shown in Figure~\ref{fig:results_preheating_4u_train}, and the corresponding temperature profiles are shown in Figure~\ref{fig:results_preheating_4u_temp}.
Similar to the mixed H/He case, the models without nuclear preheating exhibit systematic trends in $\dt$.
The addition of preheating suppresses the time required for burn-in for $\mdot = 0.2\, \mdotedd$, and the bursts stabilise after 10--20 bursts.
For $\mdot = 0.4\, \mdotedd$, however, the burn-in appears to continue even after 38 bursts, although at a reduced rate.
The burst-to-burst variation in $\dt$ for the last 10 bursts is $< \SI{1}{\%}$.

The temperature profiles for both $\mdot = 0.2$ and $0.4\, \mdotedd$ begin colder without preheating, and systematically heat up over the course of 38 bursts.
The models with preheating, however, behave differently to the H/He case.
Whereas the shallow layers ($y \lesssim \SI{e9}{\yunits}$) start hotter than the original setup and slightly heat up, the deeper layers ($y \gtrsim \SI{e9}{\yunits}$) actually cool down after 38 bursts.
The ``kink'' in each profile corresponds to the transition from the accreted material to the inert iron substrate, which represents the deep ocean of previous burst ashes (see Section~\ref{subsec:methods_kepler_setup}).
In the H/He models (Figure~\ref{fig:results_preheating_gs}), this interface is smoothed out as the simulated ashes become closer in composition to the substrate.
In the pure He models, however, this interface persists, suggesting that the iron substrate is not a good representation for the ashes of helium bursts.

\subsection{Discussion}
\label{subsec:results_preheating_discussion}
The addition of a nuclear heat source during the setup of \kepler{} models significantly shortened the systematic model burn-in.
For the mixed hydrogen/helium case tested here, the number of bursts required was reduced from $\gtrsim 60$ to $\lesssim 10$ (Figure~\ref{fig:results_preheating_gs}).
A small amount of heating persisted in the deeper envelope, although any slight trend in $\dt$ appears to be hidden by the burst-to-burst variation.
For the two sets of pure helium bursts tested, some burn-in remained even after 30--40 bursts, particularly for the higher accretion rate of $\mdot = 0.4\, \mdotedd$ (Figure~\ref{fig:results_preheating_4u_train}).
Nevertheless, the burst-to-burst variation dropped to $< \SI{1}{\%}$, but longer simulations are still required to determine how long the downward trend in $\dt$ continues.
Longer simulations could also test whether both treatments -- with and without preheating -- do eventually converge to the same burst properties.

Future work is still needed to investigate the optimal strength, location, and distribution of nuclear preheating, depending on the model parameters.
In these tests, we used a heating strength of $\qnuc = \SI{5}{\mevnuc}$ located at $y = \SI{8e7}{\yunits}$, with a Gaussian distribution of width $\sigma = \SI{8e6}{\yunits}$.
Whereas the model burn-in was essentially eliminated for H/He models, the ignition depth and total energy release could be refined for pure He models.
Additionally, the persistence of a kink in the pure He temperature profiles (Figure~\ref{fig:results_preheating_4u_temp}) suggests that a lighter composition may be needed than the existing iron substrate.
Despite its limitations, our preheating treatment can potentially save days of computation time, by reducing the number of bursts required to obtain reliable predictions.

\section{Updated Comparison to \mesa{} Burst Models}
\label{sec:results_mesa}
Comparing the predictions of different codes is an important test of code verification and reproducibility.
Although comparisons have been made in previous works \citep[e.g.,][]{jose_hydrodynamic_2010, paxton_modules_2015, meisel_consistent_2018}, the model input parameters are typically slightly different, partly due to the limited number of published models available.
No comparison yet exists for two burst codes using the same set of $\hyd$, $\cno$, $\qb$, $\mdot$, and $g$.

Discrepancies between the predictions of \kepler{} and other 1D burst codes have been noted in the literature (see Section~\ref{sec:results_opacity}).
During the course of this research, we discovered that incorrect opacities were being used in \kepler{}, which likely contributed to this inconsistency (Section~\ref{sec:results_opacity}).
To test our updated model setup, and whether the corrected opacity improves code agreement, we have computed a set of simulations for direct comparison with existing \mesa{} models.

\begin{figure}
    \centering
    \includegraphics[width=0.7\textwidth]{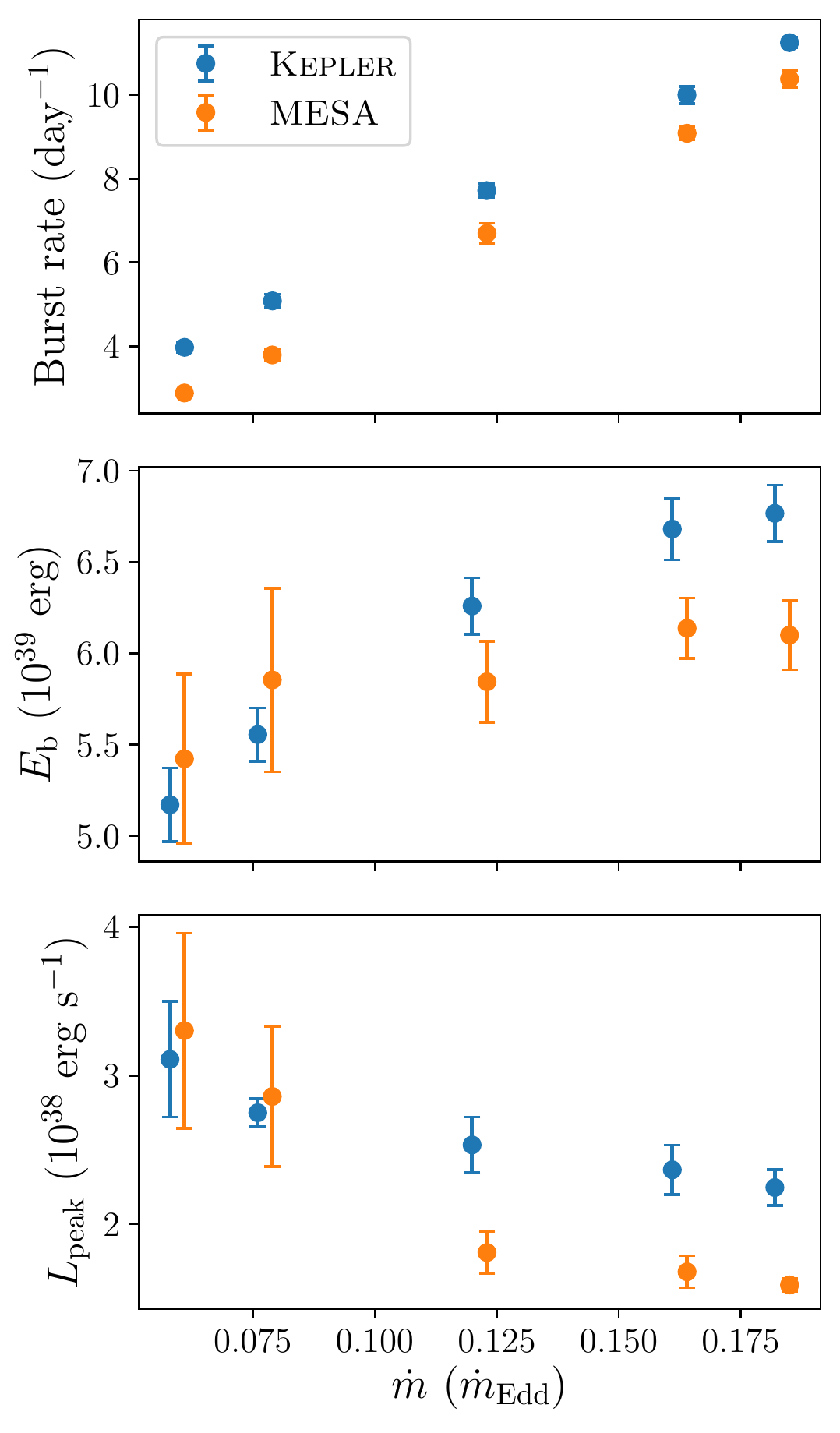}
    \caption{Comparison of burst properties predicted by \kepler{} and \mesa{} models, using the same input parameters.
    The \kepler{} values, including $\Mdot$, have been GR-corrected to the equivalent local frame of the \mesa{} models.
    The points have been slightly offset horizontally for clarity.
    }
    \label{fig:results_mesa_bprops}
\end{figure}

\subsection{Model setup}
\label{subsec:results_mesa_setup}
For our comparison, we used \mesa{} burst models of the mixed hydrogen/helium burster, \gs{}, produced by \citet{meisel_consistent_2018}.
We chose a subset of five models which were among the best fits to the observations, labelled as \textit{ma}1--\textit{ma}6 in the dataset\footnote{available at \url{https://inpp.ohio.edu/~meisel/MESA/mesaresults.html}} (excluding \textit{ma}3, which did not produce bursts).

All five models used a hydrogen mass fraction of $\hyd = 0.7$, a CNO mass fraction of $\cno = 0.02$, a crustal heating of $\qb = \SI{0.1}{\mevnuc}$, and a surface gravity of $g = \SI{1.858e14}{\gunits}$, corresponding to a gravitational mass of $\nsmass$, and a radius of $R = \SI{11.2}{km}$.
The models differed only by accretion rate, for $\Mdot = 0.061$, 0.079, 0.123, 0.164, and $0.185\, \Mdotedd$, where $\Mdotedd = \SI{1.75e-8}{\msun.yr^{-1}}$.
These values differ slightly from those reported in \citet{meisel_consistent_2018}, because \mesa{} ``settles'' into a target $\Mdot$, and we have taken the accretion rate averaged over the whole model.

We computed a set of five \kepler{} models using these same parameters, with some modifications to ensure consistency between the codes.
Because in \kepler{} Newtonian gravity, we used $\nsmass$ and $\nsradius$ to reproduce the same $g$.
With appropriate corrections for general relativity (GR), the \kepler{} models are equivalent to a neutron star of $\nsmass$ and $R = \SI{11.2}{km}$ (see Section~\ref{sec:methods_gr}).
The radius ratio is then $\xi = \gr{R} / \nw{R}$, where the subscripts ``g'' and ``k'' correspond to \mesa{} and \kepler{}, respectively.
For a given accretion rate in \mesa{}, the equivalent accretion rate used in \kepler{} is $\nw{\Mdot} = \xi^{-2} \gr{\Mdot}$.
We used V2.2 of the nuclear reaction rate library, \textsc{REACLIB} \citep{cyburt_jina_2010}, as used in the \mesa{} models.
To our knowledge, the nuclear preheating treatment we describe in Section~\ref{sec:results_preheating} is not implemented in \mesa{}, and so we disabled it for this comparison.

Sequences of $\approx 30$ bursts were computed for each model, and the average burst properties calculated using the methods described in Section~\ref{sec:methods_extracting}.
The $1\sigma$ standard deviations were taken as the model uncertainties.
To avoid potential differences in analysis techniques, we used these same routines to extract the burst properties from \mesa{}.
The Newtonian quantities predicted by \kepler{} were corrected to the equivalent GR neutron star frame of \mesa{}, using the procedure described in Section~\ref{subsec:methods_gr_bprops}.

\begin{figure}
    \centering
    \includegraphics[width=\textwidth]{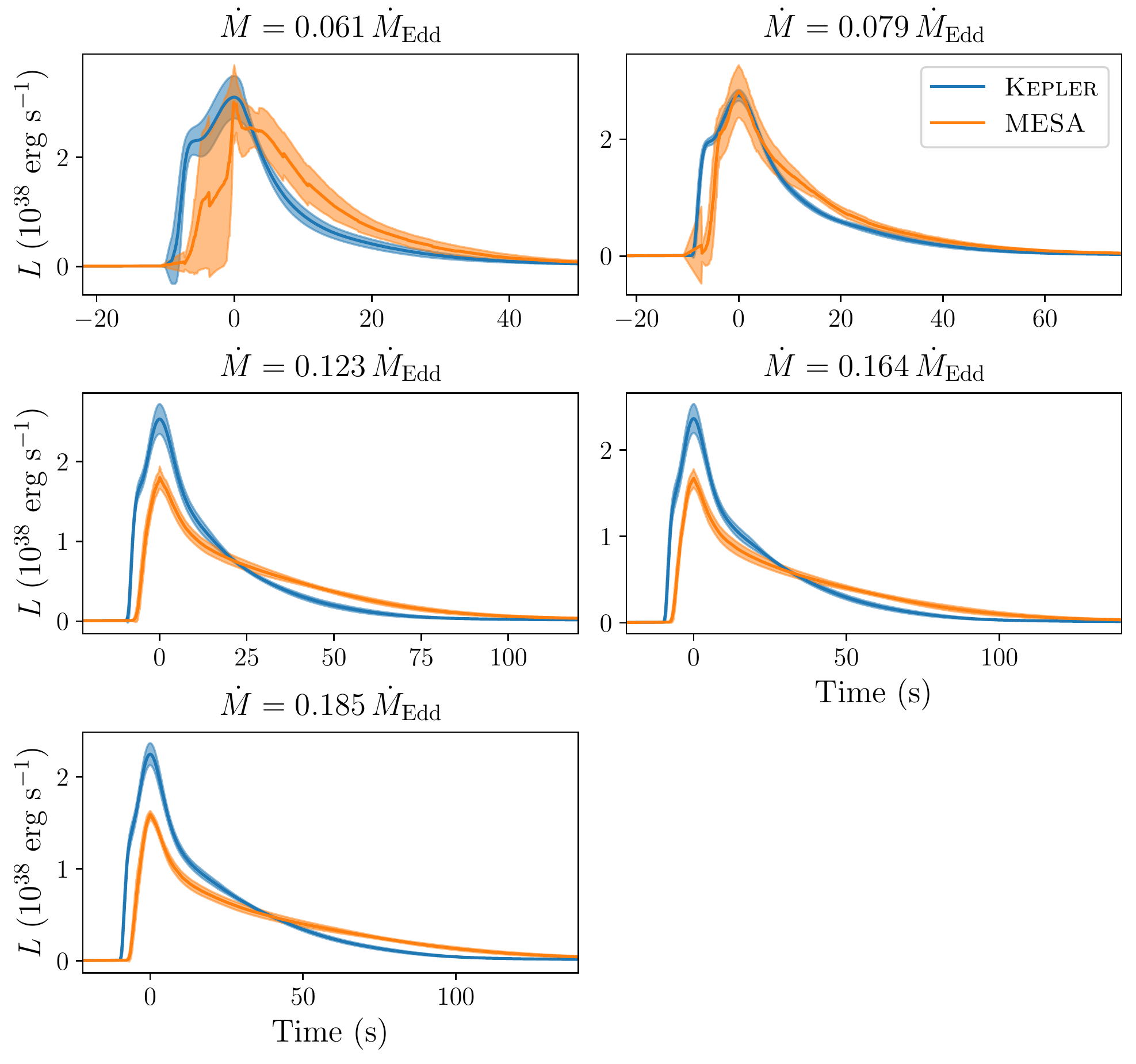}
    \caption{Comparison of the average burst lightcurves predicted by \kepler{} and \mesa{}, for five accretion rates, aligned by peak.
    The shaded regions are the $1\sigma$ standard deviations in $L$.
    The \kepler{} luminosities have been corrected for the area ratio, $\xi^2$.
    Quantities are in the local frame of the neutron star surface, and are not redshifted.
    }
    \label{fig:results_mesa_lc}
\end{figure}

\subsection{Results}
\label{subsec:results_mesa_results}
The predicted burst rate, $\brate$, burst energy, $\Eb$, and peak luminosity, $\Lpeak$, are plotted for each model in Figure~\ref{fig:results_mesa_bprops}.
Despite broad similarities for the predicted trends and values, disagreement remains between the codes.

For the burst rates, there is a persistent offset of $\approx 1$ burst per day, although the linear relationship with $\Mdot$ itself is consistent.
This relatively uniform offset suggests there may still be systematic issues affecting the codes.

The burst energies agree within uncertainties for the lowest accretion rates of $\Mdot = 0.061$ and $0.079\, \Mdotedd$, but diverge for the higher rates of $0.123$--$0.185\, \Mdotedd$, for which \kepler{} produces $\approx \SI{10}{\%}$ larger $\Eb$.
Similar to $\Eb$, the peak luminosities agree for the two lowest $\Mdot$ values, but the \kepler{} values are $\approx \SI{40}{\%}$ larger.

The average burst lightcurves are compared in Figure~\ref{fig:results_mesa_lc}, and further illustrate the pattern noted above for $\Eb$ and $\Lpeak$.
The two lowest $\Mdot$ broadly agree, whereas above $0.123\, \Mdotedd$ the \kepler{} models predict systematically stronger bursts.
On the other hand, \mesa{} yields longer lightcurve tails, which would normally indicate a larger hydrogen fraction and stronger \textit{rp}-process burning.
The longer recurrence times of the \mesa{} bursts, however, should instead result in less hydrogen at ignition due to hot CNO burning.
The \mesa{} lightcurve for $\Mdot = 0.061\, \Mdot$ exhibits erratic behaviour, suggesting an issue with the average lightcurve, perhaps from misaligned individual lightcurves.

This limited study represents the first direct comparison between 1D burst codes, using matched input parameters for $\Mdot$, $\hyd$, $\cno$, $\qb$, and $g$.
Whereas the models generally agree within uncertainties at low accretion rates, \kepler{} consistently predicts stronger bursts at higher $\Mdot$.
\kepler{} also produces larger burst rates than \mesa{} for all $\Mdot$ considered here, although this result appears to be at odds with the expected behaviour for shorter recurrence times.
Future studies are required to quantify these differences in detail, and their possible dependence on burst regime.
Additional sets of composition, crustal heating, and gravity should be explored for parameter sensitivities.
Closer inspection of the model profiles, thermal structure, and reaction networks will also help to understand the discrepancies between the codes.

\chapter{Simulating X-ray Bursts During a Transient Accretion Event}
\label{ch:paper1}

\includepdf[pages=-]{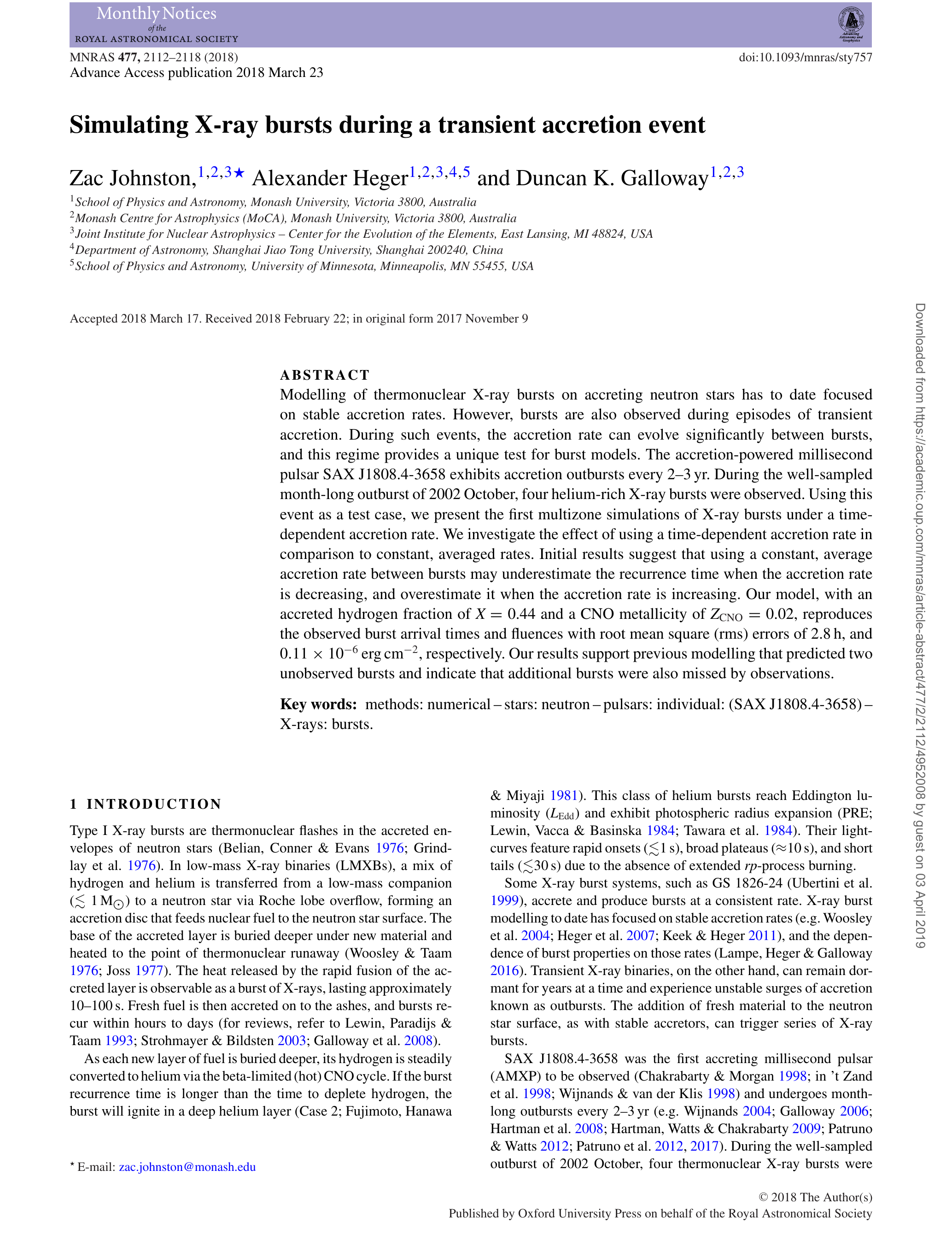}

\chapter{Multi-epoch X-ray burst modelling: MCMC with large grids of 1D simulations}
\label{ch:paper2}

\includepdf[pages=-]{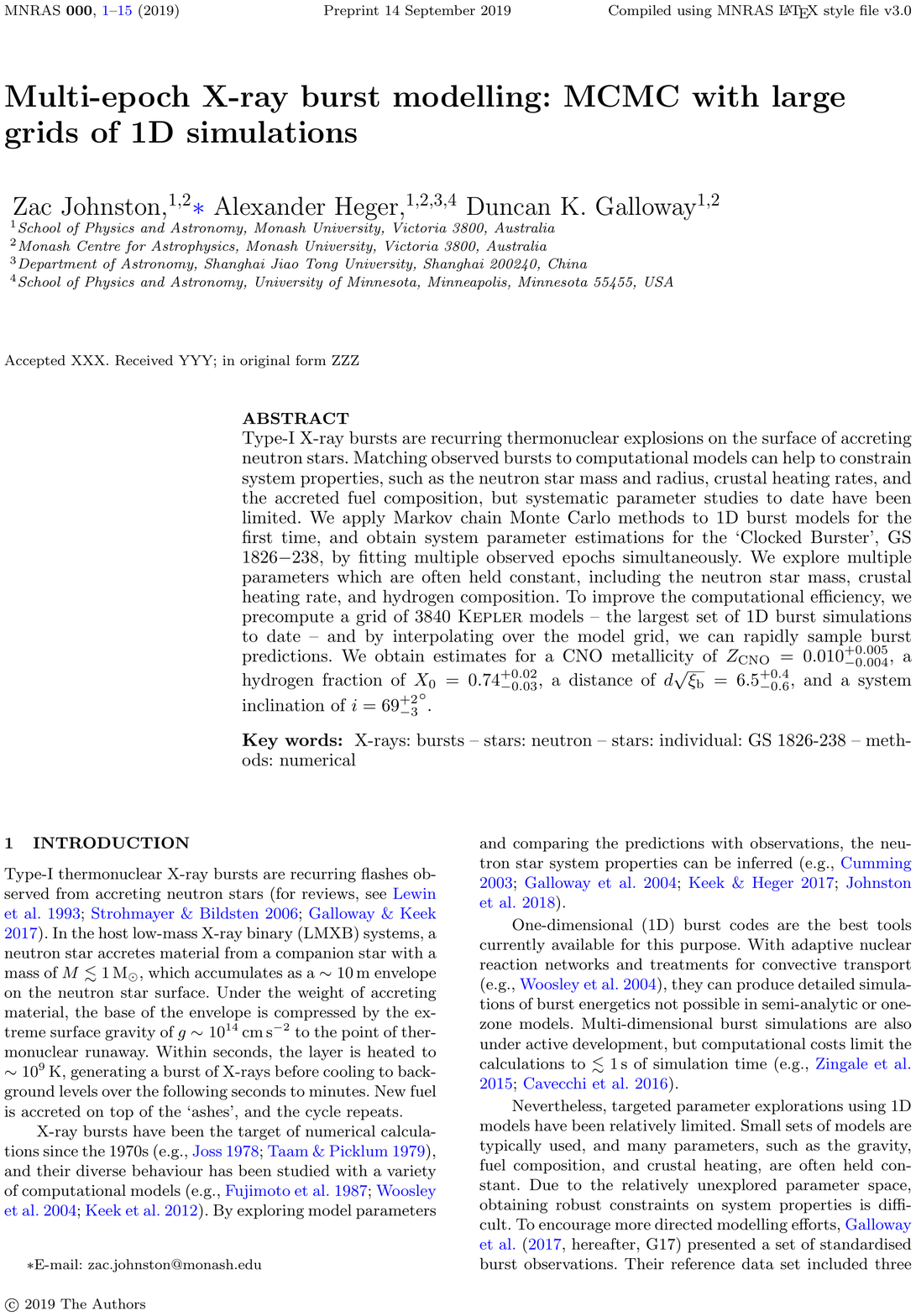}
\chapter{Multi-epoch MCMC Models of a Helium Burster}
\label{ch:4u1820}

In Chapter~\ref{ch:paper2} we applied Markov Chain Monte Carlo (MCMC) methods to one-dimensional (1D) burst models, to match multi-epoch observations of the ``Clocked Burster'', \gs{}.
We demonstrated the potential for using precomputed model grids to efficiently obtain probability distributions over system parameters.
We present here a provisional extension of these methods to \fouru, a helium-accreting system which exhibits photospheric radius-expansion (PRE) bursts.
Modelling PRE bursts poses additional challenges to mixed hydrogen/helium (H/He) bursts, and for the scope of this project, the grid parameters were kept limited in comparison to our study of \gs{}.
We therefore note that the resulting parameter estimates and model predictions should be considered provisional, and further work is needed to address the limitations discussed below.

\section{4U 1820$-$30}
\label{sec:4u1820_background}
\fouru{} is a low-mass X-ray binary (LMXB) in the globular cluster \ngc{}, and was the source in which X-ray bursts were discovered \citep{grindlay_discovery_1976}.
The companion star is a white dwarf and has one of the shortest known orbital periods of $\SI{11}{\minute}$ \citep{king_shortest_1986, stella_discovery_1987}.
This places the system in the class of ultra-compact binaries with periods of $\lesssim \SI{1}{h}$.
The compact orbit indicates that the material accreted onto the neutron star is hydrogen-poor, with previous estimates for a hydrogen mass fraction of $\hyd \lesssim 0.1$ \citep{cumming_models_2003}, and $\hyd = 0.0$ is often assumed \citep[e.g.,][]{suleimanov_basic_2017}.
The accreted fuel triggers helium bursts, in contrast to the mixed H/He bursts of \gs{} we modelled in Chapter~\ref{ch:paper2}.
Helium bursts are frequently characterised by photospheric radius-expansion (PRE), which is thought to occur when the burst luminosity reaches the local Eddington limit, $\Ledd$ \citep[e.g.,][]{kuulkers_photospheric_2003}.

Because \fouru{} resides in a globular cluster, we enjoy the benefit of independent distance measurements.
With optical observations, \citet{kuulkers_photospheric_2003} obtained a distance to \ngc{} of $d = \SI{7.6 \pm 0.4}{kpc}$, and \citet{valenti_near-infrared_2007} obtained $\SI{8.4 \pm 0.6}{kpc}$ using near-infrared measurements.

A superburst was observed from \fouru{} in September 1999 \allowbreak\citep{strohmayer_remarkable_2002}.
These rare, energetic ($\sim \SI{e42}{erg}$) bursts are thought to result from the ignition of a deep carbon ocean \citep{woosley_-ray_1976}.
The occurrence of a superburst in \fouru{} suggests that carbon is steadily accumulated during the nuclear processing of the accreted fuel \citep{cumming_carbon_2001}.

Alongside \gs{}, a multi-epoch dataset for \fouru{} was included in \citet[][hereafter G17]{galloway_thermonuclear_2017} as a target for PRE burst modelling.
The system is thus a natural choice for extending the methods from Chapter~\ref{ch:paper2}.

\renewcommand{\arraystretch}{1.3}
\begin{table}
    {\centering
    \caption{
    The multi-epoch burst data from \fouru{}, used by our MCMC routine.
    The values are adapted from Table~2 of \citetalias{galloway_thermonuclear_2017}.
    We have assumed $\Fedd$ corresponds to the observed bolometric peak flux, $\Fpeak$.
    The $\Fper$ values include the bolometric corrections from \citetalias{galloway_thermonuclear_2017}.
    Because only two bursts were observed for 2009, we have assumed an instrument timing uncertainty of \SI{1}{s}, although the burst-to-burst variation is likely larger ($\approx \SI{25}{s}$ for 1997).
    }
    \label{tab:fouru_data}
    \begin{tabular}{llll}
        \hline
        \hline
        Epoch   & $\brate$          & $\Fedd$               & $\Fper$ \\
                & $(\si{day}^{-1})$ & $(\SI{e-9}{\Funits})$ & $(\SI{e-9}{\Funits})$ \\
        \hline
        1997 May  &  $8.95 \pm 0.02$      & $61 \pm 2$     &  $5.4 \pm 0.7$  \\
        2009 June &  $12.6850 \pm 0.0019$ & $56.6 \pm 1.4$ & $8.54 \pm 0.09$ \\
        \hline
    \end{tabular}
    }
\end{table}

\section{Methods}
\label{sec:fouru_methods}
The methods for this study consist of the multi-epoch observed data (Section~\ref{subsec:fouru_data}), the construction of the model grid (Section~\ref{subsec:fouru_grid}), the interpolated multi-epoch model (Section~\ref{subsec:fouru_multi-epoch}), and the MCMC methods (Section~\ref{subsec:fouru_mcmc}).

For ease of comparison between different models and studies, all accretion rates are given as a fraction of the ``canonical'' Eddington-limited rate for solar composition, $\mdotedd = \SI{8.775e4}{\mdotunits}$.
This value assumes $\hyd = 0.7$ and Newtonian gravity for $M = \SI{1.4}{\msun}$ and $R = \SI{10}{km}$.
The ``true'' Eddington rate is a factor of 1.7 larger for pure helium, and depends on the neutron star mass and radius.

\subsection{Multi-epoch Data}
\label{subsec:fouru_data}
To model bursts from \fouru{}, we again used multi-epoch data from the reference set provided by \citetalias{galloway_thermonuclear_2017}.
The burst data was from two accretion epochs, observed on 4 May 1997 and 12 June 2009 (Table~\ref{tab:fouru_data}).
In this study we fit the observed burst rate, $\brate$, Eddington-limited flux, $\Fedd$, and persistent flux, $\Fper$.
We assumed that $\Fedd$ corresponds to the observed peak burst flux, $\Fpeak$, and that $\Fper$ corresponds to the accretion luminosity, $\Lacc$.

For this initial study, we did not fit the observed fluence, $\fluence$, because the behaviour of the neutron star atmosphere near the Eddington luminosity remains poorly understood.
It is unclear how the total burst energy, $\Eb$, translates to an observed $\fluence$, particularly because \kepler{} lacks a detailed treatment of the photosphere, and exhibits anomalous super-Eddington luminosities during PRE \citep{woosley_models_2004, johnston_simulating_2018}.
Because of this model limitation, we only used \kepler{} models to predict $\brate$, and calculated $\Fedd$ and $\Fper$ analytically (Section~\ref{subsec:fouru_multi-epoch}).

\subsection{Model Grid}
\label{subsec:fouru_grid}
To obtain burst predictions that could be sampled quickly with MCMC methods, we precomputed a grid of \kepler{} models.
Because this study represents the first extension of our MCMC methods to a new bursting regime, we reduced the total number of models by using constant values for the accreted hydrogen fraction, $\hyd$, the CNO metallicity, $\cno$, and the surface gravity $g$.
We set the hydrogen composition to $\hyd = 0.0$, which is commonly assumed for this source due to its ultra-compact orbit \citep{stella_discovery_1987}.
The CNO metallicity was set to $\cno = 0.015$, between commonly-used values of $0.01$ and $0.02$ \citep[e.g.,][]{cumming_models_2003, heger_models_2007, meisel_consistent_2018}, although the influence of $\cno$ will be reduced due to the absence of hydrogen.
The remaining mass fraction of $0.985$ was assigned to helium.
The surface gravity was set to $g = \SI{1.858e14}{\gunits}$, corresponding to a gravitational neutron star mass of $\nsmass$ and a radius of $R = \SI{11.2}{km}$, although the final choice of $M$ and $R$ remains a free parameter (Section~\ref{subsec:fouru_multi-epoch}).
Exploring variations of these parameters remains a goal for a future study (Section~\ref{subsec:fouru_discussion_future}).

We varied two model parameters: the local accretion rate, $\mdot$, and the crustal heating rate, $\qb$.
Following a limited parameter exploration, using the observed recurrence times of $1 \lesssim \dt \lesssim \SI{3}{\hour}$ as a guide, we chose a regular grid of values between $0.01 \leq \qb \leq \SI{0.4}{\mevnuc}$ and $0.175 \leq \mdot \leq 0.5\, \mdotedd$, resulting in 168 \kepler{} models (Table~\ref{tab:fouru_grid}).
Despite the reduced size of the \fouru{} model grid compared to \gs{} (Chapter~\ref{ch:paper2}), it is nevertheless the largest set of 1D models of hydrogen-poor bursts to date.

\begin{figure}
    \centering
    \includegraphics[width=0.8\linewidth]{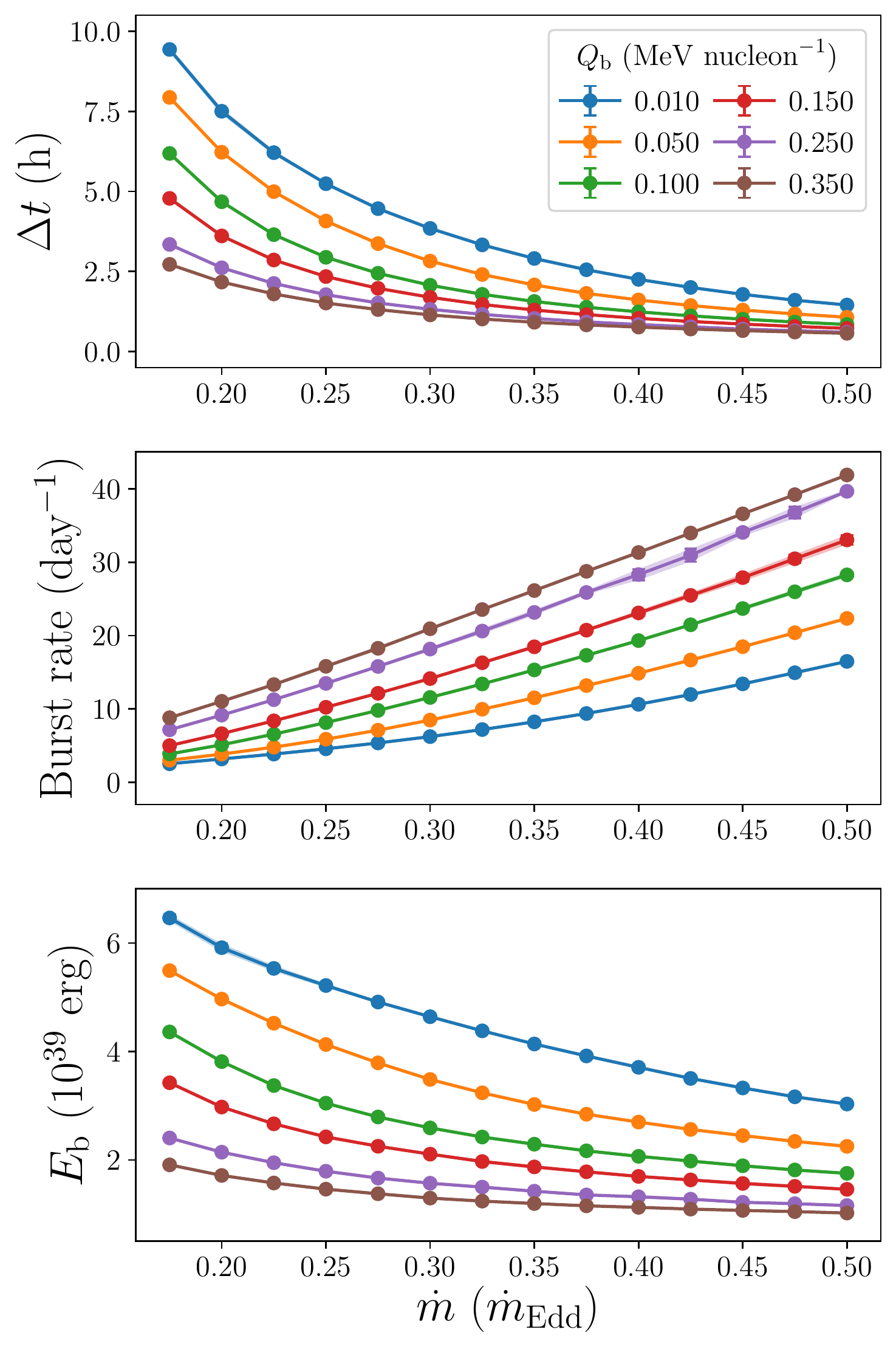}
    \caption{Burst properties for a subset of the full grid of 168 models of pure-helium bursts.
    For clarity, only every second grid point in $\qb$ is plotted.
    Each point corresponds to a \kepler{} simulation of 40--50 bursts, from which the average burst properties were calculated for the final 10--20 bursts.
    In contrast to the model grid from Chapter~\ref{ch:paper2}, we fit only the burst rate to the observed data.
    The burst energy, $\Eb$, is shown here simply for illustration, and was calculated after truncating the burst lightcurve at an assumed Eddington limit of $\Ledd = \SI{3.5e38}{\Lunits}$.
    The error bars and interpolated shaded regions are $1 \sigma$ standard deviations, but are generally too small to be visible here.
    }
    \label{fig:fouru_grid}
\end{figure}

In order to minimise the effect of model burn-in (see Section~\ref{sec:results_preheating}), a sequence of 40--50 bursts were produced for each model, and the first 30 were excluded from analysis.
The remaining 10--20 bursts were extracted using the same methods described in Section~\ref{sec:methods_extracting} and Chapter~\ref{ch:paper2}, using our \python{} package \pyburst{}.
The average burst properties were calculated, and the standard deviation was adopted as the uncertainty.
We thus obtained a tabulated set of model predictions over the grid of $\mdot$ and $\qb$ (Figure~\ref{fig:fouru_grid}).
Linear interpolation could then be used to rapidly ($\ll \SI{1}{s}$) sample burst properties anywhere across the grid.

\renewcommand{\arraystretch}{1.5}
\begin{table}
    {\centering
    \caption{The parameters of the \kepler{} model grid.
    Every combination was iterated, totalling 168 simulations.
    The $\qb$ step size of $0.025$ corresponds to the span between 0.025--0.15, and the step size of $0.05$ corresponds to the span between 0.20--0.40.
    Parameters which were held constant are $\hyd = 0.0$, $\cno = 0.015$, and $g = \SI{1.858e14}{\gunits}$.
    The accretion rates are given as a fraction of $\mdotedd = \SI{8.775e4}{\mdotunits}$ (assuming $M = \SI{1.4}{\msun}$, $R = \SI{10}{km}$, and $X = 0.7$), which is simply used as a common reference point to other models, and does not represent the ``true'' $\mdotedd$.
    }
    \label{tab:fouru_grid}
    \begin{tabular}{lllll}
        \hline
        \hline
        Parameter & Units          & Range                                 & Step size  & N\\
        \hline
        $\mdot$   & $(\mdotedd)$     &  0.175--0.500                         & 0.025      & 14 \\
        $\qb$     & $(\mathrm{MeV\, nuc}^{-1})$ &  0.01, 0.025--0.15, 0.2--0.40 & 0.025, 0.05 & 12 \\
        \hline
        \multicolumn{4}{r}{Total}      & 168 \\
        \hline
    \end{tabular}
    }
\end{table}

\subsection{Multi-epoch Modelling}
\label{subsec:fouru_multi-epoch}
To predict the multi-epoch properties of the observed data (Table~\ref{tab:fouru_data}), we used an approach similar to the multi-epoch model for \gs{} in Chapter~\ref{ch:paper2}.
For the model grid parameters of accretion rate and crustal heating we used epoch-dependent parameters, $\mdotn{1}$, $\mdotn{2}$, $\qbn{1}$, and $\qbn{2}$, where the subscripts 1 and 2 correspond to the 1997 and 2009 epochs, respectively.
The remaining ``free'' parameters were epoch-independent: the neutron star mass, $M$, the anisotropy-modified distance, $\db$, and the anisotropy ratio, $\xiratio$.
We note again that unlike our model for \gs{}, we used fixed values of $\hyd = 0.0$, $\cno = 0.015$, and $g = \SI{1.858e14}{\gunits}$ (Section~\ref{subsec:fouru_grid}).
Our multi-epoch model for \fouru{} thus consisted of seven parameters: $\mdotn{1}$, $\mdotn{2}$, $\qbn{1}$, $\qbn{2}$, $M$, $\db$, and $\xiratio$.

For a given choice of these parameters, the three observed quantities were predicted for both epochs: the burst rate, $\brate$, the Eddington flux, $\Fedd$, and the persistent flux, $\Fper$.
These observables were predicted using the same procedure described in Chapter~\ref{ch:paper2}.
The burst rate was interpolated from the model grid for the given $\mdotn{i}$ and $\qbn{i}$, for $i = 1,\, 2$, and $\Fedd$ and $\Fper$ were again calculated directly.
These calculations included GR-corrections to account for the Newtonian gravity used in \kepler{} (Section~\ref{sec:methods_gr}).

\subsection{MCMC Method}
\label{subsec:fouru_mcmc}
We used MCMC methods to sample the parameter space of our multi-epoch model (Section~\ref{subsec:fouru_multi-epoch}) and compare the predictions to the observed data (Section~\ref{subsec:fouru_data}).
The MCMC routine was adapted directly from the model applied to \gs{} in Chapter~\ref{ch:paper2}, but with modified parameters, priors, and observed data.
We again used the open-source \python{} ensemble sampler, \emcee\footnote{\url{https://emcee.readthedocs.io/en/v2.2.1}} \citep{foreman-mackey_emcee:_2013}.

The prior distribution was set to $\prior = 0$ outside the parameter boundaries.
For the parameters $\mdotn{i}$ and $\qbn{i}$, the boundaries were those of the model grid (Table~\ref{tab:fouru_grid}).
For the free parameters, we set limits of $1.0 \leq M \leq \SI{3.5}{\msun}$, $1 \leq \db \leq \SI{15}{kpc}$, and $0.1 \leq \xiratio \leq 10$.
We used a larger upper limit than $M = \SI{2.3}{\msun}$ used in Chapter~\ref{ch:paper2}, because the posterior distributions were found to be strongly truncated.
Values above this limit exceed the most massive neutron stars observed to date \citep[e.g.,][]{linares_peering_2018, cromartie_relativistic_2019}, but for this initial study of \fouru{}, we wished to explore the possible extent of the model bias towards large masses.

We applied flat (i.e.,\ uniform) prior distributions within these limits for all parameters except $\db$, for which we utilised two distance estimates for the globular cluster \ngc{}.
These estimates were $d = \SI{7.6 \pm 0.4}{kpc}$ from optical measurements \citep{kuulkers_photospheric_2003} and $d = \SI{8.4 \pm 0.6}{kpc}$ from near-infrared measurements \citep{valenti_near-infrared_2007}.
We used the joint distribution of these constraints with a flat prior for $\xib$ to obtain a Gaussian prior of $\db = \SI{7.85 \pm 0.33}{kpc}$.

Given a sample point in parameter space, the observables for $\brate$, $\Fedd$, and $\Fper$ were predicted with the multi-epoch model.
The predictions were compared with the observed data using the likelihood function given by
\begin{equation}
\label{eq:lnhood}
    \ln \left[ \likelihood \right] = -\frac{1}{2} \sum_x \left\{  \frac{(x - \sub{x}{\obslhood})^2}{\sigma^2 + \sub{\sigma}{\obslhood}^2} + \ln{\left[ 2 \pi (\sigma^2 + \sub{\sigma}{\obslhood}^2) \right]} \right\},
\end{equation}
where $x$ and its uncertainty, $\sigma$, were iterated over the predicted values for each epoch, and the subscript `\obslhood' signifies the corresponding observed values.

\begin{figure}
    \centering
    \includegraphics[width=0.8\textwidth]{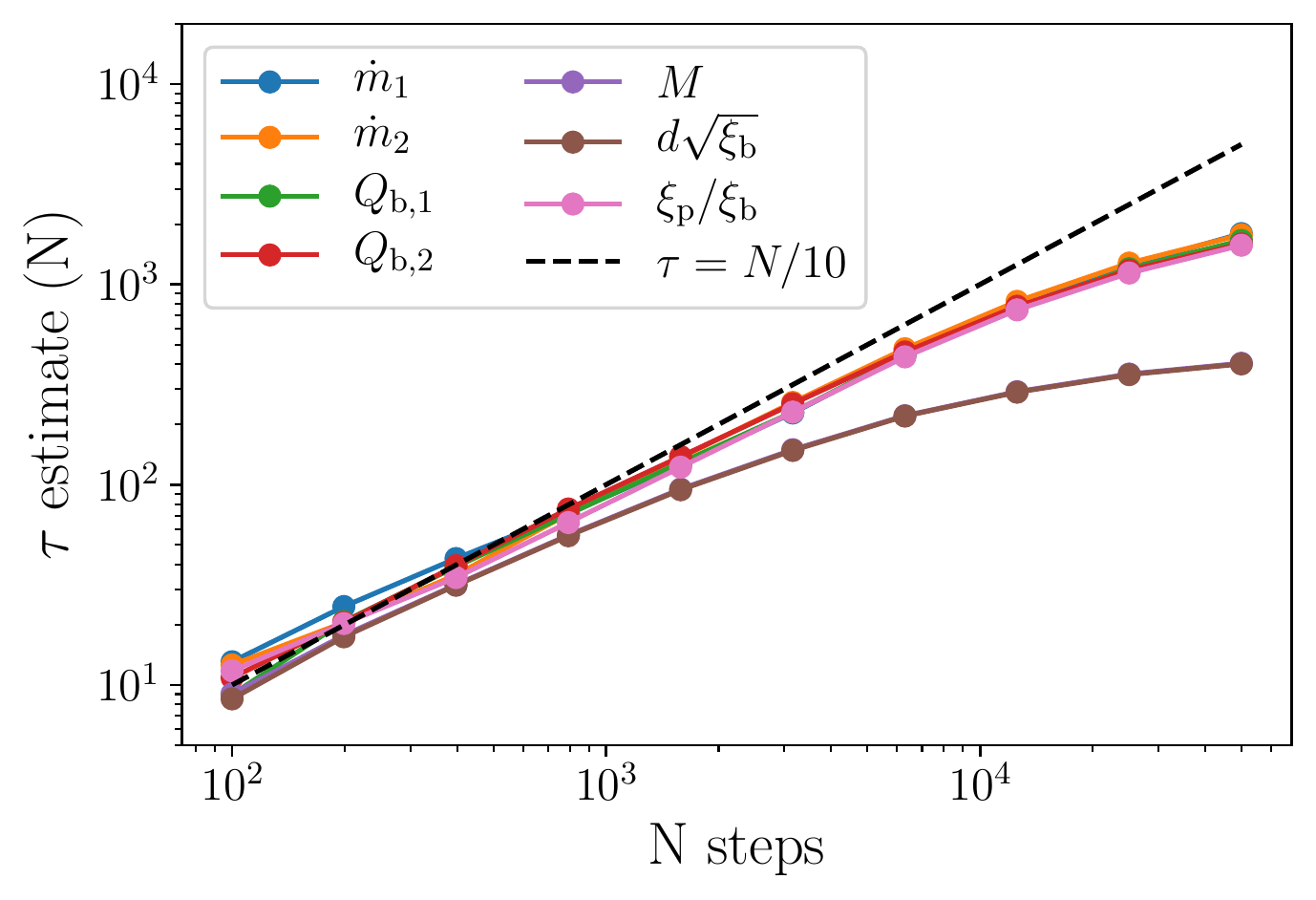}
    \caption{Estimates for the autocorrelation time, $\tau$, for each parameter at multiple points along the chain.
    Above $N \sim 10^4$, the estimates for $\tau$ are consistently growing slower than $\tau = N / 10$, indicating that the chain has progressed beyond 10 $\tau$.
    }
    \label{fig:fouru_autocorrelation}
\end{figure}

The MCMC chain consisted of 1000 walkers initialised in a small ``hyper ball'' in parameter space.
The sampler was run for 50 thousand steps, producing 50 million samples in total.
The first 1000 steps were discarded as burn-in.

To test the sampler convergence, the autocorrelation time $\tau$ was estimated for each parameter at multiple steps along the chain\footnote{using a \python{} routine adapted from \url{https://dfm.io/posts/autocorr}} (Figure~\ref{fig:fouru_autocorrelation}).
For large numbers of samples, $N \gtrsim 10^4$, the estimates for $\tau$ begin to converge toward a final value.
Although the chain is not long enough to obtain converged values for $\tau$ itself, the estimates are consistently growing slower than the $\tau = N/10$ line, indicating that the total chain length is larger than $10\, \tau$.

\renewcommand{\arraystretch}{1.5}
\begin{table}
    \centering
    \caption{The maximum likelihood estimates for the 1D marginalised posteriors, with 68\% credible intervals.
    The quantities of $d$, $i$, $\xib$, and $\xip$ were calculated using a disc anisotropy model.
    The neutron star properties of $R$ and $z$ were calculated from $M$ and the fixed value of $g = \SI{1.858e14}{\gunits}$.
    The global accretion rates, $\Mdotn{i}$, were calculated from $\mdotn{i}$ and $R$.
    The accretion rates are given as fractions of the canonical Eddington rate, $\mdotedd = \SI{8.775e4}{\mdotunits}$ and $\Mdotedd = \SI{1.75e-8}{\msun.yr^{-1}}$.
    }
    \label{tab:fouru_estimates}
    \begin{tabular}{lll}
        \hline
		Parameter & Units & Estimate \\
		\hline
		$\dot{m}_1$               & ($\mdotedd$)   & $0.208^{+0.05}_{-0.005}$ \\
		$\dot{m}_2$               & ($\mdotedd$)   & $0.37^{+0.06}_{-0.07}$ \\
		$Q_\mathrm{b,1}$          & (\si{\mevnuc}) & $0.36^{+0.04}_{-0.17}$ \\
		$Q_\mathrm{b,2}$          & (\si{\mevnuc}) & $0.04^{+0.11}_{-0.02}$ \\
		$M$                       & ($\msun$)      & $2.3^{+0.3}_{-0.2}$ \\
		$d \sqrt{\xi_\mathrm{b}}$ & (\si{kpc})     & $7.9 \pm 0.3$ \\
		$\xi_\mathrm{p} / \xi_\mathrm{b}$ & --     & $1.8^{+0.4}_{-0.3}$ \\

        $d$                       & (\si{kpc})     & $6.9 \pm 0.3$ \\
		$i$                       & (\si{deg})     & $74^{+2}_{-4}$ \\
        $\xib$                    & --             & $1.33 \pm 0.09$ \\
		$\xip$                    & --             & $2.4^{+0.7}_{-0.5}$ \\

        $R$                       & (\si{km})      & $15.1^{+0.9}_{-1}$ \\
		$z$                       & --             & $0.36 \pm 0.03$ \\

        $\Mdotn{1}$               & ($\Mdotedd$)   & $0.34^{+0.06}_{-0.04}$ \\
		$\Mdotn{2}$               & ($\Mdotedd$)   & $0.54^{+0.13}_{-0.08}$ \\
		\hline
    \end{tabular}
\end{table}

\begin{figure}
    \centering
    \includegraphics[width=\textwidth]{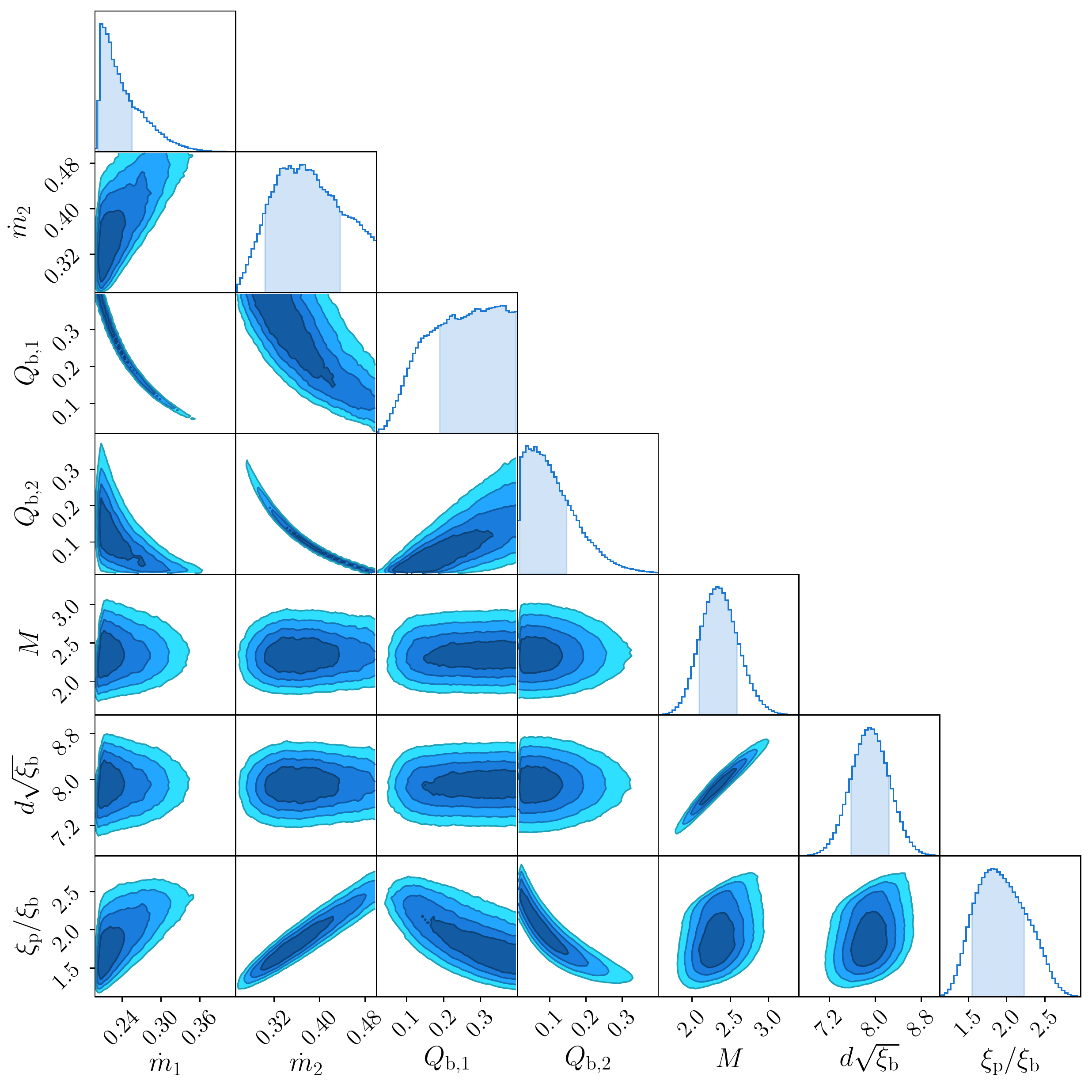}
    \caption{Marginalised posterior distributions for all seven MCMC parameters.
    The 1D posteriors are along the diagonal, with \SI{68}{\%} credible intervals shaded.
    The 2D contour levels are 38, 68, 87, and \SI{95}{\%} credible regions.
    The units for $\mdot$ are $\mdotedd = \SI{8.775e4}{\mdotunits}$, $\qb$ are \si{\mevnuc}, $M$ are \si{\msun}, and $\db$ are \si{kpc}.
    The maximum likelihood estimates for the 1D posteriors are listed in Table~\ref{tab:fouru_estimates}.
    }
    \label{fig:fouru_posteriors}
\end{figure}

\section{Results}
\label{sec:fouru_results}
The MCMC chain was analysed using the open source \python{} package \chainconsumer\footnote{\url{https://samreay.github.io/ChainConsumer}} \citep{hinton_chainconsumer_2016}.
The marginalised one-dimensional (1D) and two-dimensional (2D) posterior distributions for all seven parameters are plotted in Figure~\ref{fig:fouru_posteriors}.
The maximum likelihood estimates for the 1D posteriors are listed in Table~\ref{tab:fouru_estimates}.
We discuss here the general results of the MCMC posteriors, and the additional system properties we can derive from them.
We further discuss the specific parameter estimates and comparisons to previous works in Section~\ref{sec:fouru_discussion}.

In the H/He models of \gs{} in Chapter~\ref{ch:paper2}, we found that some of the posteriors were limited by the parameter boundaries, and the same issue is evident for the \fouru{} distributions here.
Most of the 38\% contours of the 2D posteriors for $\qbn{1}$ reach the model grid upper limit of $\SI{0.4}{\mevnuc}$, and all of the 87\% contours for $\mdotn{2}$ are truncated at the upper limit of $\mdot = 0.5 \mdotedd$.
On the other hand, the 95\% 2D contours between the free parameters of $M$, $\db$, and $\xiratio$ lie completely within the boundaries, indicating that the distributions for these parameters are not artificially constrained.

In contrast to the \gs{} models in Chapter~\ref{ch:paper2}, each epoch pair of $\mdotn{i}$ and $\qbn{i}$ are tightly correlated.
This difference may be partly due to the lack of hydrogen burning, reducing the burst-to-burst variability of the pure helium models.
The typical standard deviation of $\dt$ for the helium burst models was $\approx \SI{0.9}{\%}$, in comparison to $\approx \SI{4}{\%}$ for the H/He models.
The larger variability of the H/He bursts, and the additional influence of $\hyd$ and $\cno$ on burst ignition, possibly ``washed out'' the correlation.

The anti-correlation itself between the posteriors of $\qbn{i}$ and $\mdotn{i}$ can be understood from their combined effect on the burst ignition depth, $\yign$, and recurrence time, $\dt$.
Increasing $\qb$ decreases the depth at which nuclear burning is unstable, $\yign$.
The recurrence time is then determined by the time taken to accrete the column of material, $\dt = \yign / \mdot$.
For a given observed $\dt$ used by the MCMC model, an increase in $\mdot$ can be compensated by a decrease in $\qb$.

\subsection{Predicted Observables}
\label{subsec:fouru_results_observables}
\begin{figure}
    \centering
    \includegraphics[width=\textwidth]{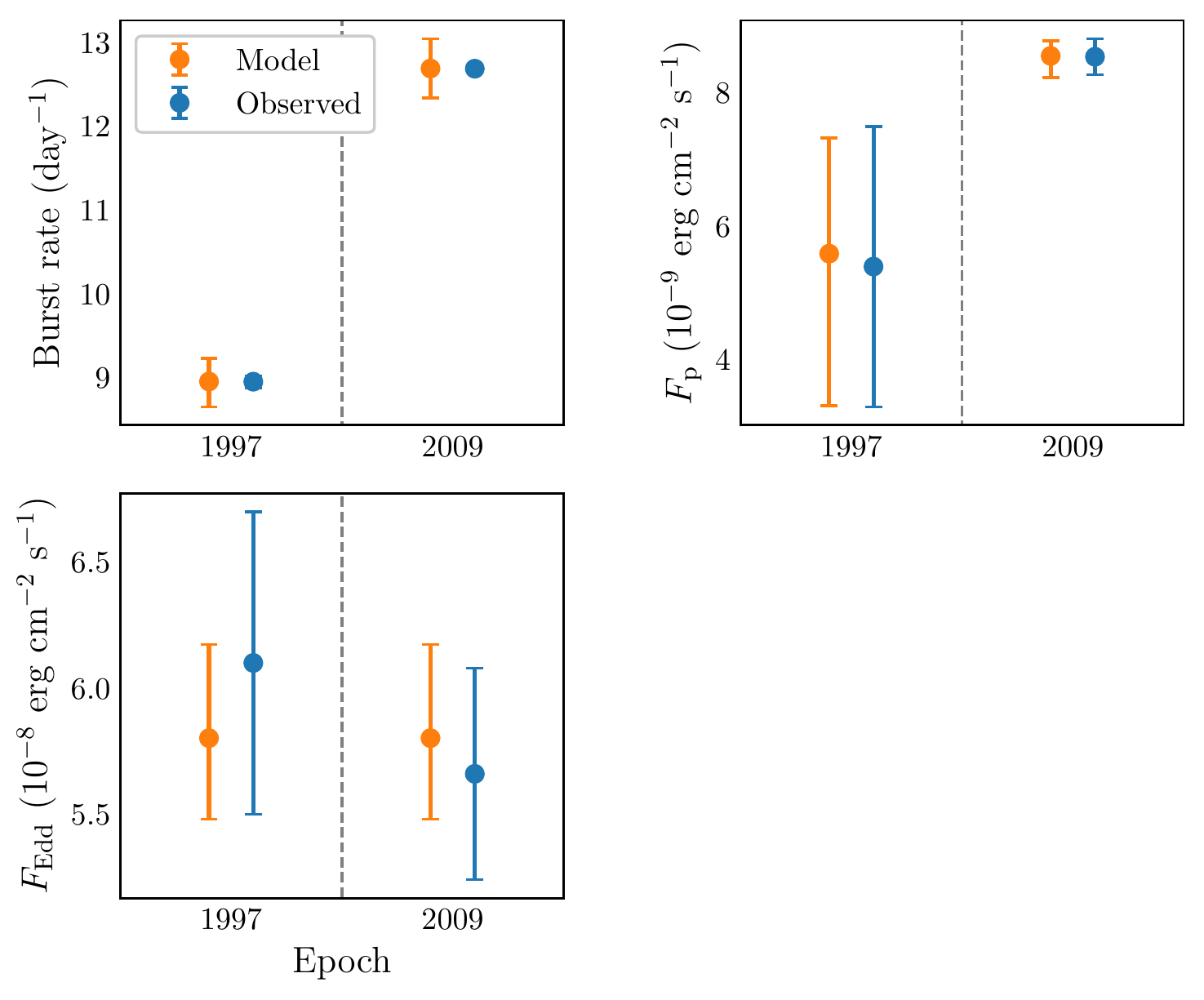}
    \caption{Distributions of the predicted burst properties (i.e., the posterior predictive distribution;
    orange points), and the observed epoch data (blue points).
    The error bars are 68\% credible intervals.
    The observed data are consistent with the predicted distributions to within uncertainties.
    }
    \label{fig:fouru_bprop_sample}
\end{figure}

The posterior predictive distribution is given by the distribution of observables predicted by the MCMC model.
As a consistency check, we can compare these predicted distributions to the original observed data to ensure the model is behaving as expected.
We took a random sample of \num{20000} points from the MCMC chain, and extracted the multi-epoch burst properties predicted by the model.
The peaks and 68\% intervals for the distributions are plotted against the observed data in Figure~\ref{fig:fouru_bprop_sample}.
The observed values are consistent with the prediction distributions within the uncertainties.
This consistency indicates that the model predictions are behaving normally in the MCMC simulation.

\subsection{Crustal heating and accretion rate}
\label{subsec:fouru_results_qb-mdot}
\begin{figure}
    \centering
    \includegraphics[width=0.8\textwidth]{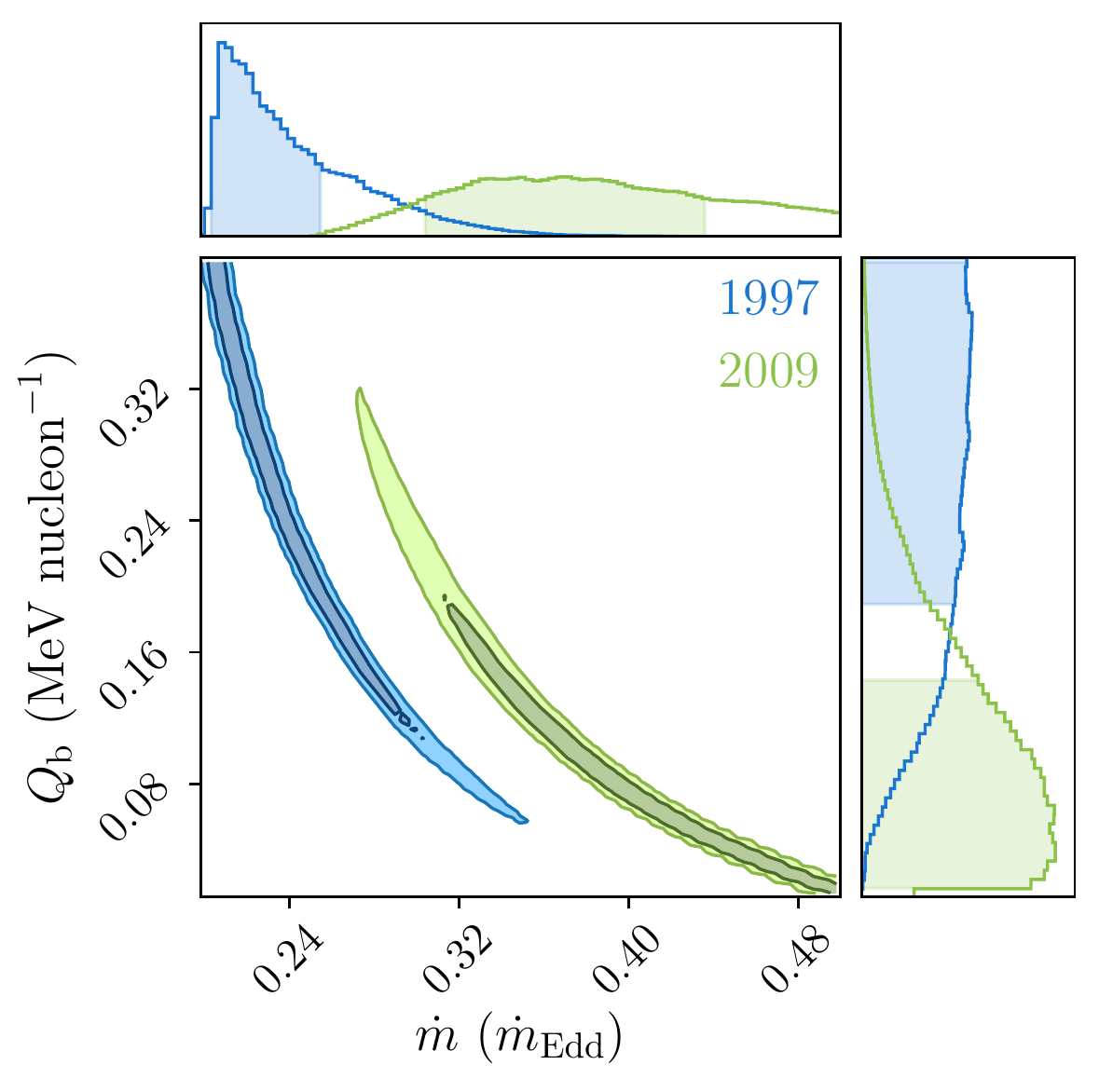}
    \caption{The posteriors for $\qb$ and $\mdot$ for each epoch.
    For clarity, only the 68 and 95\% contour levels are shown.
    The distribution for each epoch is strongly degenerate, but the 1997 epoch is overall consistent with lower accretion rates and lower crustal heating than 2009.
    }
    \label{fig:fouru_qb-mdot}
\end{figure}

The epoch-dependant crustal heating parameters, $\qbn{i}$, allow us to examine the relationship between $\qb$ and $\mdot$.
The 2D posteriors for each epoch are plotted in Figure~\ref{fig:fouru_qb-mdot}.
As discussed above, these narrow contours arise due to the degeneracy between $\qb$ and $\mdot$ for a given $\dt$, and should not be confused with the underlying relationship between the two quantities.

There is some overlap between the tails of the 1D posteriors for $\qb$, although the 1997 epoch favours larger values of $0.36^{+0.04}_{-0.17}\, \si{\mevnuc}$, whereas 2009 is consistent with smaller values of $0.04^{+0.11}_{-0.02}\, \si{\mevnuc}$.
The 1D estimates are similarly separated in $\mdot$, with lower values for 1997 of $.208^{+0.05}_{-0.005}\, \mdotedd$, and higher values for 2009 of $0.37^{+0.06}_{-0.07}\, \mdotedd$.
As with the \gs{} models in Chapter~\ref{ch:paper2}, this comparison offers a tentative indication towards an underlying relationship between $\qb$ and $\mdot$.

\subsection{Distance and Inclination}
\label{subsec:4u1820_results_distance}
\begin{figure}
    \centering
    \subfloat{\includegraphics[width=0.5\textwidth]{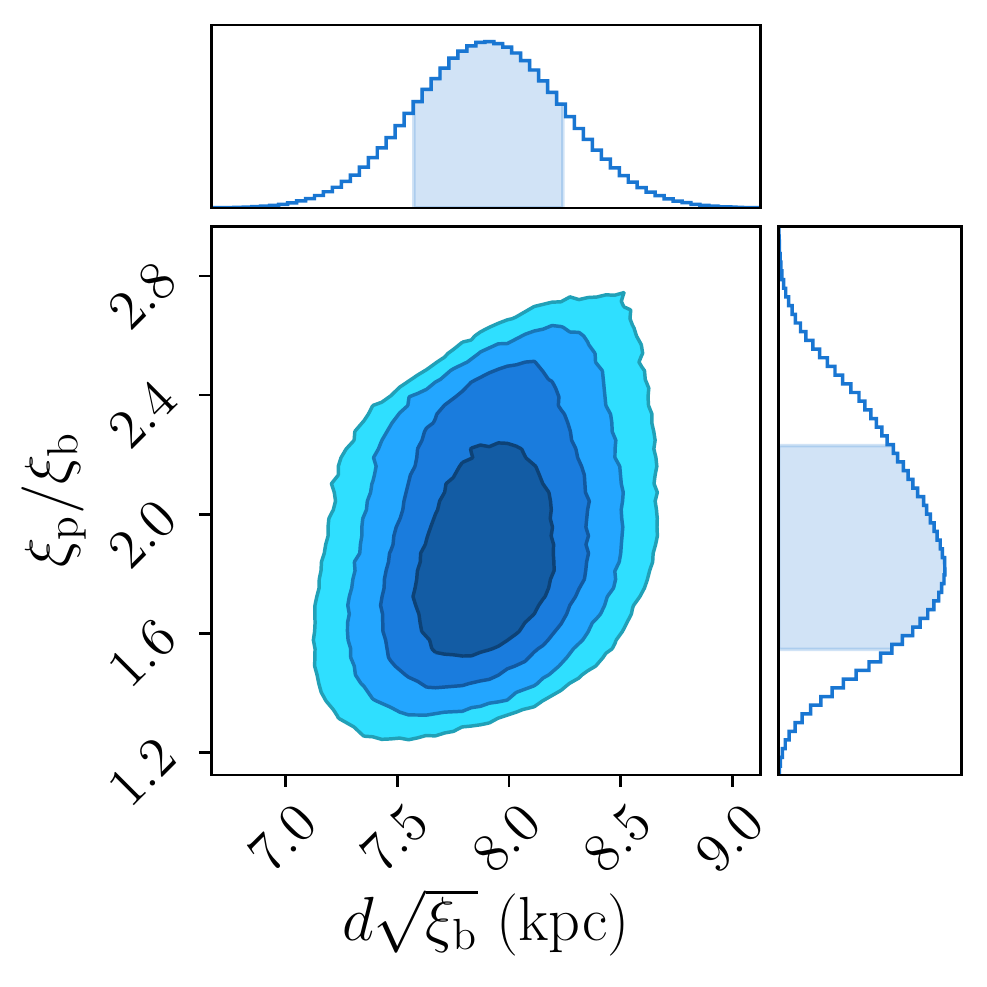}}
    \subfloat{\includegraphics[width=0.5\textwidth]{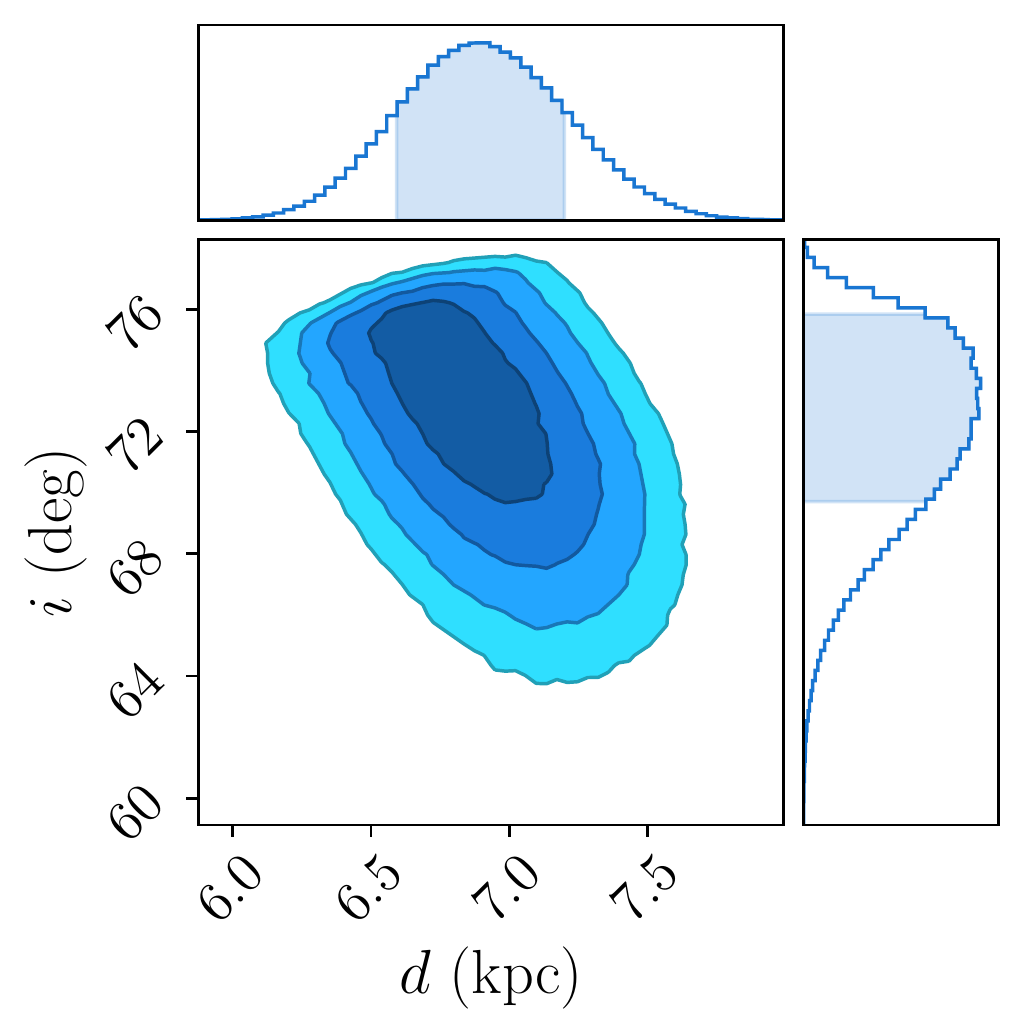}}
    \caption{Posteriors for the distance and anisotropy parameters of the MCMC model (left panel), and the inclination and absolute distance calculated from them using the \textit{disc a} anisotropy model of \citet[][right panel]{he_anisotropy_2016}.
    The 2D contour levels are  38, 68, 87, and \SI{95}{\%} credible regions.
    The 1D shaded intervals are 68\% credible intervals.
    The maximum likelihood estimates for the 1D posteriors are listed in Table~\ref{tab:fouru_estimates}.
    }
    \label{fig:fouru_disc}
\end{figure}

Using the same approach as the \gs{} models in Chapter~\ref{ch:paper2}, we obtained posteriors for the system inclination, $i$, and the absolute distance, $d$, by choosing a model for the disc anisotropy.
We used the model \textit{Disc a} for a thin flat disc from \citet[][see Figure~\ref{fig:methods_anisotropy} from Chapter~\ref{ch:methods}]{he_anisotropy_2016} to calculate these quantities from the anisotropy ratio, $\xiratio$, and the modified distance, $\db$.
The posteriors for these quantities are plotted in Figure~\ref{fig:fouru_disc}, and the 1D estimates are listed in Table~\ref{tab:fouru_estimates}, in addition to individual estimates for $\xib$ and $\xip$.

The burst anisotropy factor of $\xib = 1.33 \pm 0.09$ indicates that the burst emission is preferentially beamed \textit{away} from the observer.
This effect causes the source to appear dimmer, and at a larger inferred distance of $\db = 7.9 \pm \SI{0.3}{kpc}$ when isotropic emission is assumed.
Given this estimate of $\xib$, the actual distance is closer, at $d = 6.9 \pm \SI{+0.3}{kpc}$.

We note that these estimates are dependent on the disc model from \citet{he_anisotropy_2016}, and using other models for anisotropy could produce different estimates.

\subsection{Neutron Star Properties}
\label{subsec:4u1820_results_grav}
\begin{figure}
    \centering
    \includegraphics[width=\textwidth]{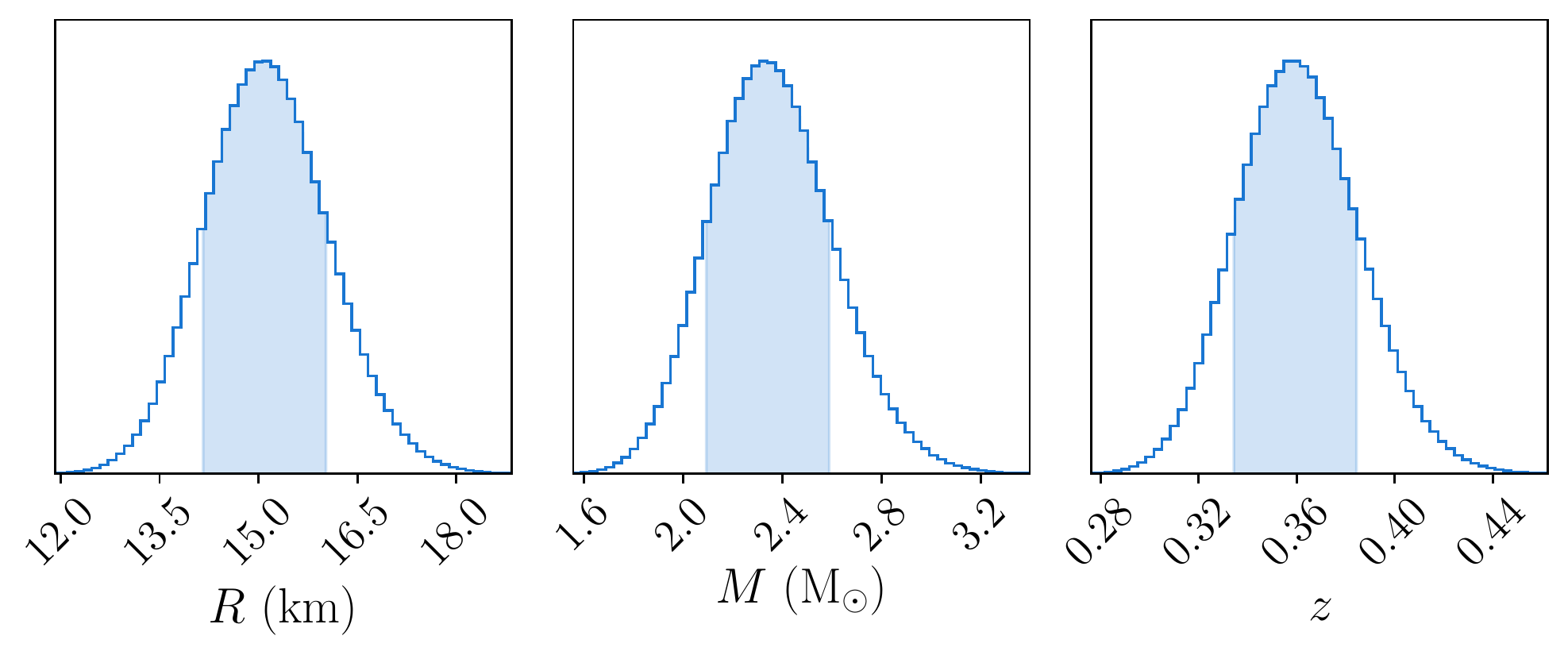}
    \caption{Posterior distributions for the neutron star properties.
    The neutron star mass, $M$, is an original parameter of the MCMC model, whereas the radius, $R$, and redshift, $z$, are calculated from $M$ and $g$.
    In contrast to the \gs{} model from Chapter~\ref{ch:paper2}, we used a fixed gravity of $g = \SI{1.858e14}{\gunits}$, and thus only a single contour line of $M$ and $R$ was explored.
    The shaded region is the 68\% credible interval.
    }
    \label{fig:fouru_grav}
\end{figure}

Using the same procedure as Chapter~\ref{ch:paper2}, we calculate the neutron star radius, $R$, and the gravitational redshift, $z$, using the MCMC parameter of $M$ and the fixed value of $g = \SI{1.858e14}{\gunits}$.
The 1D posteriors for these quantities are plotted in Figure~\ref{fig:fouru_grav}, and the estimates are listed in Table~\ref{tab:fouru_estimates}.

Due to the larger upper boundary of $M = \SI{3.5}{\msun}$, compared to $M = \SI{2.2}{\msun}$ in Chapter~\ref{ch:paper2}, the model was free to explore larger masses, resulting in a posterior of $2.3^{+0.3}_{-0.2}$.
This value is larger than the typical observed range of 1--2 $\msun$ \citep{ozel_mass_2012, miller_astrophysical_2013}, and also exceeds both the largest observed mass of $2.14^{+0.10}_{-0.09}\, \si{\msun}$ \citep{cromartie_relativistic_2019} and the maximum mass of $2.17\, \si{\msun}$ inferred from the neutron star merger GW170817 \citep{margalit_constraining_2017}.
We further discuss these results in Section~\ref{sec:fouru_discussion}.

\subsection{Global Accretion Rate}
\label{subsec:fouru_results_mdot}
We calculate the global accretion rates given by $\Mdot = 4 \pi R^2 \mdot$ for each sample of $R$ and $\mdot$.
The 1D marginalised estimates for $\Mdotn{i}$ are listed in Table~\ref{tab:fouru_estimates}, given as a fraction of the canonical Eddington-limited rate, $\Mdotedd = \SI{1.75e-8}{\msun.yr^{-1}}$, which is equivalent to $\mdotedd = \SI{1.492e5}{\mdotunits}$ assuming $R = \SI{10}{km}$.
Once again, this Eddington rate is simply used as a common reference point, and is not adjusted for each sample of $M$ and $R$.

\subsection{Lightcurve Sample}
\label{subsec:fouru_lcsample}
The MCMC routine only compared a single quantity extracted from the observed lightcurve: the peak flux, $\Fpeak$, which was assumed to be the Eddington flux, $\Fedd$.
As with our models for \gs{} in Chapter~\ref{ch:paper2}, we produced a limited sample of full burst lightcurves to check for consistency with the observations.

\begin{figure}
    \centering
    \includegraphics[width=\textwidth]{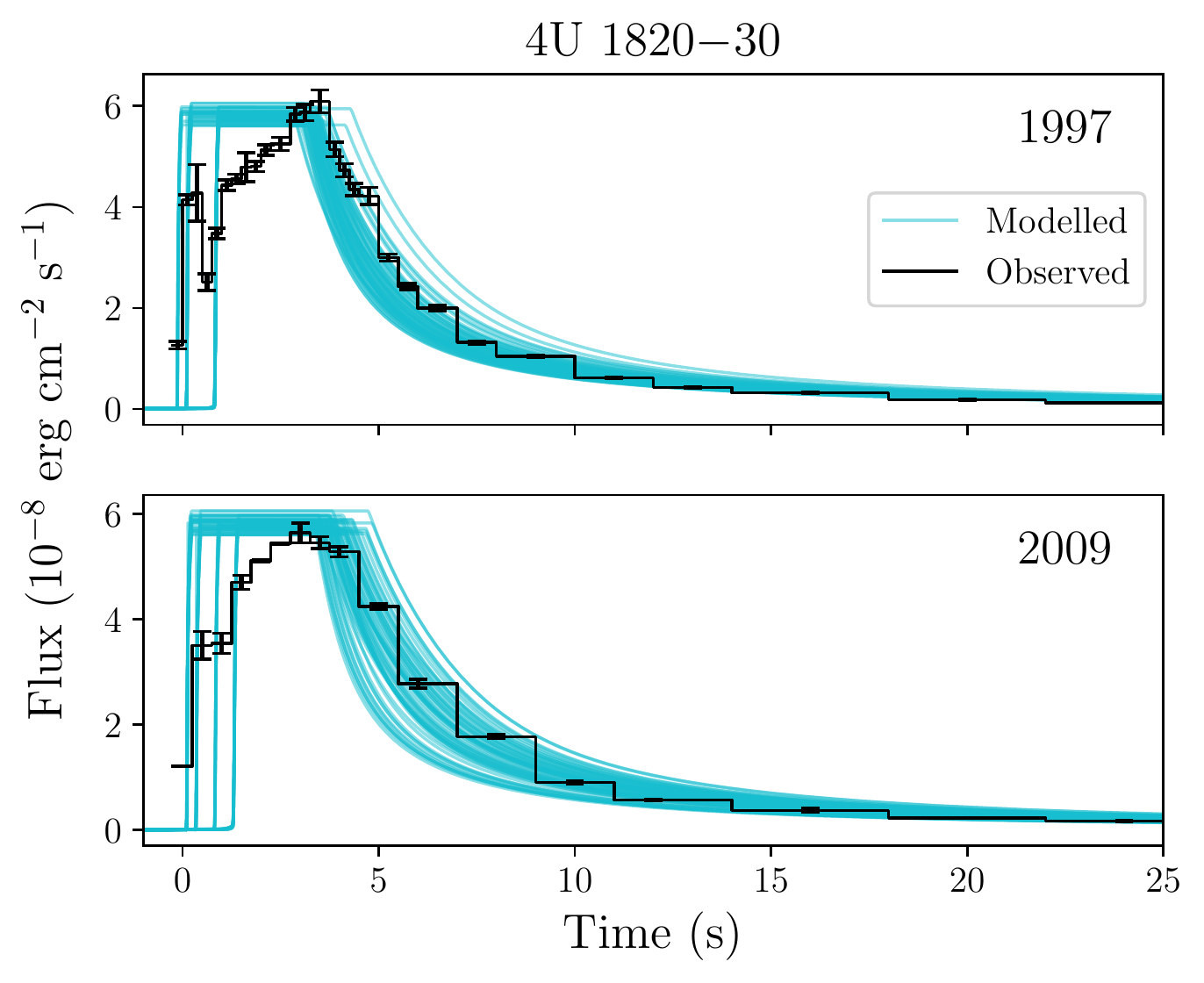}
    \caption{The full burst lightcurves from an additional 30 epoch pairs of \kepler{} models (blue curves), and the observed lightcurves (black histograms).
    The model parameters were taken from a random sample of the MCMC chain, and
    the lightcurves were truncated at $\Ledd$, because \kepler{} has limited ability to model the PRE burst phase.
    Instead of the average lightcurves used in Chapter~\ref{ch:paper2}, only the last two bursts from each model are shown.
    There is good overall agreement, particularly for the burst tail.
    The only lightcurve quantity matched by the MCMC routine was $\Fedd$, which we assumed to correspond to $\Fpeak$.
    }
    \label{fig:fouru_lc_sample}
\end{figure}

We took a random sample of 30 points from the MCMC chain, and computed a new epoch pair of \kepler{} models for each point of $\mdotn{i}$ and $\qbn{i}$, resulting in a total of 60 models.
The model luminosities were truncated at the corresponding Eddington limit, $\Ledd$, and the train of burst lightcurves was extracted.
These lightcurves were then transformed to observable fluxes using the sampled parameters of $M$ and $\db$.
Instead of calculating an average burst lightcurve for each model sequence, we simply compare the last two bursts of each model.
The total sample of 120 model lightcurves are shown in Figure~\ref{fig:fouru_lc_sample} with the epoch observations.

The collective model lightcurves are broadly consistent with the observations, and the length of the PRE phase and the cooling of the tail are reproduced.
\kepler{} has limited atmosphere physics for modelling the PRE process, and with our simple truncation at $\Ledd$, the PRE phase itself is poorly reproduced.
Nevertheless, this comparison suggests that the modelled bursts remain broadly consistent with observations, even given that $\Fedd$ is the only lightcurve quantity being matched by the MCMC routine.

\section{Discussion}
\label{sec:fouru_discussion}
We compare here the parameter estimates obtained for \fouru{} with previous studies and predictions.
Our reported values are the maximum likelihoods for the 1D marginalised posteriors, listed in Table~\ref{tab:fouru_estimates}, where the uncertainties are the 68\% credible intervals.

We obtain accretion rates of $\mdotn{1} = 0.208^{+0.05}_{-0.005}$ and $\mdotn{2} = 0.37^{+0.06}_{-0.07}\, \mdotedd$ for the 1997 and 2009 epochs, respectively.
These values are larger than the initial 0.144 and $0.226\, \mdotedd$ suggested by \citet{galloway_thermonuclear_2017}, although their values did not include anisotropy and thus implicitly assumed $\xip = 1$.
Our estimate for a persistent anisotropy of $\xip = 2.4^{+0.7}_{-0.5}$ implies larger local accretion rates, because the resulting emission is preferentially beamed away from the observer.
On the other hand, $\mdotn{1}$ is consistent with the range of $\mdot \approx \numrange{0.2}{0.26}\, \mdotedd$ reported by \citet{cumming_models_2003}, for models of an earlier epoch with a slightly longer recurrence time of $\dt = \SI{3.2}{hr}$, compared to $\dt \approx \SI{2.7}{\hour}$ for 1997.

Our estimates for crustal heating are $\qbn{1} =  0.36^{+0.04}_{-0.17}$ and $\qbn{2} = 0.04^{+0.11}_{-0.02}\, \si{\mevnuc}$.
Although broad, these values are roughly consistent with the range of 0.1--0.2 \si{\mevnuc} used by \citet{cumming_models_2003}, and the typical expected rate of $\qb \approx \SI{0.15}{\mevnuc}$ \citep[e.g.,][]{cumming_long_2006}.
The total heating in the crust is predicted to be approximately 1--2 \si{\mevnuc} \citep{haensel_models_2008}, but the outflowing flux into the envelope depends on the thermal transport properties of the crust \citep[e.g.,][]{brown_mapping_2009}, shallow heating \citep{deibel_strong_2015}, and Urca neutrino cooling \citep{schatz_strong_2014}.

Similar to the results for \gs{} in Chapter~\ref{ch:paper2}, we again obtain an unusually large neutron star mass of $M = 2.3^{+0.3}_{-0.2}\, \si{\msun}$ and a very large radius of $R = 15.1^{+0.9}_{-1}\, \si{km}$.
The mass is larger than the typical expected range of $1 \lesssim M \lesssim 2\, \msun$ \citep{miller_astrophysical_2013}, and exceeds all previous estimates for \fouru{}, including $M = 1.29^{+0.19}_{-0.07}\, \si{\msun}$ \citep{shaposhnikov_nature_2004} and $M = 1.58 \pm \SI{0.06}{\msun}$ \citep{guver_mass_2010}.
It even exceeds the two largest observed neutron star masses of $2.27^{+0.17}_{-0.15}\, \si{\msun}$ \citep{linares_peering_2018}, and $2.14^{+0.10}_{-0.09}\, \si{\msun}$ \citep{cromartie_relativistic_2019}.
The radius is unrealistically large when compared to typical values predicted from equation of state models \citep[e.g.,][]{ozel_masses_2016}, and in light of recent constraints from the first gravitational wave observations of a neutron star merger \citep{abbott_gw170817_2018, most_new_2018}.
We again emphasise the limitations of this initial study, and further investigation will be required before more robust estimates for the neutron star properties can be obtained.

Our gravitational redshift of $z = 0.36 \pm 0.03$ is similar to the redshift inferred for \gs{} of $z = 0.39 \pm 0.07$ (Chapter~\ref{ch:paper2}), and is consistent with the fit of $z = 0.36$ assuming $d = \SI{7.0}{kpc}$ from \citet{shaposhnikov_nature_2004}.
It is slightly smaller than the value of $z \approx 0.43$ from the mass and radius estimates of \citet{guver_mass_2010}.

Our distance estimate of $\db = 7.9 \pm \SI{0.3}{kpc}$ is almost unchanged from the prior distribution of $\db = 7.85 \pm \SI{0.33}{kpc}$, and is consistent with the posterior of $\db = 7.47 \pm \SI{0.38}{kpc}$ reported by \citet{ozel_dense_2016}, who also used a similar prior.
It is larger, however, than the value of $\db = 6.5 \pm \SI{0.5}{kpc}$ from \citet{suleimanov_basic_2017}.

Using a model for disc anisotropy from \citet{he_anisotropy_2016}, we obtained estimates for an inclination of $i = {74^{+2}_{-4}}^{\circ}$, and anisotropy factors of $\xib = 1.33 \pm 0.09$ and $\xip = 2.4^{+0.7}_{-0.5}$.
The inclination is larger than the earlier estimate of 35--50$^{\circ}$ inferred from ultraviolet modulations in the companion \citep{anderson_time-resolved_1997}, but is consistent with the value of 73--80$^{\circ}$ inferred from radius expansion models with varied mass and radius \citep{shaposhnikov_nature_2004}.
The persistent anisotropy is slightly larger than the range of 1.5--2 from \citet{cumming_models_2003}.

\subsection{Future Work}
\label{subsec:fouru_discussion_future}
This study is the first extension of our multi-epoch MCMC models to a helium burster, \fouru{}.
Because of the simplifications made, these results should be considered a proof-of-concept for future efforts.

To reduce the number of precomputed \kepler{} simulations in the grid, we used fixed values for $\hyd$, $\cno$, and $g$ (Section~\ref{subsec:fouru_grid}).
Although the short orbital period of \SI{11}{min} indicates a hydrogen-poor accreted composition, semi-analytic models suggest a small hydrogen fraction of $\hyd \lesssim 0.10$ may still be possible \citep{cumming_models_2003}.
Expanding our model grid beyond $\hyd = 0.0$ and $\cno = 0.015$ could improve the constraints on the accreted composition.

The particularly large neutron star mass of $M = 2.3^{+0.3}_{-0.2}\, \si{\msun}$ and radius of $R = 15.1^{+0.9}_{-1}\, \si{km}$, along with our results for \gs{} (Chapter~\ref{ch:paper2}), suggests a possible bias in the model towards large masses and radii.
A possible contributing factor is that we assumed a fixed surface gravity of $g = \SI{1.858e14}{\gunits}$, and so only a single contour of $M$ and $R$ was explored, potentially excluding more realistic values.
Using a variable parameter for $g$ when modelling \gs{} did not prevent large mass estimates, although a more typical radius of $R = 11.3 \pm 1.3\, \si{km}$ was obtained.
Another possibility is that the flat priors used for $M$ and $g$ do not appropriately ``penalise'' unrealistic values.
Future studies could incorporate existing mass constraints into the prior.
For example, the mass distribution of observed neutron star populations of $M = 1.46 \pm \SI{0.21}{\msun}$ \citep{ozel_mass_2012}, or theoretical predictions from equation of state models, as used in \citet{goodwin_bayesian_2019}.

Only the burst recurrence time, $\dt$, was predicted with the model grid, while $\Fedd$ and $\Fper$ were calculated analytically.
The dataset from \citet{galloway_thermonuclear_2017} does include burst fluences, and
modelling these in future studies may help to break the degeneracy between $\qb$ and $\mdot$ (Figure~\ref{fig:fouru_posteriors}).
However, the appropriate method to predict fluence remains unclear, because \kepler{} can only crudely reproduce PRE lightcurves (Section~\ref{subsec:fouru_lcsample}).
A simple cutoff at the Eddington luminosity could be adopted (Figure~\ref{fig:fouru_lc_sample}), but the value of $\Ledd$ relies on the choice of $M$ and $R$ during MCMC sampling.
Integrating over the model lightcurves during the MCMC routine, instead of storing precomputed fluences in the grid, could severely impact computational efficiency.

Similar to our posteriors for \gs{}, some of the parameter distributions for \fouru{} are truncated at the boundaries of the grid (Figure~\ref{fig:fouru_posteriors}).
For example, $\qbn{1}$ is limited at $\SI{0.4}{\mevnuc}$, and $\mdotn{2}$ at $0.5\, \mdotedd$.
The model grid could simply be extended, but preliminary models with larger $\qb$ and $\mdot$ transitioned to stable burning.
Because of the rectangular structure of the model grid, regions of parameter space with low probability density must be simulated, for example large pairs of $\qb$ and $\mdot$ (Figure~\ref{fig:fouru_qb-mdot}).
An irregular grid structure could instead be used for the parameter space of interest, but this approach may slow down interpolation calculations and reduce the efficiency of the MCMC routine.
Alternatively, separate grids could be used to model each epoch, each covering a smaller suitable region of parameter space.

\section{Conclusion}
\label{sec:fouru_conclusion}
We have presented the application of MCMC methods to 1D models of hydrogen-poor PRE bursts.
By comparing samples from a precomputed model grid with multi-epoch data, we obtained system parameters for the ultra-compact helium accretor, \fouru{}.
Our parameter estimates were generally consistent with previous works, although the anomalous neutron star mass of $M = 2.3^{+0.3}_{-0.2}\, \si{\msun}$ indicates that more robust priors  should be explored.
Future studies should also aim to include $\hyd$, $\cno$, and $g$ in the model grid, and extend the range of models in $\qb$ and $\mdot$.
Despite these limitations, the posterior distributions of the model predictions were consistent with the data (Figure~\ref{fig:fouru_bprop_sample}), and our comparison of model lightcurves, truncated at $\Ledd$, also revealed broad agreement (Figure~\ref{fig:fouru_lc_sample}).

This study demonstrates that the MCMC models developed in Chapter~\ref{ch:paper2} for \gs{} can be extended to other systems and bursting regimes.
With further refinement, these methods represent a promising avenue for constraining the properties of accreting neutron star systems.

\chapter{Conclusion}
\label{ch:conclusion}
The results presented in this thesis represent valuable steps forward in the modelling of X-ray bursts.
In this concluding chapter, we summarise the main results from Chapters~\ref{ch:results}, \ref{ch:paper1}, \ref{ch:paper2}, and \ref{ch:4u1820}, and discuss the possible directions of future work.

\section{Summary}
\label{sec:conclusion_summary}
In Chapter~\ref{ch:results}, we presented improvements to the setup and analysis of \kepler{} burst models.
It was discovered that previous \kepler{} burst models had inadvertently been using incorrect opacities (Section~\ref{sec:results_opacity}).
An opacity multiplication factor of $\approx 1.5$ had mistakenly remained in the setup files, which led to artificially hotter thermal profiles and increased burst rates.
This error likely contributed to discrepancies that had been noticed between the recurrence times predicted by \kepler{} and other burst codes.

Another issue we uncovered was extended model burn-in (Section~\ref{sec:results_preheating}).
Previously, models were assumed to reach a steady limit cycle within the first few bursts.
We found, however, that systematic trends in the burst properties could persist for tens of bursts, potentially affecting the entire model sequence.
To address this issue, we tested the addition of a nuclear heat source during the thermal setup of the envelope, which was previously neglected.
We found that including this nuclear ``preheating'' helped the envelopes begin closer to thermal equilibrium, and largely removed the burn-in.

Following our improvements to the \kepler{} burst models, we performed the first direct comparisons between 1D burst codes for the same input parameters (Section~\ref{sec:results_mesa}).
Using an existing set of five \mesa{} burst models, we compared the predicted burst rates, energetics, and lightcurves to \kepler{}.
Although we found generally consistent predictions for low accretion rates, at higher accretion rates \kepler{} produced energetics that were $\approx 10$--$\SI{40}{\%}$ larger.
Additionally, there was a systematic offset of 1 burst per day between the predicted burst rates.

In Chapter~\ref{ch:paper1}, we presented the first burst simulations to use time-dependent accretion rates.
By allowing the accretion rate to vary continuously with time, we modelled four observed bursts from a transient accretion episode of \saxj{}.
We successfully reproduced the observed burst timings and and fluences, and predicted additional bursts during windows when the source was not being observed, in line with previous models.
We also computed models using average accretion rates instead of continuously-varying rates, to compare with the method previously used for models of this system.
The comparison suggested a possible systematic bias when using averaged accretion rates, resulting in larger recurrence times when $\Mdot$ is increasing, and smaller recurrence times when $\Mdot$  is decreasing.

In Chapter~\ref{ch:paper2}, we presented the first application of Markov Chain Monte Carlo (MCMC) methods to large grids of burst models.
We precomputed a grid of 3840 \kepler{} simulations across five model parameters.
This was the largest set of 1D burst models to date, and varied parameters that are often left fixed, including the crustal heating, $\qb$, and the surface gravity, $g$.
By interpolating the burst predictions over the grid, we could efficiently sample the parameter space using MCMC methods.
Using multi-epoch observations of the Clocked Burster, \gs{}, we obtained posterior probability distributions for the system parameters.
With epoch-dependent crustal heating, we could test for a dependence of $\qb$ on $\mdot$, and found that lower accretion rates were consistent with larger $\qb$, although there was significant overlap between the posteriors.
This study demonstrated the possibility of constraining system properties using multi-epoch burst data.

In Chapter~\ref{ch:4u1820}, we extended our MCMC methods from Chapter~\ref{ch:paper2} to a helium bursting source, \fouru{}.
We precomputed a grid of 168 hydrogen-poor simulations, and fit observed photospheric radius expansion (PRE) bursts from two epochs of \fouru{}.
We thus obtained posterior distributions for the system parameters.
The predicted distributions of the observables were consistent with the data, and the posterior constraints on the system parameters generally agreed with previous estimates.
A sample comparison of the full burst lightcurves suggested that the overall morphology is reproduced, despite the known limitations of \kepler{} for simulating PRE lightcurves.
This initial extension of our multi-epoch MCMC routine to another system demonstrated its feasibility as a generalised approach to burst modelling.

\section{Future Work and Outlook}
\label{sec:conclusion_future}
Our improvements to the \kepler{} model setup in Chapter~\ref{ch:results} can still be developed further.
The implementation of nuclear preheating largely reduced the model burn-in, but small systematic trends in $\dt$ remained for pure He models.
The preheating setup was only tested for a depth of $y = \SI{8e7}{\yunits}$ and a strength of $\qnuc = \SI{5}{\mevnuc}$, and these values should ideally be adjusted depending on the burst conditions for each model.
Predicting the nuclear energy production prior to actually computing the \kepler{} simulation may prove difficult, and so an iterative approach may be needed.

Additional comparisons between \kepler{}, \mesa{}, and other burst codes should also be pursued.
Our comparison focused on five \mesa{} models of varying $\mdot$, with fixed values for $\hyd$, $\cno$, $\qb$, and $g$.
The discrepancies we found in Section~\ref{sec:results_mesa} may behave differently for other input parameters.
In-depth comparisons of the burst trains, thermal profiles, convective regions, and ashes composition will also help determine the fundamental differences between the burst codes.

Our simulation in Chapter~\ref{ch:paper1} demonstrated the feasibility of modelling bursts during unstable accretion episodes.
Only a single model matched to the observations was presented as a test case, but systematic parameter studies (such as that carried out in Chapter~\ref{ch:paper2}) are needed to constrain the system properties.
In particular, determining the accreted composition of $\hyd$ and $\cno$ could help to constrain the evolutionary history of the binary system.
Extending these methods to other transient accretors could also further test the
model capabilities for reproducing observed burst properties.

Our multi-epoch MCMC models in Chapter~\ref{ch:paper2} pose promising avenues for future work.
As this study was the first implementation of these methods, multiple simplifying assumptions were made.
Flat prior distributions were used for all parameters except $\cno$.
Priors could be explored which are informed by theoretical expectations, such as distributions over the neutron star mass and radius based on equation of state models.
This may help to address the unusually large posteriors for mass, which were truncated at our chosen upper limit of $M = \SI{2.2}{\msun}$.
Of the observables matched by the MCMC routine, the fluence and peak flux were derived from the burst lightcurve, but the full lightcurve itself was not compared with the models.
A challenge with implementing such a comparison, however, is the question of how to interpolate lightcurves between the model grid points.
Alternatively, additional parametrisations of the lightcurve could be used as further constraints, such as exponential or power law fits to the decay tail.
Care should be taken that these quantities behave smoothly over the model grid.
Finally, several posteriors were truncated by the boundaries of the model grid, for example $g$, $\cno$, and $\qb$.
The existing model grid can be extended in these parameters, to better span the parameter space.

Most of the improvements proposed above also apply to our extension to \fouru{} in Chapter~\ref{ch:4u1820}.
In comparison to the models of \gs{}, this study was kept to a limited scope, and lays the groundwork for an expanded project in future.
Only $\mdot$ and $\qb$ were varied for the model grid, to limit the total number of simulations.
Expanding the grid to include the parameters explored for \gs{} -- $\hyd$, $\cno$, and $g$ -- is a natural next step.
\kepler{} lightcurves exhibit super-Eddington luminosities during PRE\@.
To avoid the ambiguity regarding how to correctly extract burst fluences, we interpolated only the burst rate from the model grid.
Fitting the observed fluences may help to break the strong degeneracies between $\mdot$ and $\qb$ seen in the posteriors.
A possible first test is to simply calculate fluences after applying a flat truncation at the Eddington luminosity, as used for the sample lightcurve comparison.

In closing, we have presented multiple contributions to the modelling of thermonuclear X-ray bursts.
The methods developed here serve as a step towards obtaining robust constraints for the properties of accreting neutron stars.

\printbibliography[heading=bibintoc]

@article{foreman-mackey_emcee:_2013,
	title = {emcee: {The} {MCMC} {Hammer}},
	volume = {125},
	issn = {1538-3873},
	shorttitle = {emcee},
	url = {https://doi.org/10.1086%2F670067},
	doi = {10.1086/670067},
	language = {en},
	number = {925},
	urldate = {2019-07-22},
	journal = {Publications of the Astronomical Society of the Pacific},
	author = {Foreman-Mackey, Daniel and Hogg, David W. and Lang, Dustin and Goodman, Jonathan},
	month = mar,
	year = {2013},
	pages = {306--312}
}

@article{cavecchi_fast_2016,
	title = {Fast and slow magnetic deflagration fronts in type {I} {X}-ray bursts},
	volume = {459},
	issn = {0035-8711},
	url = {https://academic.oup.com/mnras/article/459/2/1259/2595115},
	doi = {10.1093/mnras/stw728},
	language = {en},
	number = {2},
	urldate = {2019-08-05},
	journal = {Monthly Notices of the Royal Astronomical Society},
	author = {Cavecchi, Yuri and Levin, Yuri and Watts, Anna L. and Braithwaite, Jonathan},
	month = jun,
	year = {2016},
	pages = {1259--1275}
}

@article{cornelisse_six_2003,
	title = {Six years of {BeppoSAX} {Wide} {Field} {Cameras} observations of nine galactic type {I} {X}-ray bursters},
	volume = {405},
	issn = {0004-6361},
	url = {http://adsabs.harvard.edu/abs/2003A%26A...405.1033C},
	doi = {10.1051/0004-6361:20030629},
	urldate = {2019-08-05},
	journal = {Astronomy and Astrophysics},
	author = {Cornelisse, R. and in't Zand, J. J. M. and Verbunt, F. and Kuulkers, E. and Heise, J. and den Hartog, P. R. and Cocchi, M. and Natalucci, L. and Bazzano, A. and Ubertini, P.},
	month = jul,
	year = {2003},
	pages = {1033--1042}
}

@article{sztajno_constraints_1987,
	title = {Constraints on the mass-radius relation of the neutron star in {4U} 1746-37/{NGC} 6441},
	volume = {226},
	issn = {0035-8711},
	url = {http://adsabs.harvard.edu/abs/1987MNRAS.226...39S},
	doi = {10.1093/mnras/226.1.39},
	urldate = {2019-08-05},
	journal = {Monthly Notices of the Royal Astronomical Society},
	author = {Sztajno, M. and Fujimoto, M. Y. and van Paradijs, J. and Vacca, W. D. and Lewin, W. H. G. and Penninx, W. and Trumper, J.},
	month = may,
	year = {1987},
	pages = {39--55}
}

@article{van_paradijs_relation_1988,
	title = {On the relation between {X}-ray burst properties and the persistent {X}-ray luminosity},
	volume = {233},
	issn = {0035-8711},
	url = {http://adsabs.harvard.edu/abs/1988MNRAS.233..437V},
	doi = {10.1093/mnras/233.2.437},
	urldate = {2019-08-05},
	journal = {Monthly Notices of the Royal Astronomical Society},
	author = {van Paradijs, J. and Penninx, W. and Lewin, W. H. G.},
	month = jul,
	year = {1988},
	pages = {437--450}
}

@article{brown_mapping_2009,
	title = {Mapping crustal heating with the cooling light curves of quasi-persistent transients},
	volume = {698},
	issn = {0004-637X},
	url = {https://doi.org/10.1088%2F0004-637x%2F698%2F2%2F1020},
	doi = {10.1088/0004-637X/698/2/1020},
	language = {en},
	number = {2},
	urldate = {2019-07-24},
	journal = {The Astrophysical Journal},
	author = {Brown, Edward F. and Cumming, Andrew},
	month = may,
	year = {2009},
	pages = {1020--1032}
}

@article{bildsten_thermonuclear_1997,
	title = {Thermonuclear {Burning} on {Rapidly} {Accreting} {Neutron} {Stars}},
	url = {http://arxiv.org/abs/astro-ph/9709094},
	urldate = {2019-07-03},
	journal = {arXiv:astro-ph/9709094},
	author = {Bildsten, Lars},
	month = sep,
	year = {1997},
	note = {arXiv: astro-ph/9709094}
}

@article{cumming_models_2003,
	title = {Models of {Type} {I} {X}-{Ray} {Bursts} from {4U} 1820-30},
	volume = {595},
	issn = {0004-637X},
	url = {http://adsabs.harvard.edu/abs/2003ApJ...595.1077C},
	doi = {10.1086/377446},
	urldate = {2019-05-28},
	journal = {The Astrophysical Journal},
	author = {Cumming, Andrew},
	month = oct,
	year = {2003},
	pages = {1077--1085}
}

@article{kuulkers_photospheric_2003,
	title = {Photospheric radius expansion {X}-ray bursts as standard candles},
	volume = {399},
	issn = {0004-6361},
	url = {http://adsabs.harvard.edu/abs/2003A%26A...399..663K},
	doi = {10.1051/0004-6361:20021781},
	urldate = {2019-05-28},
	journal = {Astronomy and Astrophysics},
	author = {Kuulkers, E. and den Hartog, P. R. and in't Zand, J. J. M. and Verbunt, F. W. M. and Harris, W. E. and Cocchi, M.},
	month = feb,
	year = {2003},
	pages = {663--680}
}

@article{fisker_explosive_2008,
	title = {Explosive {Hydrogen} {Burning} during {Type} {I} {X}-{Ray} {Bursts}},
	volume = {174},
	issn = {0067-0049},
	url = {https://iopscience.iop.org/article/10.1086/521104/meta},
	doi = {10.1086/521104},
	language = {en},
	number = {1},
	urldate = {2019-07-19},
	journal = {The Astrophysical Journal Supplement Series},
	author = {Fisker, Jacob Lund and Schatz, Hendrik and Thielemann, Friedrich-Karl},
	month = jan,
	year = {2008},
	pages = {261}
}

@article{jose_hydrodynamic_2010,
	title = {Hydrodynamic models of {Type} {I} {X}-ray bursts: metallicity effects},
	volume = {189},
	issn = {0067-0049},
	shorttitle = {{HYDRODYNAMIC} {MODELS} {OF} {TYPE} {I} {X}-{RAY} {BURSTS}},
	url = {https://doi.org/10.1088%2F0067-0049%2F189%2F1%2F204},
	doi = {10.1088/0067-0049/189/1/204},
	language = {en},
	number = {1},
	urldate = {2019-07-16},
	journal = {The Astrophysical Journal Supplement Series},
	author = {Jos{\'e}, Jordi and Moreno, Ferm{\'i}n and Parikh, Anuj and Iliadis, Christian},
	month = jun,
	year = {2010},
	pages = {204--239}
}

@article{schatz_rp-process_1998,
	title = {rp-{Process} {Nucleosynthesis} at {Extreme} {Temperature} and {Density} {Conditions}},
	volume = {294},
	issn = {0370-1573},
	url = {http://adsabs.harvard.edu/abs/1998PhR...294..167S},
	doi = {10.1016/S0370-1573(97)00048-3},
	urldate = {2019-07-19},
	journal = {Physics Reports},
	author = {Schatz, H. and Aprahamian, A. and Goerres, J. and Wiescher, M. and Rauscher, T. and Rembges, J. F. and Thielemann, F.-K. and Pfeiffer, B. and Moeller, P. and Kratz, K.-L. and Herndl, H. and Brown, B. A. and Rebel, H.},
	month = feb,
	year = {1998}
}

@article{valenti_near-infrared_2007,
	title = {Near-{Infrared} {Properties} of 24 {Globular} {Clusters} in the {Galactic} {Bulge}*},
	volume = {133},
	issn = {1538-3881},
	url = {https://iopscience.iop.org/article/10.1086/511271/meta},
	doi = {10.1086/511271},
	language = {en},
	number = {4},
	urldate = {2019-07-22},
	journal = {The Astronomical Journal},
	author = {Valenti, E. and Ferraro, F. R. and Origlia, L.},
	month = feb,
	year = {2007},
	pages = {1287}
}

@article{paxton_modules_2015,
	title = {Modules for experiments in stellar astrophysics ({MESA}): binaries, pulsations, and explosions},
	volume = {220},
	issn = {0067-0049},
	shorttitle = {{MODULES} {FOR} {EXPERIMENTS} {IN} {S}TEL{LAR} {ASTROPHYSICS} ({MESA})},
	url = {https://doi.org/10.1088%2F0067-0049%2F220%2F1%2F15},
	doi = {10.1088/0067-0049/220/1/15},
	language = {en},
	number = {1},
	urldate = {2019-07-16},
	journal = {The Astrophysical Journal Supplement Series},
	author = {Paxton, Bill and Marchant, Pablo and Schwab, Josiah and Bauer, Evan B. and Bildsten, Lars and Cantiello, Matteo and Dessart, Luc and Farmer, R. and Hu, H. and Langer, N. and Townsend, R. H. D. and Townsley, Dean M. and Timmes, F. X.},
	month = sep,
	year = {2015},
	pages = {15}
}

@article{hinton_chainconsumer_2016,
	title = {{ChainConsumer}},
	volume = {1},
	url = {http://adsabs.harvard.edu/abs/2016JOSS....1...45H},
	doi = {10.21105/joss.00045},
	urldate = {2019-07-14},
	journal = {The Journal of Open Source Software},
	author = {Hinton, Samuel R.},
	month = aug,
	year = {2016},
	pages = {00045}
}

@article{heger_how_2003,
	title = {How {Massive} {Single} {Stars} {End} {Their} {Life}},
	volume = {591},
	issn = {0004-637X},
	url = {http://adsabs.harvard.edu/abs/2003ApJ...591..288H},
	doi = {10.1086/375341},
	urldate = {2019-07-04},
	journal = {The Astrophysical Journal},
	author = {Heger, A. and Fryer, C. L. and Woosley, S. E. and Langer, N. and Hartmann, D. H.},
	month = jul,
	year = {2003},
	pages = {288--300}
}

@article{goodwin_bayesian_2019,
	title = {A {Bayesian} {Approach} to {Matching} {Thermonuclear} {X}-ray {Burst} {Observations} with {Models}},
	url = {http://arxiv.org/abs/1907.00996},
	urldate = {2019-07-03},
	journal = {arXiv:1907.00996 [astro-ph]},
	author = {Goodwin, A. J. and Galloway, D. K. and Heger, A. and Cumming, A. and Johnston, Z.},
	month = jul,
	year = {2019},
	note = {arXiv: 1907.00996}
}

@article{guver_mass_2010,
	title = {The {Mass} and {Radius} of the {Neutron} {Star} in {4U} 1820-30},
	volume = {719},
	issn = {0004-637X},
	url = {http://adsabs.harvard.edu/abs/2010ApJ...719.1807G},
	doi = {10.1088/0004-637X/719/2/1807},
	urldate = {2019-05-28},
	journal = {The Astrophysical Journal},
	author = {G{\"u}ver, Tolga and Wroblewski, Patricia and Camarota, Larry and {\"O}zel, Feryal},
	month = aug,
	year = {2010},
	pages = {1807--1812}
}

@article{ozel_dense_2016,
	title = {The dense matter equation of state from neutron star radius and mass measurements},
	volume = {820},
	issn = {0004-637X},
	url = {https://doi.org/10.3847%2F0004-637x%2F820%2F1%2F28},
	doi = {10.3847/0004-637X/820/1/28},
	language = {en},
	number = {1},
	urldate = {2019-05-28},
	journal = {The Astrophysical Journal},
	author = {{\"O}zel, Feryal and Psaltis, Dimitrios and G{\"u}ver, Tolga and Baym, Gordon and Heinke, Craig and Guillot, Sebastien},
	month = mar,
	year = {2016},
	pages = {28}
}

@article{deibel_strong_2015,
	title = {A {Strong} {Shallow} {Heat} {Source} in the {Accreting} {Neutron} {Star} {MAXI} {J0556}-332},
	volume = {809},
	issn = {0004-637X},
	url = {http://adsabs.harvard.edu/abs/2015ApJ...809L..31D},
	doi = {10.1088/2041-8205/809/2/L31},
	urldate = {2017-10-18},
	journal = {The Astrophysical Journal Letters},
	author = {Deibel, Alex and Cumming, Andrew and Brown, Edward F. and Page, Dany},
	month = aug,
	year = {2015},
	pages = {L31}
}

@article{tawara_very_1984,
	title = {A very long {X}-ray burst with a precursor from {XB} 1715-321},
	volume = {276},
	issn = {0004-637X},
	url = {http://adsabs.harvard.edu/abs/1984ApJ...276L..41T},
	doi = {10.1086/184184},
	language = {en},
	urldate = {2017-10-05},
	journal = {The Astrophysical Journal},
	author = {Tawara, Y. and Kii, T. and Hayakawa, S. and Kunieda, H. and Masai, K. and Nagase, F. and Inoue, H. and Koyama, K. and Makino, F. and Makishima, K. and Matsuoka, M. and Murakami, T. and Oda, M. and Ogawara, Y. and Ohashi, T. and Shibazaki, N. and Tanaka, Y. and Miyamoto, S. and Tsunemi, H. and Yamashita, K. and Kondo, I.},
	month = jan,
	year = {1984},
	pages = {L41--L44}
}

@article{lewin_precursors_1984,
	title = {Precursors to {X}-ray bursts - {The} result of expansion and subsequent contraction of the neutron star's photosphere},
	volume = {277},
	issn = {0004-637X},
	url = {http://adsabs.harvard.edu/abs/1984ApJ...277L..57L},
	doi = {10.1086/184202},
	language = {en},
	urldate = {2017-10-05},
	journal = {The Astrophysical Journal},
	author = {Lewin, W. H. G. and Vacca, W. D. and Basinska, E. M.},
	month = feb,
	year = {1984},
	pages = {L57--L60}
}

@article{wallace_explosive_1981,
	title = {Explosive hydrogen burning},
	volume = {45},
	issn = {0067-0049},
	url = {http://adsabs.harvard.edu/abs/1981ApJS...45..389W},
	doi = {10.1086/190717},
	urldate = {2017-10-02},
	journal = {The Astrophysical Journal Supplement Series},
	author = {Wallace, R. K. and Woosley, S. E.},
	month = feb,
	year = {1981},
	pages = {389--420}
}

@article{peng_sedimentation_2007,
	title = {Sedimentation and {Type} {I} {X}-{Ray} {Bursts} at {Low} {Accretion} {Rates}},
	volume = {654},
	issn = {0004-637X},
	url = {http://iopscience.iop.org/0004-637X/654/2/1022},
	doi = {10.1086/509628},
	language = {en},
	number = {2},
	urldate = {2015-06-12},
	journal = {The Astrophysical Journal},
	author = {Peng, Fang and Brown, Edward F. and Truran, James W.},
	month = jan,
	year = {2007},
	pages = {1022}
}

@article{galloway_thermonuclear_2008,
	title = {Thermonuclear ({Type} {I}) {X}-{Ray} {Bursts} {Observed} by the {Rossi} {X}-{Ray} {Timing} {Explorer}},
	volume = {179},
	issn = {0067-0049},
	url = {http://iopscience.iop.org/0067-0049/179/2/360},
	doi = {10.1086/592044},
	language = {en},
	number = {2},
	urldate = {2015-06-15},
	journal = {The Astrophysical Journal Supplement Series},
	author = {Galloway, Duncan K. and Muno, Michael P. and Hartman, Jacob M. and Psaltis, Dimitrios and Chakrabarty, Deepto},
	month = dec,
	year = {2008},
	pages = {360}
}

@article{galloway_thermonuclear_2017,
	title = {Thermonuclear {Burst} {Observations} for {Model} {Comparisons}: {A} {Reference} {Sample}},
	volume = {34},
	issn = {1323-3580},
	shorttitle = {Thermonuclear {Burst} {Observations} for {Model} {Comparisons}},
	url = {http://adsabs.harvard.edu/abs/2017PASA...34...19G},
	doi = {10.1017/pasa.2017.12},
	urldate = {2017-05-04},
	journal = {Publications of the Astronomical Society of Australia},
	author = {Galloway, Duncan K. and Goodwin, Adelle J. and Keek, Laurens},
	month = apr,
	year = {2017},
	pages = {e019}
}

@article{heger_models_2007,
	title = {Models of {Type} {I} {X}-{Ray} {Bursts} from {GS} 1826-24: {A} {Probe} of rp-{Process} {Hydrogen} {Burning}},
	volume = {671},
	issn = {0004-637X},
	shorttitle = {Models of {Type} {I} {X}-{Ray} {Bursts} from {GS} 1826-24},
	url = {http://adsabs.harvard.edu/abs/2007ApJ...671L.141H},
	doi = {10.1086/525522},
	urldate = {2015-06-01},
	journal = {The Astrophysical Journal Letters},
	author = {Heger, Alexander and Cumming, Andrew and Galloway, Duncan K. and Woosley, Stanford E.},
	month = dec,
	year = {2007},
	pages = {L141--L144}
}

@article{lampe_influence_2016,
	title = {The {Influence} of {Accretion} {Rate} and {Metallicity} on {Thermonuclear} {Bursts}: {Predictions} from {KEPLER} {Models}},
	volume = {819},
	issn = {0004-637X},
	shorttitle = {The {Influence} of {Accretion} {Rate} and {Metallicity} on {Thermonuclear} {Bursts}},
	url = {http://adsabs.harvard.edu/abs/2016ApJ...819...46L},
	doi = {10.3847/0004-637X/819/1/46},
	urldate = {2016-10-30},
	journal = {The Astrophysical Journal},
	author = {Lampe, Nathanael and Heger, Alexander and Galloway, Duncan K.},
	month = mar,
	year = {2016},
	pages = {46}
}

@article{lewin_x-ray_1993,
	title = {X-ray bursts},
	volume = {62},
	issn = {0038-6308, 1572-9672},
	url = {http://link.springer.com/article/10.1007/BF00196124},
	doi = {10.1007/BF00196124},
	language = {en},
	number = {3-4},
	urldate = {2015-06-11},
	journal = {Space Science Reviews},
	author = {Lewin, Walter H. G. and Paradijs, Jan Van and Taam, Ronald E.},
	month = sep,
	year = {1993},
	pages = {223--389}
}

@article{he_anisotropy_2016,
	title = {Anisotropy of {X}-{Ray} {Bursts} from {Neutron} {Stars} with {Concave} {Accretion} {Disks}},
	volume = {819},
	issn = {0004-637X},
	url = {http://adsabs.harvard.edu/abs/2016ApJ...819...47H},
	doi = {10.3847/0004-637X/819/1/47},
	urldate = {2016-08-29},
	journal = {The Astrophysical Journal},
	author = {He, C.-C. and Keek, L.},
	month = mar,
	year = {2016},
	pages = {47}
}

@article{galloway_thermonuclear_2017-1,
	title = {Thermonuclear {X}-ray bursts},
	url = {http://arxiv.org/abs/1712.06227},
	urldate = {2018-01-09},
	journal = {arXiv:1712.06227 [astro-ph]},
	author = {Galloway, Duncan K. and Keek, Laurens},
	month = dec,
	year = {2017},
	note = {arXiv: 1712.06227}
}

@article{rauscher_nucleosynthesis_2002,
	title = {Nucleosynthesis in {Massive} {Stars} with {Improved} {Nuclear} and {Stellar} {Physics}},
	volume = {576},
	issn = {0004-637X},
	url = {http://adsabs.harvard.edu/abs/2002ApJ...576..323R},
	doi = {10.1086/341728},
	urldate = {2017-11-06},
	journal = {The Astrophysical Journal},
	author = {Rauscher, T. and Heger, A. and Hoffman, R. D. and Woosley, S. E.},
	month = sep,
	year = {2002},
	pages = {323--348}
}

@article{woosley_evolution_2002,
	title = {The evolution and explosion of massive stars},
	volume = {74},
	issn = {0034-6861},
	url = {http://adsabs.harvard.edu/abs/2002RvMP...74.1015W},
	doi = {10.1103/RevModPhys.74.1015},
	urldate = {2016-11-29},
	journal = {Reviews of Modern Physics},
	author = {Woosley, S. E. and Heger, A. and Weaver, T. A.},
	month = nov,
	year = {2002},
	pages = {1015--1071}
}

@article{belian_discovery_1976,
	title = {The discovery of {X}-ray bursts from a region in the constellation {Norma}},
	volume = {206},
	issn = {0004-637X},
	url = {http://adsabs.harvard.edu/abs/1976ApJ...206L.135B},
	doi = {10.1086/182151},
	urldate = {2016-08-12},
	journal = {The Astrophysical Journal Letters},
	author = {Belian, R. D. and Conner, J. P. and Evans, W. D.},
	month = jun,
	year = {1976},
	pages = {L135--L138}
}

@article{weaver_presupernova_1978,
	title = {Presupernova evolution of massive stars},
	volume = {225},
	issn = {0004-637X},
	url = {http://adsabs.harvard.edu/abs/1978ApJ...225.1021W},
	doi = {10.1086/156569},
	urldate = {2015-11-09},
	journal = {The Astrophysical Journal},
	author = {Weaver, T. A. and Zimmerman, G. B. and Woosley, S. E.},
	month = nov,
	year = {1978},
	pages = {1021--1029}
}

@article{grindlay_discovery_1976,
	title = {Discovery of intense {X}-ray bursts from the globular cluster {NGC} 6624},
	volume = {205},
	issn = {0004-637X},
	url = {http://adsabs.harvard.edu/abs/1976ApJ...205L.127G},
	doi = {10.1086/182105},
	urldate = {2016-08-12},
	journal = {The Astrophysical Journal Letters},
	author = {Grindlay, J. and Gursky, H. and Schnopper, H. and Parsignault, D. R. and Heise, J. and Brinkman, A. C. and Schrijver, J.},
	month = may,
	year = {1976},
	pages = {L127--L130}
}

@article{wiescher_cold_2010,
	title = {The {Cold} and {Hot} {CNO} {Cycles}},
	volume = {60},
	url = {http://dx.doi.org/10.1146/annurev.nucl.012809.104505},
	doi = {10.1146/annurev.nucl.012809.104505},
	number = {1},
	urldate = {2016-07-14},
	journal = {Annual Review of Nuclear and Particle Science},
	author = {Wiescher, M. and G{\"o}rres, J. and Uberseder, E. and Imbriani, G. and Pignatari, M.},
	year = {2010},
	pages = {381--404}
}

@article{miller_astrophysical_2013,
	title = {Astrophysical {Constraints} on {Dense} {Matter} in {Neutron} {Stars}},
	url = {http://arxiv.org/abs/1312.0029},
	urldate = {2016-03-11},
	journal = {arXiv:1312.0029 [astro-ph, physics:nucl-th]},
	author = {Miller, M. Coleman},
	month = nov,
	year = {2013},
	note = {arXiv: 1312.0029}
}

@article{fujimoto_shell_1981,
	title = {Shell flashes on accreting neutron stars and {X}-ray bursts},
	volume = {247},
	issn = {0004-637X},
	url = {http://adsabs.harvard.edu/abs/1981ApJ...247..267F},
	doi = {10.1086/159034},
	urldate = {2015-11-03},
	journal = {The Astrophysical Journal},
	author = {Fujimoto, M. Y. and Hanawa, T. and Miyaji, S.},
	month = jul,
	year = {1981},
	pages = {267--278}
}

@article{zingale_comparisons_2015,
	title = {Comparisons of {Two}- and {Three}-{Dimensional} {Convection} in {Type} {I} {X}-{Ray} {Bursts}},
	volume = {807},
	issn = {0004-637X},
	url = {http://adsabs.harvard.edu/abs/2015ApJ...807...60Z},
	doi = {10.1088/0004-637X/807/1/60},
	urldate = {2015-11-01},
	journal = {The Astrophysical Journal},
	author = {Zingale, M. and Malone, C. M. and Nonaka, A. and Almgren, A. S. and Bell, J. B.},
	month = jul,
	year = {2015},
	pages = {60}
}

@article{woosley_-ray_1976,
	title = {$\gamma$-ray bursts from thermonuclear explosions on neutron stars},
	volume = {263},
	copyright = {{\textcopyright} 1976 Nature Publishing Group},
	url = {http://www.nature.com/nature/journal/v263/n5573/abs/263101a0.html},
	doi = {10.1038/263101a0},
	language = {en},
	number = {5573},
	urldate = {2015-06-15},
	journal = {Nature},
	author = {Woosley, S. E. and Taam, Ronald E.},
	month = sep,
	year = {1976},
	pages = {101--103}
}

@article{strohmayer_remarkable_2002,
	title = {A {Remarkable} 3 {Hour} {Thermonuclear} {Burst} from {4U} 1820-30},
	volume = {566},
	issn = {0004-637X},
	url = {http://adsabs.harvard.edu/abs/2002ApJ...566.1045S},
	doi = {10.1086/338337},
	urldate = {2015-06-15},
	journal = {The Astrophysical Journal},
	author = {Strohmayer, Tod E. and Brown, Edward F.},
	month = feb,
	year = {2002},
	pages = {1045--1059}
}

@article{cavecchi_flame_2013,
	title = {Flame propagation on the surfaces of rapidly rotating neutron stars during {Type} {I} {X}-ray bursts},
	volume = {434},
	issn = {0035-8711},
	url = {http://adsabs.harvard.edu/abs/2013MNRAS.434.3526C},
	doi = {10.1093/mnras/stt1273},
	urldate = {2015-06-14},
	journal = {Monthly Notices of the Royal Astronomical Society},
	author = {Cavecchi, Yuri and Watts, Anna L. and Braithwaite, Jonathan and Levin, Yuri},
	month = oct,
	year = {2013},
	pages = {3526--3541}
}

@article{cavecchi_rotational_2015,
	title = {Rotational effects in thermonuclear type {I} bursts: equatorial crossing and directionality of flame spreading},
	volume = {448},
	issn = {0035-8711},
	shorttitle = {Rotational effects in thermonuclear type {I} bursts},
	url = {http://adsabs.harvard.edu/abs/2015MNRAS.448..445C},
	doi = {10.1093/mnras/stu2764},
	urldate = {2015-06-14},
	journal = {Monthly Notices of the Royal Astronomical Society},
	author = {Cavecchi, Yuri and Watts, Anna L. and Levin, Yuri and Braithwaite, Jonathan},
	month = mar,
	year = {2015},
	pages = {445--455}
}

@article{schatz_strong_2014,
	title = {Strong neutrino cooling by cycles of electron capture and $\beta$- decay in neutron star crusts},
	volume = {505},
	copyright = {{\textcopyright} 2013 Nature Publishing Group, a division of Macmillan Publishers Limited. All Rights Reserved.},
	issn = {0028-0836},
	url = {http://www.nature.com/nature/journal/v505/n7481/full/nature12757.html},
	doi = {10.1038/nature12757},
	language = {en},
	number = {7481},
	urldate = {2015-06-10},
	journal = {Nature},
	author = {Schatz, H. and Gupta, S. and M{\"o}ller, P. and Beard, M. and Brown, E. F. and Deibel, A. T. and Gasques, L. R. and Hix, W. R. and Keek, L. and Lau, R. and Steiner, A. W. and Wiescher, M.},
	month = jan,
	year = {2014},
	pages = {62--65}
}

@article{cyburt_jina_2010,
	title = {The {JINA} {REACLIB} {Database}: {Its} {Recent} {Updates} and {Impact} on {Type}-{I} {X}-ray {Bursts}},
	volume = {189},
	issn = {0067-0049},
	shorttitle = {The {JINA} {REACLIB} {Database}},
	url = {http://adsabs.harvard.edu/abs/2010ApJS..189..240C},
	doi = {10.1088/0067-0049/189/1/240},
	urldate = {2015-06-01},
	journal = {The Astrophysical Journal Supplement Series},
	author = {Cyburt, Richard H. and Amthor, A. Matthew and Ferguson, Ryan and Meisel, Zach and Smith, Karl and Warren, Scott and Heger, Alexander and Hoffman, R. D. and Rauscher, Thomas and Sakharuk, Alexander and Schatz, Hendrik and Thielemann, F. K. and Wiescher, Michael},
	month = jul,
	year = {2010},
	pages = {240--252}
}

@article{keek_superburst_2012,
	title = {Superburst {Models} for {Neutron} {Stars} with {Hydrogen}- and {Helium}-rich {Atmospheres}},
	volume = {752},
	issn = {0004-637X},
	url = {http://adsabs.harvard.edu/abs/2012ApJ...752..150K},
	doi = {10.1088/0004-637X/752/2/150},
	urldate = {2015-06-01},
	journal = {The Astrophysical Journal},
	author = {Keek, L. and Heger, A. and in't Zand, J. J. M.},
	month = jun,
	year = {2012},
	pages = {150}
}

@article{keek_multi-zone_2011,
	title = {Multi-zone {Models} of {Superbursts} from {Accreting} {Neutron} {Stars}},
	volume = {743},
	issn = {0004-637X},
	url = {http://iopscience.iop.org/0004-637X/743/2/189},
	doi = {10.1088/0004-637X/743/2/189},
	language = {en},
	number = {2},
	urldate = {2015-06-01},
	journal = {The Astrophysical Journal},
	author = {Keek, L. and Heger, A.},
	month = dec,
	year = {2011},
	pages = {189}
}

@article{woosley_models_2004,
	title = {Models for {Type} {I} {X}-{Ray} {Bursts} with {Improved} {Nuclear} {Physics}},
	volume = {151},
	issn = {0067-0049},
	url = {http://adsabs.harvard.edu/abs/2004ApJS..151...75W},
	doi = {10.1086/381533},
	urldate = {2015-06-01},
	journal = {The Astrophysical Journal Supplement Series},
	author = {Woosley, S. E. and Heger, A. and Cumming, A. and Hoffman, R. D. and Pruet, J. and Rauscher, T. and Fisker, J. L. and Schatz, H. and Brown, B. A. and Wiescher, M.},
	month = mar,
	year = {2004},
	pages = {75--102}
}

@article{cumming_carbon_2001,
	title = {Carbon {Flashes} in the {Heavy}-{Element} {Ocean} on {Accreting} {Neutron} {Stars}},
	volume = {559},
	issn = {0004-637X},
	url = {http://adsabs.harvard.edu/abs/2001ApJ...559L.127C},
	doi = {10.1086/323937},
	urldate = {2015-06-01},
	journal = {The Astrophysical Journal Letters},
	author = {Cumming, Andrew and Bildsten, Lars},
	month = oct,
	year = {2001},
	pages = {L127--L130}
}

@article{liebendorfer_adaptive_2002,
	title = {An {Adaptive} {Grid}, {Implicit} {Code} for {Spherically} {Symmetric}, {General} {Relativistic} {Hydrodynamics} in {Comoving} {Coordinates}},
	volume = {141},
	issn = {0067-0049},
	url = {http://adsabs.harvard.edu/abs/2002ApJS..141..229L},
	doi = {10.1086/339872},
	urldate = {2017-05-24},
	journal = {The Astrophysical Journal Supplement Series},
	author = {Liebend{\"o}rfer, Matthias and Rosswog, Stephan and Thielemann, Friedrich-Karl},
	month = jul,
	year = {2002},
	pages = {229--246}
}

@article{meisel_constraints_2017,
	title = {Constraints on {Bygone} {Nucleosynthesis} of {Accreting} {Neutron} {Stars}},
	volume = {837},
	issn = {0004-637X},
	url = {http://adsabs.harvard.edu/abs/2017ApJ...837...73M},
	doi = {10.3847/1538-4357/aa618d},
	urldate = {2017-03-15},
	journal = {The Astrophysical Journal},
	author = {Meisel, Zach and Deibel, Alex},
	month = mar,
	year = {2017},
	pages = {73}
}

@inproceedings{galloway_thermonuclear_2004,
	address = {eprint: arXiv:astro-ph/0404449},
	title = {Thermonuclear burst physics with {RXTE}},
	volume = {714},
	url = {http://adsabs.harvard.edu/abs/2004AIPC..714..266G},
	doi = {10.1063/1.1781039},
	urldate = {2017-02-28},
	author = {Galloway, Duncan K. and Chakrabarty, Deepto and Cumming, Andrew and Kuulkers, Erik and Bildsten, Lars and Rothschild, Richard},
	month = jul,
	year = {2004},
	pages = {266--272}
}

@article{cyburt_dependence_2016,
	title = {Dependence of {X}-{Ray} {Burst} {Models} on {Nuclear} {Reaction} {Rates}},
	volume = {830},
	issn = {0004-637X},
	url = {http://adsabs.harvard.edu/abs/2016ApJ...830...55C},
	doi = {10.3847/0004-637X/830/2/55},
	urldate = {2017-01-31},
	journal = {The Astrophysical Journal},
	author = {Cyburt, R. H. and Amthor, A. M. and Heger, A. and Johnson, E. and Keek, L. and Meisel, Z. and Schatz, H. and Smith, K.},
	month = oct,
	year = {2016},
	pages = {55}
}

@article{cumming_long_2006,
	title = {Long {Type} {I} {X}-{Ray} {Bursts} and {Neutron} {Star} {Interior} {Physics}},
	volume = {646},
	issn = {0004-637X},
	url = {http://adsabs.harvard.edu/abs/2006ApJ...646..429C},
	doi = {10.1086/504698},
	urldate = {2016-11-28},
	journal = {The Astrophysical Journal},
	author = {Cumming, Andrew and Macbeth, Jared and in 't Zand, J. J. M. and Page, Dany},
	month = jul,
	year = {2006},
	pages = {429--451}
}

@article{lapidus_angular_1985,
	title = {Angular distribution and polarization of {X}-ray-burster radiation (during stationary and flash phases)},
	volume = {217},
	issn = {0035-8711},
	url = {http://adsabs.harvard.edu/abs/1985MNRAS.217..291L},
	doi = {10.1093/mnras/217.2.291},
	urldate = {2016-11-17},
	journal = {Monthly Notices of the Royal Astronomical Society},
	author = {Lapidus, I. I. and Sunyaev, R. A.},
	month = nov,
	year = {1985},
	pages = {291--303}
}

@article{fujimoto_angular_1988,
	title = {Angular distribution of radiation from low-mass {X}-ray binaries},
	volume = {324},
	issn = {0004-637X},
	url = {http://adsabs.harvard.edu/abs/1988ApJ...324..995F},
	doi = {10.1086/165955},
	urldate = {2016-11-17},
	journal = {The Astrophysical Journal},
	author = {Fujimoto, Masayuki Y.},
	month = jan,
	year = {1988},
	pages = {995--1000}
}

@article{joss_x-ray_1977,
	title = {X-ray bursts and neutron-star thermonuclear flashes},
	volume = {270},
	issn = {0028-0836},
	url = {http://adsabs.harvard.edu/abs/1977Natur.270..310J},
	doi = {10.1038/270310a0},
	urldate = {2016-10-30},
	journal = {Nature},
	author = {Joss, P. C.},
	month = nov,
	year = {1977},
	pages = {310--314}
}

@article{joss_helium-burning_1978,
	title = {Helium-burning flashes on an accreting neutron star - {A} model for {X}-ray burst sources},
	volume = {225},
	issn = {0004-637X},
	url = {http://adsabs.harvard.edu/abs/1978ApJ...225L.123J},
	doi = {10.1086/182808},
	urldate = {2016-10-30},
	journal = {The Astrophysical Journal Letters},
	author = {Joss, P. C.},
	month = nov,
	year = {1978},
	pages = {L123--L127}
}

@article{taam_x-ray_1980,
	title = {X-ray bursts from thermonuclear runaways on accreting neutron stars},
	volume = {241},
	issn = {0004-637X},
	url = {http://adsabs.harvard.edu/abs/1980ApJ...241..358T},
	doi = {10.1086/158348},
	urldate = {2016-10-30},
	journal = {The Astrophysical Journal},
	author = {Taam, R. E.},
	month = oct,
	year = {1980},
	pages = {358--366}
}

@article{heger_millihertz_2007,
	title = {Millihertz {Quasi}-periodic {Oscillations} from {Marginally} {Stable} {Nuclear} {Burning} on an {Accreting} {Neutron} {Star}},
	volume = {665},
	issn = {0004-637X},
	url = {http://adsabs.harvard.edu/abs/2007ApJ...665.1311H},
	doi = {10.1086/517491},
	urldate = {2016-10-26},
	journal = {The Astrophysical Journal},
	author = {Heger, Alexander and Cumming, Andrew and Woosley, S. E.},
	month = aug,
	year = {2007},
	pages = {1311--1320}
}

@article{haensel_models_2008,
	title = {Models of crustal heating in accreting neutron stars},
	volume = {480},
	issn = {0004-6361},
	url = {http://adsabs.harvard.edu/abs/2008A%26A...480..459H},
	doi = {10.1051/0004-6361:20078578},
	urldate = {2016-09-06},
	journal = {Astronomy and Astrophysics},
	author = {Haensel, P. and Zdunik, J. L.},
	month = mar,
	year = {2008},
	pages = {459--464}
}

@article{watts_thermonuclear_2012,
	title = {Thermonuclear {Burst} {Oscillations}},
	volume = {50},
	url = {https://doi.org/10.1146/annurev-astro-040312-132617},
	doi = {10.1146/annurev-astro-040312-132617},
	number = {1},
	urldate = {2018-05-29},
	journal = {Annual Review of Astronomy and Astrophysics},
	author = {Watts, Anna L.},
	year = {2012},
	pages = {609--640}
}

@article{johnston_simulating_2018,
	title = {Simulating {X}-ray bursts during a transient accretion event},
	volume = {477},
	issn = {0035-8711, 1365-2966},
	url = {https://academic.oup.com/mnras/article/477/2/2112/4952008},
	doi = {10.1093/mnras/sty757},
	language = {en},
	number = {2},
	urldate = {2018-09-30},
	journal = {Monthly Notices of the Royal Astronomical Society},
	author = {Johnston, Zac and Heger, Alexander and Galloway, Duncan K},
	month = jun,
	year = {2018},
	pages = {2112--2118}
}

@article{meisel_consistent_2018,
	title = {Consistent {Modeling} of {GS} 1826-24 {X}-{Ray} {Bursts} for {Multiple} {Accretion} {Rates} {Demonstrates} the {Possibility} of {Constraining} \textit{rp} -process {Reaction} {Rates}},
	volume = {860},
	issn = {1538-4357},
	url = {http://stacks.iop.org/0004-637X/860/i=2/a=147?key=crossref.79f751c8f258d911473eb9b0a25a0b85},
	doi = {10.3847/1538-4357/aac3d3},
	language = {en},
	number = {2},
	urldate = {2018-09-30},
	journal = {The Astrophysical Journal},
	author = {Meisel, Zach},
	month = jun,
	year = {2018},
	pages = {147}
}

@article{fisker_importance_2006,
	title = {The {Importance} of {15O}($\alpha$,$\gamma$){19Ne} to {X}-{Ray} {Bursts} and {Superbursts}},
	volume = {650},
	url = {https://ui.adsabs.harvard.edu/abs/2006ApJ...650..332F/abstract},
	doi = {10.1086/507083},
	language = {en},
	number = {1},
	urldate = {2019-10-10},
	journal = {The Astrophysical Journal},
	author = {Fisker, Jacob Lund and G{\"o}rres, Joachim and Wiescher, Michael and Davids, Barry},
	month = oct,
	year = {2006},
	pages = {332}
}

@article{meisel_influence_2019,
	title = {Influence of {Nuclear} {Reaction} {Rate} {Uncertainties} on {Neutron} {Star} {Properties} {Extracted} from {X}-{Ray} {Burst} {Model}{\textendash}{Observation} {Comparisons}},
	volume = {872},
	issn = {1538-4357},
	url = {http://stacks.iop.org/0004-637X/872/i=1/a=84?key=crossref.b19254c80c2d22963aa448e3ade3f35f},
	doi = {10.3847/1538-4357/aafede},
	number = {1},
	urldate = {2019-10-10},
	journal = {The Astrophysical Journal},
	author = {Meisel, Zach and Merz, Grant and Medvid, Sophia},
	month = feb,
	year = {2019},
	pages = {84}
}

@article{hanawa_thermal_1984,
	title = {Thermal response of neutron stars to shell flashes},
	volume = {36},
	issn = {0004-6264},
	url = {https://ui.adsabs.harvard.edu/abs/1984PASJ...36..199H/abstract},
	language = {en},
	number = {2},
	urldate = {2019-10-09},
	journal = {Publications of the Astronomical Society of Japan},
	author = {Hanawa, T. and Fujimoto, M. Y.},
	year = {1984},
	pages = {199}
}

@article{schatz_end_2001,
	title = {End {Point} of the rp {Process} on {Accreting} {Neutron} {Stars}},
	volume = {86},
	url = {https://ui.adsabs.harvard.edu/abs/2001PhRvL..86.3471S/abstract},
	doi = {10.1103/PhysRevLett.86.3471},
	language = {en},
	number = {16},
	urldate = {2019-10-09},
	journal = {Physical Review Letters},
	author = {Schatz, H. and Aprahamian, A. and Barnard, V. and Bildsten, L. and Cumming, A. and Ouellette, M. and Rauscher, T. and Thielemann, F.-K. and Wiescher, M.},
	month = apr,
	year = {2001},
	pages = {3471}
}

@article{wallace_thermonuclear_1982,
	title = {The thermonuclear model for {X}-ray transients},
	volume = {258},
	doi = {10.1086/160119},
	journal = {The Astrophysical Journal},
	author = {Wallace, R. K. and Woosley, S. E. and Weaver, T. A.},
	month = jul,
	year = {1982},
	pages = {696--715}
}

@article{johnston_multi-epoch_2019,
	title = {Multi-epoch {X}-ray burst modelling: {MCMC} with large grids of {1D} simulations},
	shorttitle = {Multi-epoch {X}-ray burst modelling},
	url = {http://arxiv.org/abs/1909.07977},
	urldate = {2019-10-04},
	journal = {arXiv:1909.07977 [astro-ph]},
	author = {Johnston, Zac and Heger, Alexander and Galloway, Duncan K.},
	month = sep,
	year = {2019},
	note = {arXiv: 1909.07977}
}

@article{jose_nucleosynthesis_1998,
	title = {Nucleosynthesis in {Classical} {Novae}: {CO} versus {ONe} {White} {Dwarfs}},
	volume = {494},
	issn = {0004-637X},
	shorttitle = {Nucleosynthesis in {Classical} {Novae}},
	url = {http://adsabs.harvard.edu/abs/1998ApJ...494..680J},
	doi = {10.1086/305244},
	urldate = {2019-10-03},
	journal = {The Astrophysical Journal},
	author = {Jos{\'e}, Jordi and Hernanz, Margarita},
	month = feb,
	year = {1998},
	pages = {680--690}
}

@article{rakavy_carbon_1967,
	title = {Carbon and {Oxygen} {Burning} {Stars} and {Pre}-{Supernova} {Models}},
	volume = {150},
	doi = {10.1086/149318},
	journal = {The Astrophysical Journal},
	author = {Rakavy, G. and Shaviv, G. and Zinamon, Z.},
	month = oct,
	year = {1967},
	pages = {131}
}

@article{taam_nuclear_1978,
	title = {Nuclear fusion and carbon flashes on neutron stars},
	volume = {224},
	issn = {0004-637X, 1538-4357},
	url = {http://adsabs.harvard.edu/doi/10.1086/156367},
	doi = {10.1086/156367},
	language = {en},
	urldate = {2019-09-30},
	journal = {The Astrophysical Journal},
	author = {Taam, R. E. and Picklum, R. E.},
	month = aug,
	year = {1978},
	pages = {210}
}

@article{fujimoto_asymptotic_1979,
	title = {Asymptotic {Strength} of {Thermal} {Pulses} in the {Helium} {Shell} {Burning}},
	volume = {31},
	journal = {Publications of the Astronomical Society of Japan},
	author = {Fujimoto, M. Y. and Sugimoto, D.},
	month = jan,
	year = {1979},
	pages = {1--10}
}

@article{sugimoto_general_1978,
	title = {A {General} {Theory} for {Thermal} {Pulses} of {Finite} {Amplitude} in {Nuclear} {Shell}- {Burnings}},
	volume = {30},
	journal = {Publications of the Astronomical Society of Japan},
	author = {Sugimoto, D. and Fujimoto, M. Y.},
	month = jan,
	year = {1978},
	pages = {467--482}
}

@article{lamb_nuclear_1978,
	title = {Nuclear burning in accreting neutron stars and {X}-ray bursts},
	volume = {220},
	issn = {0004-637X},
	url = {http://adsabs.harvard.edu/abs/1978ApJ...220..291L},
	doi = {10.1086/155905},
	urldate = {2019-09-30},
	journal = {The Astrophysical Journal},
	author = {Lamb, D. Q. and Lamb, F. K.},
	month = feb,
	year = {1978},
	pages = {291--302}
}

@inproceedings{maraschi_x-ray_1977,
	title = {X-ray bursts of nuclear origin?},
	volume = {4},
	url = {https://ui.adsabs.harvard.edu/abs/1977HiA.....4A..71V/abstract},
	booktitle = {Highlights in {Astronomy}},
	author = {Maraschi, L. and Cavaliere, A.},
	year = {1977},
	pages = {127}
}

@article{keek_carbon_2016,
	title = {Carbon production on accreting neutron stars in a new regime of stable nuclear burning},
	volume = {456},
	issn = {1745-3925},
	url = {https://academic.oup.com/mnrasl/article/456/1/L11/2589543},
	doi = {10.1093/mnrasl/slv167},
	language = {en},
	number = {1},
	urldate = {2019-09-30},
	journal = {Monthly Notices of the Royal Astronomical Society: Letters},
	author = {Keek, L. and Heger, A.},
	month = feb,
	year = {2016},
	pages = {L11--L15}
}

@article{hoffman_dual_1978,
	title = {Dual character of the rapid burster and a classification of {X}-ray bursts},
	volume = {271},
	copyright = {1978 Nature Publishing Group},
	issn = {1476-4687},
	url = {https://www-nature-com.ezproxy.lib.monash.edu.au/articles/271630a0},
	doi = {10.1038/271630a0},
	language = {en},
	number = {5646},
	urldate = {2019-09-29},
	journal = {Nature},
	author = {Hoffman, Jeffrey A. and Marshall, Herman L. and Lewin, Walter H. G.},
	month = feb,
	year = {1978},
	pages = {630--633}
}

@article{schwarzschild_thermal_1965,
	title = {Thermal {Instability} in {Non}-{Degenerate} {Stars}.},
	volume = {142},
	url = {https://ui.adsabs.harvard.edu/abs/1965ApJ...142..855S/abstract},
	doi = {10.1086/148358},
	language = {en},
	urldate = {2019-09-29},
	journal = {The Astrophysical Journal},
	author = {Schwarzschild, M. and H{\"a}rm, R.},
	month = oct,
	year = {1965},
	pages = {855}
}

@article{hansen_steady-state_1975,
	title = {Steady-state nuclear fusion in accreting neutron-star envelopes},
	volume = {195},
	issn = {0004-637X, 1538-4357},
	url = {http://adsabs.harvard.edu/doi/10.1086/153375},
	doi = {10.1086/153375},
	language = {en},
	urldate = {2019-09-29},
	journal = {The Astrophysical Journal},
	author = {Hansen, C. J. and van Horn, H. M.},
	month = feb,
	year = {1975},
	pages = {735}
}

@article{clark_recurrent_1976,
	title = {Recurrent brief {X}-ray bursts from the globular cluster {NGC} 6624},
	volume = {207},
	issn = {0004-637X},
	url = {http://adsabs.harvard.edu/abs/1976ApJ...207L.105C},
	doi = {10.1086/182190},
	language = {en},
	urldate = {2019-09-29},
	journal = {The Astrophysical Journal},
	author = {Clark, G. W. and Jernigan, J. G. and Bradt, H. and Canizares, C. and Lewin, W. H. G. and Li, F. K. and Mayer, W. and McClintock, J. and Schnopper, H.},
	month = jul,
	year = {1976},
	pages = {L105--L108}
}

@article{stella_discovery_1987,
	title = {The discovery of a 685 second orbital period from the {X}-ray source {4U} 1820 - 30 in the globular cluster {NGC} 6624},
	volume = {312},
	issn = {0004-637X},
	url = {http://adsabs.harvard.edu/abs/1987ApJ...312L..17S},
	doi = {10.1086/184811},
	urldate = {2019-09-26},
	journal = {The Astrophysical Journal Letters},
	author = {Stella, L. and Priedhorsky, W. and White, N. E.},
	month = jan,
	year = {1987},
	pages = {L17--L21}
}

@article{anderson_time-resolved_1997,
	title = {Time-{Resolved} {Ultraviolet} {Observations} of the {Globular} {Cluster} {X}-{Ray} {Source} in {NGC} 6624: {The} {Shortest} {Known} {Period} {Binary} {System}},
	volume = {482},
	issn = {0004-637X},
	shorttitle = {Time-{Resolved} {Ultraviolet} {Observations} of the {Globular} {Cluster} {X}-{Ray} {Source} in {NGC} 6624},
	url = {http://adsabs.harvard.edu/abs/1997ApJ...482L..69A},
	doi = {10.1086/310672},
	urldate = {2019-09-23},
	journal = {The Astrophysical Journal Letters},
	author = {Anderson, Scott F. and Margon, Bruce and Deutsch, Eric W. and Downes, Ronald A. and Allen, Richard G.},
	month = jun,
	year = {1997},
	pages = {L69--L72}
}

@article{cromartie_relativistic_2019,
	title = {Relativistic {Shapiro} delay measurements of an extremely massive millisecond pulsar},
	copyright = {2019 The Author(s), under exclusive licence to Springer Nature Limited},
	issn = {2397-3366},
	url = {https://www.nature.com/articles/s41550-019-0880-2},
	doi = {10.1038/s41550-019-0880-2},
	language = {en},
	urldate = {2019-09-22},
	journal = {Nature Astronomy},
	author = {Cromartie, H. T. and Fonseca, E. and Ransom, S. M. and Demorest, P. B. and Arzoumanian, Z. and Blumer, H. and Brook, P. R. and DeCesar, M. E. and Dolch, T. and Ellis, J. A. and Ferdman, R. D. and Ferrara, E. C. and Garver-Daniels, N. and Gentile, P. A. and Jones, M. L. and Lam, M. T. and Lorimer, D. R. and Lynch, R. S. and McLaughlin, M. A. and Ng, C. and Nice, D. J. and Pennucci, T. T. and Spiewak, R. and Stairs, I. H. and Stovall, K. and Swiggum, J. K. and Zhu, W. W.},
	month = sep,
	year = {2019},
	pages = {1--5}
}

@article{margalit_constraining_2017,
	title = {Constraining the {Maximum} {Mass} of {Neutron} {Stars} from {Multi}-messenger {Observations} of {GW170817}},
	volume = {850},
	issn = {2041-8205},
	url = {https://doi.org/10.3847%2F2041-8213%2Faa991c},
	doi = {10.3847/2041-8213/aa991c},
	language = {en},
	number = {2},
	urldate = {2019-09-22},
	journal = {The Astrophysical Journal},
	author = {Margalit, Ben and Metzger, Brian D.},
	month = nov,
	year = {2017},
	pages = {L19}
}

@article{shaposhnikov_nature_2004,
	title = {On the {Nature} of the {Flux} {Variability} during an {Expansion} {Stage} of a {Type} {I} {X}-{Ray} {Burst}: {Constraints} on {Neutron} {Star} {Parameters} for {4U} 1820{\textendash}30},
	volume = {606},
	issn = {1538-4357},
	shorttitle = {On the {Nature} of the {Flux} {Variability} during an {Expansion} {Stage} of a {Type} {I} {X}-{Ray} {Burst}},
	url = {https://iopscience.iop.org/article/10.1086/421015/meta},
	doi = {10.1086/421015},
	language = {en},
	number = {1},
	urldate = {2019-09-18},
	journal = {The Astrophysical Journal Letters},
	author = {Shaposhnikov, Nickolai and Titarchuk, Lev},
	month = apr,
	year = {2004},
	pages = {L57}
}

@article{suleimanov_basic_2017,
	title = {Basic parameters of the helium-accreting {X}-ray bursting neutron star in {4U} 1820-30},
	volume = {472},
	issn = {0035-8711},
	url = {https://academic.oup.com/mnras/article/472/4/3905/4102338},
	doi = {10.1093/mnras/stx2234},
	language = {en},
	number = {4},
	urldate = {2019-09-18},
	journal = {Monthly Notices of the Royal Astronomical Society},
	author = {Suleimanov, Valery F. and Kajava, Jari J. E. and Molkov, Sergey V. and N{\"a}ttil{\"a}, Joonas and Lutovinov, Alexander A. and Werner, Klaus and Poutanen, Juri},
	month = dec,
	year = {2017},
	pages = {3905--3913}
}

@article{ozel_mass_2012,
	title = {The mass and radius of the neutron star in the bulge low-mass {X}-ray binary {KS} 1731{\textendash}260},
	volume = {748},
	issn = {0004-637X},
	url = {https://doi.org/10.1088%2F0004-637x%2F748%2F1%2F5},
	doi = {10.1088/0004-637X/748/1/5},
	language = {en},
	number = {1},
	urldate = {2019-09-13},
	journal = {The Astrophysical Journal},
	author = {{\"O}zel, Feryal and Gould, Andrew and G{\"u}ver, Tolga},
	month = feb,
	year = {2012},
	pages = {5}
}

@article{ozel_masses_2016,
	title = {Masses, {Radii}, and the {Equation} of {State} of {Neutron} {Stars}},
	volume = {54},
	issn = {0066-4146},
	url = {https://www.annualreviews.org/doi/10.1146/annurev-astro-081915-023322},
	doi = {10.1146/annurev-astro-081915-023322},
	number = {1},
	urldate = {2019-09-13},
	journal = {Annual Review of Astronomy and Astrophysics},
	author = {{\"O}zel, Feryal and Freire, Paulo},
	month = sep,
	year = {2016},
	pages = {401--440}
}

@article{linares_peering_2018,
	title = {Peering into the {Dark} {Side}: {Magnesium} {Lines} {Establish} a {Massive} {Neutron} {Star} in {PSR} {J2215}+5135},
	volume = {859},
	issn = {0004-637X},
	shorttitle = {Peering into the {Dark} {Side}},
	url = {https://doi.org/10.3847%2F1538-4357%2Faabde6},
	doi = {10.3847/1538-4357/aabde6},
	language = {en},
	number = {1},
	urldate = {2019-09-11},
	journal = {The Astrophysical Journal},
	author = {Linares, M. and Shahbaz, T. and Casares, J.},
	month = may,
	year = {2018},
	pages = {54}
}

@article{gupta_heating_2007,
	title = {Heating in the {Accreted} {Neutron} {Star} {Ocean}: {Implications} for {Superburst} {Ignition}},
	volume = {662},
	issn = {0004-637X},
	shorttitle = {Heating in the {Accreted} {Neutron} {Star} {Ocean}},
	url = {https://iopscience.iop.org/article/10.1086/517869/meta},
	doi = {10.1086/517869},
	language = {en},
	number = {2},
	urldate = {2019-09-11},
	journal = {The Astrophysical Journal},
	author = {Gupta, Sanjib and Brown, Edward F. and Schatz, Hendrik and M{\"o}ller, Peter and Kratz, Karl-Ludwig},
	month = jun,
	year = {2007},
	pages = {1188}
}

@article{king_shortest_1986,
	title = {The shortest period binary star?},
	volume = {323},
	url = {https://ui.adsabs.harvard.edu/abs/1986Natur.323..105K/abstract},
	doi = {10.1038/323105a0},
	language = {en},
	number = {6084},
	urldate = {2019-09-09},
	journal = {Nature},
	author = {King, A. R. and Watson, M. G.},
	month = sep,
	year = {1986},
	pages = {105}
}

@incollection{strohmayer_new_2006,
	title = {New {Views} of {Thermonuclear} {Bursts}},
	volume = {39},
	isbn = {978-0-521-82659-4},
	booktitle = {Compact {Stellar} {X}-ray {Sources}},
	publisher = {Cambridge University Press},
	author = {Strohmayer, Tod and Bildsten, Lars},
	editor = {Lewin, Walter and van der Klis, Michiel},
	month = apr,
	year = {2006},
	pages = {113--156}
}

@article{lewin_discovery_1976,
	title = {Discovery of {X}-ray bursts from several sources near the galactic centre},
	volume = {177},
	issn = {0035-8711},
	url = {http://adsabs.harvard.edu/abs/1976MNRAS.177P..83L},
	doi = {10.1093/mnras/177.1.83P},
	urldate = {2019-08-10},
	journal = {Monthly Notices of the Royal Astronomical Society},
	author = {Lewin, W. H. G. and Hoffman, J. A. and Doty, J. and Hearn, D. R. and Clark, G. W. and Jernigan, J. G. and Li, F. K. and McClintock, J. E. and Richardson, J.},
	month = dec,
	year = {1976},
	pages = {83P--92P}
}

@article{brown_ocean_1998,
	title = {The {Ocean} and {Crust} of a {Rapidly} {Accreting} {Neutron} {Star}: {Implications} for {Magnetic} {Field} {Evolution} and {Thermonuclear} {Flashes}},
	volume = {496},
	issn = {0004-637X},
	shorttitle = {The {Ocean} and {Crust} of a {Rapidly} {Accreting} {Neutron} {Star}},
	url = {https://iopscience.iop.org/article/10.1086/305419/meta},
	doi = {10.1086/305419},
	language = {en},
	number = {2},
	urldate = {2019-07-24},
	journal = {The Astrophysical Journal},
	author = {Brown, Edward F. and Bildsten, Lars},
	month = apr,
	year = {1998},
	pages = {915}
}

@article{parikh_effects_2008,
	title = {The {Effects} of {Variations} in {Nuclear} {Processes} on {Type} {I} {X}-{Ray} {Burst} {Nucleosynthesis}},
	volume = {178},
	issn = {0067-0049},
	url = {http://adsabs.harvard.edu/abs/2008ApJS..178..110P},
	doi = {10.1086/589879},
	urldate = {2020-01-31},
	journal = {The Astrophysical Journal Supplement Series},
	author = {Parikh, Anuj and Jos{\'e}, Jordi and Moreno, Ferm{\'i}n and Iliadis, Christian},
	month = sep,
	year = {2008},
	pages = {110--136}
}

@article{parikh_impact_2009,
	title = {Impact of uncertainties in reaction {Q} values on nucleosynthesis in {Type} {I} {X}-ray bursts},
	volume = {79},
	issn = {0556-2813},
	url = {http://adsabs.harvard.edu/abs/2009PhRvC..79d5802P},
	doi = {10.1103/PhysRevC.79.045802},
	urldate = {2020-01-31},
	journal = {Physical Review C},
	author = {Parikh, A. and Jos{\'e}, J. and Iliadis, C. and Moreno, F. and Rauscher, T.},
	month = apr,
	year = {2009},
	pages = {045802}
}

@article{koike_rapid_1999,
	title = {Rapid proton capture on accreting neutron stars - effects of uncertainty in the nuclear process},
	volume = {342},
	issn = {0004-6361},
	url = {http://adsabs.harvard.edu/abs/1999A%26A...342..464K},
	urldate = {2020-01-29},
	journal = {Astronomy and Astrophysics},
	author = {Koike, O. and Hashimoto, M. and Arai, K. and Wanajo, S.},
	month = feb,
	year = {1999},
	pages = {464--473}
}

@article{koike_final_2004,
	title = {Final {Products} of the rp-{Process} on {Accreting} {Neutron} {Stars}},
	volume = {603},
	issn = {0004-637X},
	url = {http://adsabs.harvard.edu/abs/2004ApJ...603..242K},
	doi = {10.1086/381354},
	urldate = {2020-01-29},
	journal = {The Astrophysical Journal},
	author = {Koike, Osamu and Hashimoto, Masa-aki and Kuromizu, Reiko and Fujimoto, Shin-ichirou},
	month = mar,
	year = {2004},
	pages = {242--251}
}

@article{babushkina_hard_1975,
	title = {Hard {X}-ray bursts in {June} 1971},
	volume = {1},
	url = {http://adsabs.harvard.edu/abs/1975SvAL....1...32B},
	urldate = {2020-01-29},
	journal = {Soviet Astronomy Letters},
	author = {Babushkina, O. P. and Kudriavtsev, M. I. and Melioranskii, A. S. and Savenko, I. A. and Iushkov, B. Iu. and Bratoliubova-Tsulukidze, L. S.},
	month = feb,
	year = {1975},
	pages = {32--34}
}

@article{shara_localized_1982,
	title = {Localized thermonuclear runaways and volcanoes on degenerate dwarf stars},
	volume = {261},
	issn = {0004-637X},
	url = {http://adsabs.harvard.edu/abs/1982ApJ...261..649S},
	doi = {10.1086/160376},
	urldate = {2020-02-11},
	journal = {The Astrophysical Journal},
	author = {Shara, M. M.},
	month = oct,
	year = {1982},
	pages = {649--660}
}

@article{abbott_gw170817_2018,
	title = {{GW170817}: {Measurements} of {Neutron} {Star} {Radii} and {Equation} of {State}},
	volume = {121},
	issn = {0031-9007},
	shorttitle = {{GW170817}},
	url = {http://adsabs.harvard.edu/abs/2018PhRvL.121p1101A},
	doi = {10.1103/PhysRevLett.121.161101},
	urldate = {2020-02-16},
	journal = {Physical Review Letters},
	author = {Abbott, B. P. and Abbott, R. and Abbott, T. D. and Acernese, F. and Ackley, K. and Adams, C. and Adams, T. and Addesso, P. and Adhikari, R. X. and Adya, V. B. and Affeldt, C. and Agarwal, B. and Agathos, M. and Agatsuma, K. and Aggarwal, N. and Aguiar, O. D. and Aiello, L. and Ain, A. and Ajith, P. and Allen, B. and Allen, G. and Allocca, A. and Aloy, M. A. and Altin, P. A. and Amato, A. and Ananyeva, A. and Anderson, S. B. and Anderson, W. G. and Angelova, S. V. and Antier, S. and Appert, S. and Arai, K. and Araya, M. C. and Areeda, J. S. and Ar{\`e}ne, M. and Arnaud, N. and Arun, K. G. and Ascenzi, S. and Ashton, G. and Ast, M. and Aston, S. M. and Astone, P. and Atallah, D. V. and Aubin, F. and Aufmuth, P. and Aulbert, C. and AultONeal, K. and Austin, C. and Avila-Alvarez, A. and Babak, S. and Bacon, P. and Badaracco, F. and Bader, M. K. M. and Bae, S. and Baker, P. T. and Baldaccini, F. and Ballardin, G. and Ballmer, S. W. and Banagiri, S. and Barayoga, J. C. and Barclay, S. E. and Barish, B. C. and Barker, D. and Barkett, K. and Barnum, S. and Barone, F. and Barr, B. and Barsotti, L. and Barsuglia, M. and Barta, D. and Bartlett, J. and Bartos, I. and Bassiri, R. and Basti, A. and Batch, J. C. and Bawaj, M. and Bayley, J. C. and Bazzan, M. and B{\'e}csy, B. and Beer, C. and Bejger, M. and Belahcene, I. and Bell, A. S. and Beniwal, D. and Bensch, M. and Berger, B. K. and Bergmann, G. and Bernuzzi, S. and Bero, J. J. and Berry, C. P. L. and Bersanetti, D. and Bertolini, A. and Betzwieser, J. and Bhandare, R. and Bilenko, I. A. and Bilgili, S. A. and Billingsley, G. and Billman, C. R. and Birch, J. and Birney, R. and Birnholtz, O. and Biscans, S. and Biscoveanu, S. and Bisht, A. and Bitossi, M. and Bizouard, M. A. and Blackburn, J. K. and Blackman, J. and Blair, C. D. and Blair, D. G. and Blair, R. M. and Bloemen, S. and Bock, O. and Bode, N. and Boer, M. and Boetzel, Y. and Bogaert, G. and Bohe, A. and Bondu, F. and Bonilla, E. and Bonnand, R. and Booker, P. and Boom, B. A. and Booth, C. D. and Bork, R. and Boschi, V. and Bose, S. and Bossie, K. and Bossilkov, V. and Bosveld, J. and Bouffanais, Y. and Bozzi, A. and Bradaschia, C. and Brady, P. R. and Bramley, A. and Branchesi, M. and Brau, J. E. and Briant, T. and Brighenti, F. and Brillet, A. and Brinkmann, M. and Brisson, V. and Brockill, P. and Brooks, A. F. and Brown, D. D. and Brunett, S. and Buchanan, C. C. and Buikema, A. and Bulik, T. and Bulten, H. J. and Buonanno, A. and Buskulic, D. and Buy, C. and Byer, R. L. and Cabero, M. and Cadonati, L. and Cagnoli, G. and Cahillane, C. and Calder{\'o}n Bustillo, J. and Callister, T. A. and Calloni, E. and Camp, J. B. and Canepa, M. and Canizares, P. and Cannon, K. C. and Cao, H. and Cao, J. and Capano, C. D. and Capocasa, E. and Carbognani, F. and Caride, S. and Carney, M. F. and Carullo, G. and Casanueva Diaz, J. and Casentini, C. and Caudill, S. and Cavagli{\`a}, M. and Cavalier, F. and Cavalieri, R. and Cella, G. and Cepeda, C. B. and Cerd{\'a}-Dur{\'a}n, P. and Cerretani, G. and Cesarini, E. and Chaibi, O. and Chamberlin, S. J. and Chan, M. and Chao, S. and Charlton, P. and Chase, E. and Chassande-Mottin, E. and Chatterjee, D. and Chatziioannou, K. and Cheeseboro, B. D. and Chen, H. Y. and Chen, X. and Chen, Y. and Cheng, H.-P. and Chia, H. Y. and Chincarini, A. and Chiummo, A. and Chmiel, T. and Cho, H. S. and Cho, M. and Chow, J. H. and Christensen, N. and Chu, Q. and Chua, A. J. K. and Chua, S. and Chung, K. W. and Chung, S. and Ciani, G. and Ciobanu, A. A. and Ciolfi, R. and Cipriano, F. and Cirelli, C. E. and Cirone, A. and Clara, F. and Clark, J. A. and Clearwater, P. and Cleva, F. and Cocchieri, C. and Coccia, E. and Cohadon, P.-F. and Cohen, D. and Colla, A. and Collette, C. G. and Collins, C. and Cominsky, L. R. and Constancio, M. and Conti, L. and Cooper, S. J. and Corban, P. and Corbitt, T. R. and Cordero-Carri{\'o}n, I. and Corley, K. R. and Cornish, N. and Corsi, A. and Cortese, S. and Costa, C. A. and Cotesta, R. and Coughlin, M. W. and Coughlin, S. B. and Coulon, J.-P. and Countryman, S. T. and Couvares, P. and Covas, P. B. and Cowan, E. E. and Coward, D. M. and Cowart, M. J. and Coyne, D. C. and Coyne, R. and Creighton, J. D. E. and Creighton, T. D. and Cripe, J. and Crowder, S. G. and Cullen, T. J. and Cumming, A. and Cunningham, L. and Cuoco, E. and Canton, T. Dal and D{\'a}lya, G. and Danilishin, S. L. and D'Antonio, S. and Danzmann, K. and Dasgupta, A. and Da Silva Costa, C. F. and Dattilo, V. and Dave, I. and Davier, M. and Davis, D. and Daw, E. J. and Day, B. and DeBra, D. and Deenadayalan, M. and Degallaix, J. and De Laurentis, M. and Del{\'e}glise, S. and Del Pozzo, W. and Demos, N. and Denker, T. and Dent, T. and De Pietri, R. and Derby, J. and Dergachev, V. and De Rosa, R. and De Rossi, C. and DeSalvo, R. and de Varona, O. and Dhurandhar, S. and D{\'i}az, M. C. and Dietrich, T. and Di Fiore, L. and Di Giovanni, M. and Di Girolamo, T. and Di Lieto, A. and Ding, B. and Di Pace, S. and Di Palma, I. and Di Renzo, F. and Dmitriev, A. and Doctor, Z. and Dolique, V. and Donovan, F. and Dooley, K. L. and Doravari, S. and Dorrington, I. and Dovale {\'A}lvarez, M. and Downes, T. P. and Drago, M. and Dreissigacker, C. and Driggers, J. C. and Du, Z. and Dupej, P. and Dwyer, S. E. and Easter, P. J. and Edo, T. B. and Edwards, M. C. and Effler, A. and Eggenstein, H.-B. and Ehrens, P. and Eichholz, J. and Eikenberry, S. S. and Eisenmann, M. and Eisenstein, R. A. and Essick, R. C. and Estelles, H. and Estevez, D. and Etienne, Z. B. and Etzel, T. and Evans, M. and Evans, T. M. and Fafone, V. and Fair, H. and Fairhurst, S. and Fan, X. and Farinon, S. and Farr, B. and Farr, W. M. and Fauchon-Jones, E. J. and Favata, M. and Fays, M. and Fee, C. and Fehrmann, H. and Feicht, J. and Fejer, M. M. and Feng, F. and Fernandez-Galiana, A. and Ferrante, I. and Ferreira, E. C. and Ferrini, F. and Fidecaro, F. and Fiori, I. and Fiorucci, D. and Fishbach, M. and Fisher, R. P. and Fishner, J. M. and Fitz-Axen, M. and Flaminio, R. and Fletcher, M. and Fong, H. and Font, J. A. and Forsyth, P. W. F. and Forsyth, S. S. and Fournier, J.-D. and Frasca, S. and Frasconi, F. and Frei, Z. and Freise, A. and Frey, R. and Frey, V. and Fritschel, P. and Frolov, V. V. and Fulda, P. and Fyffe, M. and Gabbard, H. A. and Gadre, B. U. and Gaebel, S. M. and Gair, J. R. and Gammaitoni, L. and Ganija, M. R. and Gaonkar, S. G. and Garcia, A. and Garc{\'i}a-Quir{\'o}s, C. and Garufi, F. and Gateley, B. and Gaudio, S. and Gaur, G. and Gayathri, V. and Gemme, G. and Genin, E. and Gennai, A. and George, D. and George, J. and Gergely, L. and Germain, V. and Ghonge, S. and Ghosh, Abhirup and Ghosh, Archisman and Ghosh, S. and Giacomazzo, B. and Giaime, J. A. and Giardina, K. D. and Giazotto, A. and Gill, K. and Giordano, G. and Glover, L. and Goetz, E. and Goetz, R. and Goncharov, B. and Gonz{\'a}lez, G. and Gonzalez Castro, J. M. and Gopakumar, A. and Gorodetsky, M. L. and Gossan, S. E. and Gosselin, M. and Gouaty, R. and Grado, A. and Graef, C. and Granata, M. and Grant, A. and Gras, S. and Gray, C. and Greco, G. and Green, A. C. and Green, R. and Gretarsson, E. M. and Groot, P. and Grote, H. and Grunewald, S. and Gruning, P. and Guidi, G. M. and Gulati, H. K. and Guo, X. and Gupta, A. and Gupta, M. K. and Gushwa, K. E. and Gustafson, E. K. and Gustafson, R. and Halim, O. and Hall, B. R. and Hall, E. D. and Hamilton, E. Z. and Hamilton, H. F. and Hammond, G. and Haney, M. and Hanke, M. M. and Hanks, J. and Hanna, C. and Hannam, M. D. and Hannuksela, O. A. and Hanson, J. and Hardwick, T. and Harms, J. and Harry, G. M. and Harry, I. W. and Hart, M. J. and Haster, C.-J. and Haughian, K. and Healy, J. and Heidmann, A. and Heintze, M. C. and Heitmann, H. and Hello, P. and Hemming, G. and Hendry, M. and Heng, I. S. and Hennig, J. and Heptonstall, A. W. and Hernandez, F. J. and Heurs, M. and Hild, S. and Hinderer, T. and Ho, W. C. G. and Hoak, D. and Hochheim, S. and Hofman, D. and Holland, N. A. and Holt, K. and Holz, D. E. and Hopkins, P. and Horst, C. and Hough, J. and Houston, E. A. and Howell, E. J. and Hreibi, A. and Huerta, E. A. and Huet, D. and Hughey, B. and Hulko, M. and Husa, S. and Huttner, S. H. and Huynh-Dinh, T. and Iess, A. and Indik, N. and Ingram, C. and Inta, R. and Intini, G. and Irwin, B. S. and Isa, H. N. and Isac, J.-M. and Isi, M. and Iyer, B. R. and Izumi, K. and Jacqmin, T. and Jani, K. and Jaranowski, P. and Johnson, D. S. and Johnson, W. W. and Jones, D. I. and Jones, R. and Jonker, R. J. G. and Ju, L. and Junker, J. and Kalaghatgi, C. V. and Kalogera, V. and Kamai, B. and Kandhasamy, S. and Kang, G. and Kanner, J. B. and Kapadia, S. J. and Karki, S. and Karvinen, K. S. and Kasprzack, M. and Katolik, M. and Katsanevas, S. and Katsavounidis, E. and Katzman, W. and Kaufer, S. and Kawabe, K. and Keerthana, N. V. and K{\'e}f{\'e}lian, F. and Keitel, D. and Kemball, A. J. and Kennedy, R. and Key, J. S. and Khalili, F. Y. and Khamesra, B. and Khan, H. and Khan, I. and Khan, S. and Khan, Z. and Khazanov, E. A. and Kijbunchoo, N. and Kim, Chunglee and Kim, J. C. and Kim, K. and Kim, W. and Kim, W. S. and Kim, Y.-M. and King, E. J. and King, P. J. and Kinley-Hanlon, M. and Kirchhoff, R. and Kissel, J. S. and Kleybolte, L. and Klimenko, S. and Knowles, T. D. and Koch, P. and Koehlenbeck, S. M. and Koley, S. and Kondrashov, V. and Kontos, A. and Korobko, M. and Korth, W. Z. and Kowalska, I. and Kozak, D. B. and Kr{\"a}mer, C. and Kringel, V. and Krishnan, B. and Kr{\'o}lak, A. and Kuehn, G. and Kumar, P. and Kumar, R. and Kumar, S. and Kuo, L. and Kutynia, A. and Kwang, S. and Lackey, B. D. and Lai, K. H. and Landry, M. and Landry, P. and Lang, R. N. and Lange, J. and Lantz, B. and Lanza, R. K. and Lartaux-Vollard, A. and Lasky, P. D. and Laxen, M. and Lazzarini, A. and Lazzaro, C. and Leaci, P. and Leavey, S. and Lee, C. H. and Lee, H. K. and Lee, H. M. and Lee, H. W. and Lee, K. and Lehmann, J. and Lenon, A. and Leonardi, M. and Leroy, N. and Letendre, N. and Levin, Y. and Li, J. and Li, T. G. F. and Li, X. and Linker, S. D. and Littenberg, T. B. and Liu, J. and Liu, X. and Lo, R. K. L. and Lockerbie, N. A. and London, L. T. and Longo, A. and Lorenzini, M. and Loriette, V. and Lormand, M. and Losurdo, G. and Lough, J. D. and Lousto, C. O. and Lovelace, G. and L{\"u}ck, H. and Lumaca, D. and Lundgren, A. P. and Lynch, R. and Ma, Y. and Macas, R. and Macfoy, S. and Machenschalk, B. and MacInnis, M. and Macleod, D. M. and Maga{\~n}a Hernandez, I. and Maga{\~n}a-Sandoval, F. and Maga{\~n}a Zertuche, L. and Magee, R. M. and Majorana, E. and Maksimovic, I. and Man, N. and Mandic, V. and Mangano, V. and Mansell, G. L. and Manske, M. and Mantovani, M. and Marchesoni, F. and Marion, F. and M{\'a}rka, S. and M{\'a}rka, Z. and Markakis, C. and Markosyan, A. S. and Markowitz, A. and Maros, E. and Marquina, A. and Martelli, F. and Martellini, L. and Martin, I. W. and Martin, R. M. and Martynov, D. V. and Mason, K. and Massera, E. and Masserot, A. and Massinger, T. J. and Masso-Reid, M. and Mastrogiovanni, S. and Matas, A. and Matichard, F. and Matone, L. and Mavalvala, N. and Mazumder, N. and McCann, J. J. and McCarthy, R. and McClelland, D. E. and McCormick, S. and McCuller, L. and McGuire, S. C. and McIver, J. and McManus, D. J. and McRae, T. and McWilliams, S. T. and Meacher, D. and Meadors, G. D. and Mehmet, M. and Meidam, J. and Mejuto-Villa, E. and Melatos, A. and Mendell, G. and Mendoza-Gandara, D. and Mercer, R. A. and Mereni, L. and Merilh, E. L. and Merzougui, M. and Meshkov, S. and Messenger, C. and Messick, C. and Metzdorff, R. and Meyers, P. M. and Miao, H. and Michel, C. and Middleton, H. and Mikhailov, E. E. and Milano, L. and Miller, A. L. and Miller, A. and Miller, B. B. and Miller, J. and Millhouse, M. and Mills, J. and Milovich-Goff, M. C. and Minazzoli, O. and Minenkov, Y. and Ming, J. and Mishra, C. and Mitra, S. and Mitrofanov, V. P. and Mitselmakher, G. and Mittleman, R. and Moffa, D. and Mogushi, K. and Mohan, M. and Mohapatra, S. R. P. and Montani, M. and Moore, C. J. and Moraru, D. and Moreno, G. and Morisaki, S. and Mours, B. and Mow-Lowry, C. M. and Mueller, G. and Muir, A. W. and Mukherjee, Arunava and Mukherjee, D. and Mukherjee, S. and Mukund, N. and Mullavey, A. and Munch, J. and Mu{\~n}iz, E. A. and Muratore, M. and Murray, P. G. and Nagar, A. and Napier, K. and Nardecchia, I. and Naticchioni, L. and Nayak, R. K. and Neilson, J. and Nelemans, G. and Nelson, T. J. N. and Nery, M. and Neunzert, A. and Nevin, L. and Newport, J. M. and Ng, K. Y. and Ng, S. and Nguyen, P. and Nguyen, T. T. and Nichols, D. and Nielsen, A. B. and Nissanke, S. and Nitz, A. and Nocera, F. and Nolting, D. and North, C. and Nuttall, L. K. and Obergaulinger, M. and Oberling, J. and O'Brien, B. D. and O'Dea, G. D. and Ogin, G. H. and Oh, J. J. and Oh, S. H. and Ohme, F. and Ohta, H. and Okada, M. A. and Oliver, M. and Oppermann, P. and Oram, Richard J. and O'Reilly, B. and Ormiston, R. and Ortega, L. F. and O'Shaughnessy, R. and Ossokine, S. and Ottaway, D. J. and Overmier, H. and Owen, B. J. and Pace, A. E. and Pagano, G. and Page, J. and Page, M. A. and Pai, A. and Pai, S. A. and Palamos, J. R. and Palashov, O. and Palomba, C. and Pal-Singh, A. and Pan, Howard and Pan, Huang-Wei and Pang, B. and Pang, P. T. H. and Pankow, C. and Pannarale, F. and Pant, B. C. and Paoletti, F. and Paoli, A. and Papa, M. A. and Parida, A. and Parker, W. and Pascucci, D. and Pasqualetti, A. and Passaquieti, R. and Passuello, D. and Patil, M. and Patricelli, B. and Pearlstone, B. L. and Pedersen, C. and Pedraza, M. and Pedurand, R. and Pekowsky, L. and Pele, A. and Penn, S. and Perego, A. and Perez, C. J. and Perreca, A. and Perri, L. M. and Pfeiffer, H. P. and Phelps, M. and Phukon, K. S. and Piccinni, O. J. and Pichot, M. and Piergiovanni, F. and Pierro, V. and Pillant, G. and Pinard, L. and Pinto, I. M. and Pirello, M. and Pitkin, M. and Poggiani, R. and Popolizio, P. and Porter, E. K. and Possenti, L. and Post, A. and Powell, J. and Prasad, J. and Pratt, J. W. W. and Pratten, G. and Predoi, V. and Prestegard, T. and Principe, M. and Privitera, S. and Prodi, G. A. and Prokhorov, L. G. and Puncken, O. and Punturo, M. and Puppo, P. and P{\"u}rrer, M. and Qi, H. and Quetschke, V. and Quintero, E. A. and Quitzow-James, R. and Raab, F. J. and Rabeling, D. S. and Radkins, H. and Raffai, P. and Raja, S. and Rajan, C. and Rajbhandari, B. and Rakhmanov, M. and Ramirez, K. E. and Ramos-Buades, A. and Rana, Javed and Rapagnani, P. and Raymond, V. and Razzano, M. and Read, J. and Regimbau, T. and Rei, L. and Reid, S. and Reitze, D. H. and Ren, W. and Ricci, F. and Ricker, P. M. and Riemenschneider, G. M. and Riles, K. and Rizzo, M. and Robertson, N. A. and Robie, R. and Robinet, F. and Robson, T. and Rocchi, A. and Rolland, L. and Rollins, J. G. and Roma, V. J. and Romano, R. and Romel, C. L. and Romie, J. H. and Rosi{\'n}ska, D. and Ross, M. P. and Rowan, S. and R{\"u}diger, A. and Ruggi, P. and Rutins, G. and Ryan, K. and Sachdev, S. and Sadecki, T. and Sakellariadou, M. and Salconi, L. and Saleem, M. and Salemi, F. and Samajdar, A. and Sammut, L. and Sampson, L. M. and Sanchez, E. J. and Sanchez, L. E. and Sanchis-Gual, N. and Sandberg, V. and Sanders, J. R. and Sarin, N. and Sassolas, B. and Sathyaprakash, B. S. and Saulson, P. R. and Sauter, O. and Savage, R. L. and Sawadsky, A. and Schale, P. and Scheel, M. and Scheuer, J. and Schmidt, P. and Schnabel, R. and Schofield, R. M. S. and Sch{\"o}nbeck, A. and Schreiber, E. and Schuette, D. and Schulte, B. W. and Schutz, B. F. and Schwalbe, S. G. and Scott, J. and Scott, S. M. and Seidel, E. and Sellers, D. and Sengupta, A. S. and Sentenac, D. and Sequino, V. and Sergeev, A. and Setyawati, Y. and Shaddock, D. A. and Shaffer, T. J. and Shah, A. A. and Shahriar, M. S. and Shaner, M. B. and Shao, L. and Shapiro, B. and Shawhan, P. and Shen, H. and Shoemaker, D. H. and Shoemaker, D. M. and Siellez, K. and Siemens, X. and Sieniawska, M. and Sigg, D. and Silva, A. D. and Singer, L. P. and Singh, A. and Singhal, A. and Sintes, A. M. and Slagmolen, B. J. J. and Slaven-Blair, T. J. and Smith, B. and Smith, J. R. and Smith, R. J. E. and Somala, S. and Son, E. J. and Sorazu, B. and Sorrentino, F. and Souradeep, T. and Spencer, A. P. and Srivastava, A. K. and Staats, K. and Steinke, M. and Steinlechner, J. and Steinlechner, S. and Steinmeyer, D. and Steltner, B. and Stevenson, S. P. and Stocks, D. and Stone, R. and Stops, D. J. and Strain, K. A. and Stratta, G. and Strigin, S. E. and Strunk, A. and Sturani, R. and Stuver, A. L. and Summerscales, T. Z. and Sun, L. and Sunil, S. and Suresh, J. and Sutton, P. J. and Swinkels, B. L. and Szczepa{\'n}czyk, M. J. and Tacca, M. and Tait, S. C. and Talbot, C. and Talukder, D. and Tanner, D. B. and T{\'a}pai, M. and Taracchini, A. and Tasson, J. D. and Taylor, J. A. and Taylor, R. and Tewari, S. V. and Theeg, T. and Thies, F. and Thomas, E. G. and Thomas, M. and Thomas, P. and Thorne, K. A. and Thrane, E. and Tiwari, S. and Tiwari, V. and Tokmakov, K. V. and Toland, K. and Tonelli, M. and Tornasi, Z. and Torres-Forn{\'e}, A. and Torrie, C. I. and T{\"o}yr{\"a}, D. and Travasso, F. and Traylor, G. and Trinastic, J. and Tringali, M. C. and Trovato, A. and Trozzo, L. and Tsang, K. W. and Tse, M. and Tso, R. and Tsuna, D. and Tsukada, L. and Tuyenbayev, D. and Ueno, K. and Ugolini, D. and Urban, A. L. and Usman, S. A. and Vahlbruch, H. and Vajente, G. and Valdes, G. and van Bakel, N. and van Beuzekom, M. and van den Brand, J. F. J. and Van Den Broeck, C. and Vander-Hyde, D. C. and van der Schaaf, L. and van Heijningen, J. V. and van Veggel, A. A. and Vardaro, M. and Varma, V. and Vass, S. and Vas{\'u}th, M. and Vecchio, A. and Vedovato, G. and Veitch, J. and Veitch, P. J. and Venkateswara, K. and Venugopalan, G. and Verkindt, D. and Vetrano, F. and Vicer{\'e}, A. and Viets, A. D. and Vinciguerra, S. and Vine, D. J. and Vinet, J.-Y. and Vitale, S. and Vo, T. and Vocca, H. and Vorvick, C. and Vyatchanin, S. P. and Wade, A. R. and Wade, L. E. and Wade, M. and Walet, R. and Walker, M. and Wallace, L. and Walsh, S. and Wang, G. and Wang, H. and Wang, J. Z. and Wang, W. H. and Wang, Y. F. and Ward, R. L. and Warner, J. and Was, M. and Watchi, J. and Weaver, B. and Wei, L.-W. and Weinert, M. and Weinstein, A. J. and Weiss, R. and Wellmann, F. and Wen, L. and Wessel, E. K. and We{\ss}els, P. and Westerweck, J. and Wette, K. and Whelan, J. T. and Whiting, B. F. and Whittle, C. and Wilken, D. and Williams, D. and Williams, R. D. and Williamson, A. R. and Willis, J. L. and Willke, B. and Wimmer, M. H. and Winkler, W. and Wipf, C. C. and Wittel, H. and Woan, G. and Woehler, J. and Wofford, J. K. and Wong, W. K. and Worden, J. and Wright, J. L. and Wu, D. S. and Wysocki, D. M. and Xiao, S. and Yam, W. and Yamamoto, H. and Yancey, C. C. and Yang, L. and Yap, M. J. and Yazback, M. and Yu, Hang and Yu, Haocun and Yvert, M. and Zadro{\.Z}ny, A. and Zanolin, M. and Zelenova, T. and Zendri, J.-P. and Zevin, M. and Zhang, J. and Zhang, L. and Zhang, M. and Zhang, T. and Zhang, Y.-H. and Zhao, C. and Zhou, M. and Zhou, Z. and Zhu, S. J. and Zhu, X. J. and Zimmerman, A. B. and Zlochower, Y. and Zucker, M. E. and Zweizig, J. and {LIGO Scientific Collaboration} and {Virgo Collaboration}},
	month = oct,
	year = {2018},
	pages = {161101}
}

@article{most_new_2018,
	title = {New {Constraints} on {Radii} and {Tidal} {Deformabilities} of {Neutron} {Stars} from {GW170817}},
	volume = {120},
	issn = {0031-9007},
	url = {http://adsabs.harvard.edu/abs/2018PhRvL.120z1103M},
	doi = {10.1103/PhysRevLett.120.261103},
	urldate = {2020-02-16},
	journal = {Physical Review Letters},
	author = {Most, Elias R. and Weih, Lukas R. and Rezzolla, Luciano and Schaffner-Bielich, J{\"u}rgen},
	month = jun,
	year = {2018},
	pages = {261103}
}

\end{document}